%% file: HotQCDWhitePaper.tex
\documentclass[12pt]{article}      	

\usepackage{graphicx}
\usepackage{amsmath}
\usepackage{amsfonts}
\usepackage{amssymb}
\usepackage{bm}
\usepackage{cite}
\usepackage{microtype}
\usepackage{parskip}
\usepackage{caption}
\usepackage{subcaption}
\usepackage{enumitem}
\usepackage{float}
\usepackage{color}
\usepackage{fancyhdr}

\usepackage[pagebackref=true,colorlinks=true]{hyperref}
\hypersetup{
     unicode=true,          
     pdfnewwindow=true,      
     bookmarksnumbered=true,
     colorlinks=true,       
     linkcolor=blue,          
     citecolor=darkgreen,        
     filecolor=magenta,      
     urlcolor=cyan,           
     menucolor=blue,
     linktocpage=true
}

\input{custom_defs.tex}

\begin{document}

\centerline{\Large \bf The Hot QCD White Paper:}
\vspace{5mm}
\centerline{\Large \bf Exploring the Phases of QCD at RHIC and the LHC}
\vspace*{10mm}
\begin{center}
\large
 A White Paper on Future Opportunities in Relativistic Heavy Ion Physics
\vspace{5mm}
\hrule
\vspace{22mm}
\normalsize
\today
\end{center}

\vspace*{3cm}
\begin{abstract}
The past decade has seen huge advances in experimental measurements made in heavy ion collisions at the Relativistic Heavy Ion Collider (RHIC) and more recently at the Large Hadron Collider (LHC).
These new data, in combination with theoretical advances from calculations made in a variety of frameworks, have led to a broad and deep knowledge of the properties of thermal QCD matter. 
Increasingly quantitative descriptions of the
quark-gluon plasma (QGP) created in these collisions have established that the QGP is a strongly coupled liquid with the lowest value of specific viscosity ever measured.
However, much remains to be learned about the precise nature of the initial state from which this liquid forms, how its properties vary across its phase diagram and how, at a microscopic level, the collective properties of this liquid emerge from the interactions among the individual quarks and gluons that must be visible if the liquid is probed with sufficiently high resolution. This white paper, prepared by the Hot QCD Writing Group as part of the U.S. Long Range Plan for Nuclear Physics, reviews the recent progress in the field of hot QCD and outlines the scientific opportunities in the next decade for resolving the outstanding issues in the field.
\end{abstract}

\pagebreak

\input{tex/Authors.tex}

\eject

\definecolor{darkgreen}{rgb}{0,.7,0}.

\vspace*{-1.5cm}                                                       
\renewcommand{\baselinestretch}{0.90}\normalsize 
\tableofcontents
\renewcommand{\baselinestretch}{1.0}\normalsize
\eject

\input{tex/Acronyms}
\eject

\listoffigures
\addcontentsline{toc}{section}{\protect\numberline{}List of Figures}
\eject

\pagestyle{fancy}
\renewcommand{\sectionmark}[1]{ \markright{#1}{} }

\input{tex/ExecutiveSummary}

\eject

\input{tex/Introduction}

\input{tex/ProgressAndStatus}

\input{tex/FutureProspects}

\pagebreak
\input{tex/Summary.tex}

\pagebreak
\input{tex/Acknowledgements.tex}

\clearpage
\addcontentsline{toc}{section}{References}
\bibliographystyle{tex/atlasnote}  
\bibliography{HotQCDWhitePaper}

\end{document}

%% file: custom_defs.tex
\newcommand{\pp}{\mbox{p+p}}
\newcommand{\AplusA}{\mbox{A+A}}
\newcommand{\pA}{\mbox{p+A}}
\newcommand{\AuAu}{\mbox{Au+Au}}
\newcommand{\HeAu}{\mbox{${}^3$He+Au}}
\newcommand{\CuAu}{\mbox{Cu+Au}}
\newcommand{\dA}{\mbox{d+A}}
\newcommand{\dAu}{\mbox{d+Au}}
\newcommand{\pAl}{\mbox{p+Al}}
\newcommand{\pAu}{\mbox{p+Au}}
\newcommand{\UU}{\mbox{${}^{238}$U+${}^{238}$}}
\newcommand{\PbPb}{\mbox{Pb+Pb}}
\newcommand{\pPb}{\mbox{p+Pb}}

\newcommand {\rootsNN}  {\ensuremath{\sqrt{s_{_{\rm NN}}}}}

\newcommand {\GeV}       {\ensuremath{\,\mathrm{GeV}}}

\newcommand{\Pgg}{\ensuremath{\gamma}}

\newcommand {\ptg}    {\ensuremath{p_{\mathrm{T}}^{\Pgg}}}

\providecommand{\ptg}{\ensuremath{p_{\mathrm{T},\Pgg}}}

\newcommand{\Jpsi}{J/$\psi$}

\newcommand{\JPsi}{J/$\psi$}

\newcommand{\qs}{\ensuremath{Q_{\mathrm{s}}}}
\newcommand{\qssqr}{\ensuremath{Q_{\mathrm{s}}^2}}
\newcommand{\as}{\ensuremath{\alpha_{\mathrm{s}}}}
\newcommand{\qhat}{\ensuremath{\hat{q}}}
\newcommand{\ehat}{\ensuremath{\hat{e}}}
\newcommand {\pT} {\ensuremath{p_{\rm{T}}}}
\newcommand {\pt} {\ensuremath{p_{\rm{T}}}}

%% file: tex/Authors.tex
{\bf \large Principal Authors, representing the US Heavy-Ion Community:}
\vspace{5mm}

\begin{minipage}[t]{0.5\linewidth}

{\bf Yasuyuki Akiba}\\
RIKEN Nishina Center\\ 
RIKEN\\

{\bf Aaron Angerami}\\
Department of Physics\\
Columbia University\\

{\bf Helen Caines}\\
Department of Physics\\
Yale University\\

{\bf Anthony Frawley}\\
Physics Department\\
Florida State University\\

{\bf  Ulrich Heinz}\\
Department of Physics\\
Ohio State University\\

{\bf Barbara Jacak}\\
Physics Department\\
University of California, Berkeley\\

{\bf  Jiangyong Jia}\\
Chemistry Department\\
Stony Brook University\\

{\bf Tuomas Lappi}\\
Department of Physics\\
Jyv\"askyl\"a University\\

{\bf Wei Li}\\
Physics Department\\
Rice University\\

{\bf Abhijit Majumder}\\
Department of Physics\\
Wayne State University\\

\end{minipage}
\hfill
\begin{minipage}[t]{0.5\linewidth}

{\bf David Morrison}\\
Physics Department\\
Brookhaven National Laboratory\\

{\bf Mateusz Ploskon}\\
Nuclear Science Division\\
Lawrence Berkeley National Laboratory\\

{\bf Joern Putschke}\\
Department of Physics\\
Wayne State University\\

{\bf Krishna Rajagopal}\\
Physics Department\\
Massachusetts Institute of Technology\\

{\bf Ralf Rapp}\\
Physics Department\\
Texas A\&M University\\

{\bf Gunther Roland}\\
Physics Department\\
Massachusetts Institute of Technology\\

{\bf Paul Sorensen}\\
Physics Department\\
Brookhaven National Laboratory\\

{\bf Urs Wiedemann}\\
Theory Division\\
CERN\\

{\bf Nu Xu}\\
Nuclear Science Division\\
Lawrence Berkeley National Laboratory\\

{\bf W.A. Zajc}\footnote{Writing Committee Chair}\\
Department of Physics\\
Columbia University\\

\end{minipage}

%% file: tex/Acronyms.tex
\section*{\center{Acronyms and Abbreviations}}
\addcontentsline{toc}{section}{\protect\numberline{}Acronyms and Abbreviations}
\label{Sec:AandA}

Acronyms and abbreviations used in this white paper:
\begin{itemize}

\item[\bf AdS:] Anti de Sitter space
\item[\bf ALICE:] A Large Ion Collider Experiment (at the LHC)
\item[\bf AMPT:] A Multi-Phase Transport (theoretical model)
\item[\bf ASW:] Armesto, Salgado, and Wiedemann (theoretical formalism)
\item[\bf ATLAS:] A Toroidal LHC ApparatuS (experiment at the LHC)
\item[\bf BES:] Beam Energy Scan (at RHIC)
\item[\bf BNL:] Brookhaven National Laboratory
\item[\bf BT:] Braaten-Thoma (as used in Figure~\ref{Fig:non-photonic-suppression})
\item[\bf BW:] Brookhaven-Wuppertal (theoretical collaboration)
\item[\bf CERN:] Historical acronym in current use for European Organization for Nuclear Research
\item[\bf CFT:] Conformal Field Theory
\item[\bf CGC:] Color Glass Condensate
\item[\bf CL:] Confidence Limit
\item[\bf CME:] Chiral Magnetic Effect
\item[\bf CMS:] Compact Muon System (experiment at the LHC)
\item[\bf CNM:] Cold Nuclear Matter
\item[\bf CUJET:] Columbia University Jet and Electromagnetic Tomography (theoretical formalism)
\item[\bf CVE:] Chiral Vortical Effect
\item[\bf DOE:] Department of Energy
\item[\bf EIC:] Electron Ion Collider
\item[\bf EM:] ElectroMagnetic (radiation)
\item[\bf EMCAL:] ElectroMagnetic CALorimeter
\item[\bf EOS:] Equation Of State
\item[\bf EPD:] Event Plane Detector (STAR)
\item[\bf FCS:] Forward Calorimetric System (STAR)
\item[\bf FF:] Fragmentation Function
\item[\bf FRIB:] Facility for Rare Isotope Beams
\item[\bf FTS:] Forward Tracking System (STAR)
\item[\bf GEM:] Gas Electron Multiplier
\item[\bf GLV:] Gyulassy-Levai-Vitev (theoretical formalism)
\item[\bf HFT:] Heavy Flavor Tracker (STAR)
\item[\bf HQ:] Heavy Quark
\item[\bf HT-BW:] Higher Twist Berkeley-Wuhan (theoretical formalism)
\item[\bf HT-M:] Higher Twist Majumder (theoretical formalism)
\item[\bf IP-Glasma:] Impact Parameter dependent saturation - color Glass condensate plasma
\item[\bf iTPC:] Inner TPC (STAR)
\item[\bf JET:] Jet and Electromagnetic Tomography (theory-experiment topical group)
\item[\bf JEWEL:] Jet Evolution With Energy Loss (Monte Carlo model)
\item[\bf LANL:] Los Alamos National Laboratory
\item[\bf LHC:] Large Hadron Collider
\item[\bf LO:] Leading Order (in QCD)
\item[\bf L1:] Level 1 (experimental trigger level)
\item[\bf LS1:] Long Shutdown 1 (LHC)
\item[\bf LS2:] Long Shutdown 2 (LHC)
\item[\bf MADAI:] Models and Data Initiative (experimental/ theoretical collaboration)
\item[\bf MARTINI:] Modular Algorithm for Relativistic Treatment of heavy IoN Interactions (Monte Carlo model)
\item[\bf MIE:] Major Item of Equipment (DOE)
\item[\bf MTD:] Muon Telescope Detector (STAR)
\item[\bf MUSIC:] Monotonic Upstream-centered Scheme for Conservation laws for Ion Collisions
\item[\bf NASA:] National Aeronautics and Space Administration
\item[\bf NLO:] Next to Leading Order (in QCD)
\item[\bf PHENIX:]  Pioneering High Energy Nuclear Interaction experiment (at RHIC)
\item[\bf pQCD:] perturbative QCD
\item[\bf PS:] Proton Synchrotron (CERN)
\item[\bf PYQUEN:] PYthia QUENched (Monte Carlo model)
\item[\bf QCD:] Quantum Chromodynamics
\item[\bf QM:] Qin-Majumder (as used in Figure~\ref{Fig:non-photonic-suppression})
\item[\bf RHIC:] Relativistic Heavy Ion Collider
\item[\bf SCET:] Soft Collinear Effective Theory 
\item[\bf sPHENIX:] Super PHENIX
\item[\bf SLAC:] Stanford Linear ACcelerator 
\item[\bf SM:] Standard Model
\item[\bf SPS:] Super Proton Synchrotron (CERN)
\item[\bf STAR:]  Solenoidal Tracker at RHIC
\item[\bf TAMU:] Texas A\&M University
\item[\bf TECHQM:] Theory meets Experiment Collaboration in Hot QCD Matter
\item[\bf TG:] Thoma-Gyulassy (as used in Figure~\ref{Fig:non-photonic-suppression})
\item[\bf TPC:] Time Projection Chamber
\item[\bf VTX:] (silicon) VerTeX tracker (PHENIX)
\item[\bf YaJEM:] Yet another Jet Energy-loss Model
\item[\bf WHDG:]  Wicks, Horowitz, Djordjevic, and Gyulassy (theoretical formalism)

\end{itemize}

%% file: tex/ExecutiveSummary.tex
\section{Executive Summary}
\label{Sec:ExecSummary}

\noindent 
Over the past decade a panoply of measurements made in heavy ion collisions at the Relativistic Heavy Ion Collider (RHIC) and the Large Hadron Collider (LHC), combined with theoretical advances from calculations made in a variety frameworks, have led to a broad and deep knowledge of the properties of hot QCD matter. 
However, this recently established knowledge of what thermal QCD matter does in turn raises 
new questions about how QCD works in this environment. 
High energy nuclear collisions create exploding little droplets of the hottest matter seen anywhere in the universe since it was a few microseconds old. We have increasingly quantitative empirical descriptions of the phenomena manifest in these explosions, and of some key material properties of the matter created in these ``Little Bangs''. In particular, we have determined that the quark-gluon plasma (QGP) created in these collisions is a strongly coupled liquid with the lowest value of specific viscosity ever measured. However, we do not know the precise nature of the initial state from which this liquid forms, and know very little about how the properties of this liquid vary across its phase diagram or how, at a microscopic level, the collective properties of this liquid emerge from the interactions among the individual quarks and gluons that we know must be visible if the liquid is probed with sufficiently high resolution. These findings lead us to the following recommendations:

\noindent {\bf \large Recommendation \#1:}

\noindent
{\bf 
The discoveries of the past decade have posed or sharpened questions that are central to understanding the nature, structure, and origin of the hottest liquid form of matter that the universe has ever seen. As our highest priority we recommend a program to complete the search for the critical point in the QCD phase diagram and to exploit the newly realized potential of exploring the QGPÕs  structure at multiple length scales with jets at RHIC and LHC energies. This requires 
\begin{itemize}
\item implementation of new capabilities of the RHIC facility needed to complete its scientific mission: a state-of-the-art jet detector such as sPHENIX and luminosity upgrades for running at low energies,
\item continued strong U.S. participation in the LHC heavy-ion program, and 
\item strong investment  in a broad range of theoretical efforts employing various analytical and computational methods.
\end{itemize}
}

The goals of this program are to 1) measure the temperature and chemical potential dependence of transport properties especially near the phase boundary, 2) explore the phase structure of the nuclear matter phase diagram, 3) probe the microscopic picture of the perfect liquid, and 4) image the high density gluon fields of the incoming nuclei and study their fluctuation spectrum.
These efforts will firmly establish our understanding of thermal QCD matter over a broad range of temperature. However, the precise mechanism by which those temperatures are achieved beginning from the ground state of nuclear matter requires a new initiative dedicated to the study of dense gluon fields in nuclei, which leads to our second recommendation\footnote{This recommendation is made jointly with the Cold QCD Working Group}:

\noindent {\bf \large Recommendation \#2:}

\noindent
{\bf 
A high luminosity, high-energy polarized Electron Ion Collider (EIC) is the U.S. QCD Community's highest priority for future construction
after FRIB.}

The EIC will, for the first time, precisely image the gluons and sea quarks in the proton and nuclei, resolve the protonÕs internal structure including the origin of its spin, and explore a new QCD frontier of ultra-dense gluon fields in nuclei at high energy. These advances are made possible by the EICÕs unique capability to collide polarized electrons with polarized protons and light ions at unprecedented luminosity and with heavy nuclei at high energy. The EIC is absolutely essential to maintain U.S. leadership in fundamental nuclear physics research in the coming decades.

%% file: tex/Introduction.tex
\section{Introduction}
\label{Sec:Introduction}

QCD theory and modeling, benefitting from continuous experimental guidance, have led to the development of a standard model to describe the dynamic space-time evolution of the Little Bangs created in high energy nuclear collisions\cite{Heinz:2013wva}. Collective behavior observed via correlations among the particles produced in the debris of these explosions led to the discovery that, soon after the initial collision, dense QCD matter thermalizes with a very high initial temperature and forms a strongly coupled quark-gluon plasma (QGP). Surprisingly, this matter that filled the early universe turns out to be a liquid with a specific viscosity (the ratio of viscosity to entropy density)  $\eta/s$ smaller than that of any other known substance\cite{Csernai:2006zz,Gale:2012rq} and very near a limiting value for this quantity that is characteristic of plasmas in infinitely strongly interacting gauge theories with a dual gravitational description\cite{Kovtun:2004de}.

Despite continued progress, estimates of the $\eta/s$ of QGP generally remain upper bounds, due to systematic uncertainties arising from an incomplete knowledge of the initial state. However, the unanticipated recent discovery that ripples in the near-perfect QGP liquid bring information about nucleonic and sub- nucleonic gluon fluctuations in the initial state into the final state\cite{Gale:2012in} has opened new possibilities to study the dense gluon fields and their quantum fluctuations in the colliding nuclei via correlations between final state particles. Mapping the transverse and longitudinal dependence of the initial gluon fluctuation spectrum will provide both a test for QCD calculations in a high gluon density regime as well as the description of the initial state necessary to further improve the determination of $\eta/s$.

Understanding strongly coupled or strongly correlated systems is at the intellectual forefront of multiple areas of physics. One example is the physics of ultra-cold fermionic atoms, where application of a magnetic field excites a strong resonance. In an atomic trap, such atoms form a degenerate Fermi liquid, which can be studied in exquisite detail\cite{O'Hara:2002zz}. At temperatures below $\sim 0.1\ \mu\mathrm{K}$ the atoms interact via a Feshbach resonance to form a superfluid\cite{Kinast:2004zza}. Strongly correlated electron systems in condensed matter provide another strongly coupled system\cite{Rameau:2014gma}. Here, the elementary interaction is not strong, but its role is amplified by the large number of interacting particles and their ability to dynamically correlate their quantum wave functions. In conventional plasma physics strong coupling is observed in warm, dense matter and dusty plasmas\cite{Chan:2004} residing in astrophysical environments, such as the rings of Saturn, as well as in thermonuclear fusion.

A unique feature of QCD matter compared to the other strongly coupled systems is that the interaction is specified directly at the Lagrangian level by a fundamental theory. Thus we have a chance to understand how a strongly coupled fluid emerges from a microscopic theory that is precisely known. The high temperature achieved in nuclear collisions, combined with the precision achieved by the numerical solutions of lattice gauge theory, permits {\em ab initio} calculation of equilibrium properties of hot QCD matter without any model assumptions or approximations.

At the same time, the discovery of strongly coupled QGP poses many questions. How do its properties vary over a broad range of temperature and chemical potential?
Is there a critical point in the QCD phase diagram where the hadron gas to QGP phase transition becomes first-order?  What is the smallest droplet of hot QCD matter whose behavior is liquid-like? What are the initial conditions that lead to hydrodynamic behavior, and can they be extracted from the experimental data? Can the underlying degrees of freedom in the liquid be resolved with jets and heavy flavor probes? An understanding of how the properties of the liquid emerge requires probing the matter at varied, more microscopic, length scales than those studied to date, as well as examining its production in systems of different size over a range of energies.

Answering these and other questions will depend also on an intensive modeling and computational effort to
simultaneously determine the set of key parameters needed for a multi-scale characterization of the QGP medium as well as the initial state from which it emerges. This phenomenological effort requires broad experimental input from a diverse set of measurements, including but not limited to
\begin{enumerate}

\item A more extensive set of heavy quark measurements to determine the diffusion coefficient of heavy quarks.

\item Energy scans to map the phase diagram of QCD and the dependence of transport coefficients on the temperature and chemical potential.

\item  Collisions of nuclei with varied sizes, including p+A and very high multiplicity p+p collisions, to study the emergence of collective phenomena and parton interactions with the nuclear medium.

\item The quantitative characterization of the electromagnetic radiation emitted by the Little Bangs and its spectral anisotropies.

\item A search for chiral symmetry restoration through measurements of lepton pair masses.

\item Detailed investigation of medium effects on the production rates and internal structure of jets of hadrons, for multi-scale tomographic studies of the medium. 
\end{enumerate}
This program will illuminate how ``more'' becomes ``different'' in matter governed by the equations of QCD.

RHIC and the LHC, together provide an unprecedented opportunity to pursue the program listed above and to thereby resolve the open questions in thermal properties of QCD matter. While collisions at the LHC create temperatures well above those needed for the creation of QGP and may thus be able to explore the expected transition from a strongly coupled liquid to a weakly coupled gaseous phase at higher temperatures, the RHIC program uniquely enables  research at temperatures close to the phase transition where the coupling is strongest. Moreover, the unparalleled flexibility of RHIC provides collisions between a variety of different ion species over a broad range in energy. The combined programs permit a comprehensive exploration of the QCD phase diagram, together with detailed studies of how initial conditions affect the creation and dynamical expansion of hot QCD matter and of the microscopic structure of the strongly coupled QGP liquid.

This varied program at these two colliders, covering three orders of magnitude in center of mass energy, has already led to an array of paradigmatic discoveries. Asymmetric \CuAu\ and \HeAu\ collisions\cite{Huang:2012sc}, and collisions between deformed uranium nuclei at RHIC\cite{Wang:2014qxa} are helping to constrain the initial fluctuation spectrum and to eliminate some initial energy deposition models. The recently discovered unexpected collectivity of anisotropic flow signatures observed in
\pPb\ collisions at the LHC\cite{Abelev:2012ola,Aad:2013fja} suggests that similar signatures seen in very-high-multiplicity \pp\ collisions at the LHC\cite{Khachatryan:2010gv} and in a recent re-analysis of 
\dAu\ collisions at RHIC\cite{Adare:2013piz} might also be of collective origin. How collectivity develops in such small systems cries out for explanation. 
The unavoidable question ``What is the smallest size and density of a droplet of QCD matter that behaves like a liquid?'' can only be answered by systematically exploiting RHIC's flexibility to collide atomic nuclei of any size over a wide range of energies.
The additional running in 2015 of \pAu\ and \pAl\ collisions at RHIC will augment the \dAu\ and \HeAu\ data and represents 
an important first step in this direction.

Future precision measurements made possible by increases in the energy and luminosity of the LHC will quantify thermodynamic and microscopic properties of the strongly coupled plasma at temperatures well above the transition temperature $T_C$. At the same time, RHIC  will be the only facility capable of both of providing the experimental lever arm needed to establish the temperature dependence of these parameters while also extending present knowledge of the properties of deconfined matter to larger values of the baryon chemical potential $\mu_B$, where a critical point and first
order phase transition may be awaiting discovery. There is no single facility in the short- or long-term future that could come close to duplicating what RHIC and the LHC, operating in concert, will teach us about Nature.

Both RHIC and the LHC are also capable of probing new, unmeasured physics phenomena at low longitudinal momentum fraction $x$. Proton-lead collisions at the LHC allow the study of previously unreachable regions of phase space in the search for parton saturation effects. However, a complete exploration of parton dynamics at low $x$ will require an Electron-Ion Collider (EIC). 
Accordingly, a cost-effective plan has been developed for a future transition of the RHIC facility to an EIC. 
While forward rapidity studies in \pA\ and \AplusA\ collisions at RHIC and the LHC can provide access to low-$x$ physics in a complementary kinematic range,
an EIC will be needed to deliver crucially missing precise information on the nuclear parton distribution functions within the most desirable kinematic regime. However, this white paper focuses on the future opportunities in the realm of thermal QCD, noting where relevant the need for our knowledge to be extended by the compelling insights to be derived from a future EIC\cite{Accardi:2012qut}.

The remainder of this white paper presents experimental and theoretical progress since the last Long Range Plan in Section~\ref{Sec:Progress} in order to assess the current status of the field and its facilities. Section~\ref{Sec:Future} then describes future prospects for advancing our understanding of thermal QCD and thereby addressing the deep intellectual questions posed by recent discoveries.

%% file: tex/ProgressAndStatus.tex
\section{Progress Since the Previous Long-Range Plan; Current Status}
\label{Sec:Progress}
Enormous progress has been made in the experimental and theoretical study of hot QCD matter since the 
2007 Long Range Plan\cite{LRP:2007}. A new energy frontier at the LHC, in concert with a wide variety
of experimental tools honed at RHIC, enabled rapid yet refined
analyses of the highest energy heavy ion collisions studied to date.
At the same time, upgrades to both RHIC and its experiments allowed
not only exploration of a new low-energy regime but also greatly improved
experimental sensitivity at RHIC's top energy. In combination with increasingly sophisticated
theoretical models, these developments have led to new insights into the behavior
of nuclear matter under extreme conditions. This section details
the advances in both the experimental facilities
and the results derived from them since the last Long Range Plan.

\input{tex/Facilities.tex}

\input{tex/Flow.tex}

\input{tex/Jets.tex}

\input{tex/Saturation}

\input{tex/HFandQuarkonia.tex}

\input{tex/Dileptons.tex}

\input{tex/EnergyScan.tex}

\input{tex/Exotica}

\input{tex/CrossFertilization}

%% file: tex/Facilities.tex
\subsection{Facilities Status}
\label{Sec:Facilities}
Collisions of heavy nuclei at ultra-relativistic energies are studied at two facilities: the Relativistic Heavy Ion Collider (RHIC) at Brookhaven National Laboratory, and (since 2010) the Large Hadron Collider (LHC) in Geneva, Switzerland. The impressive experimental progress in the field since the 2007 Long Range Plan that is presented in what follows has been made possible by the outstanding performance of these two facilities. 

\begin{figure}[ht]
\centerline{
\includegraphics[width=0.60\textwidth]{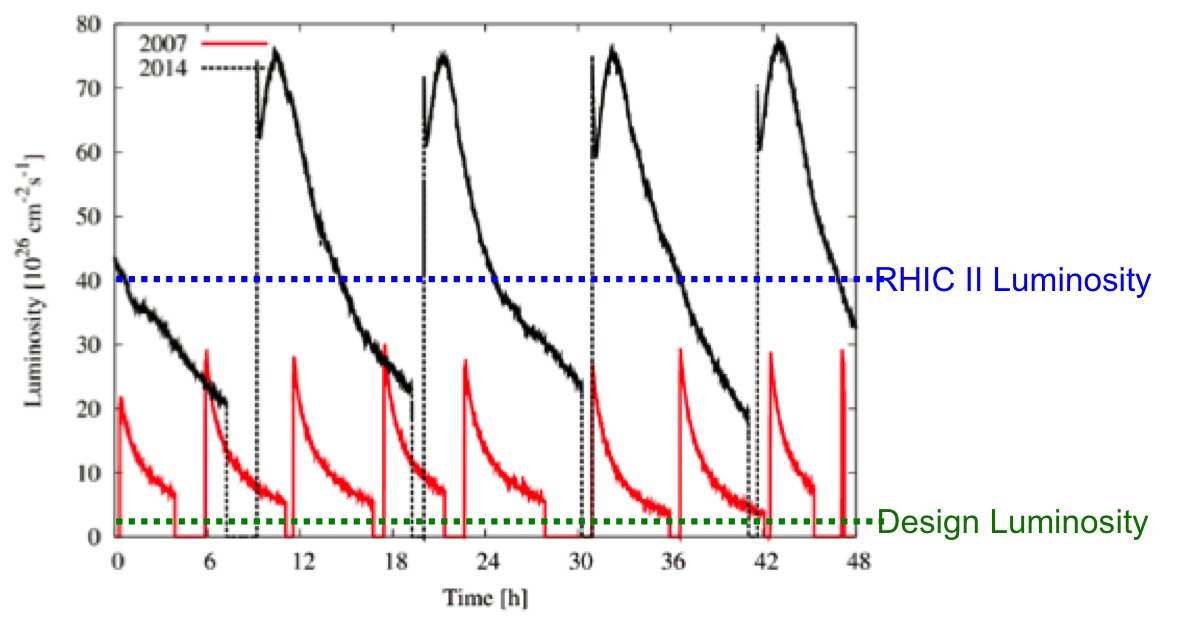}
\includegraphics[width=0.40\textwidth]{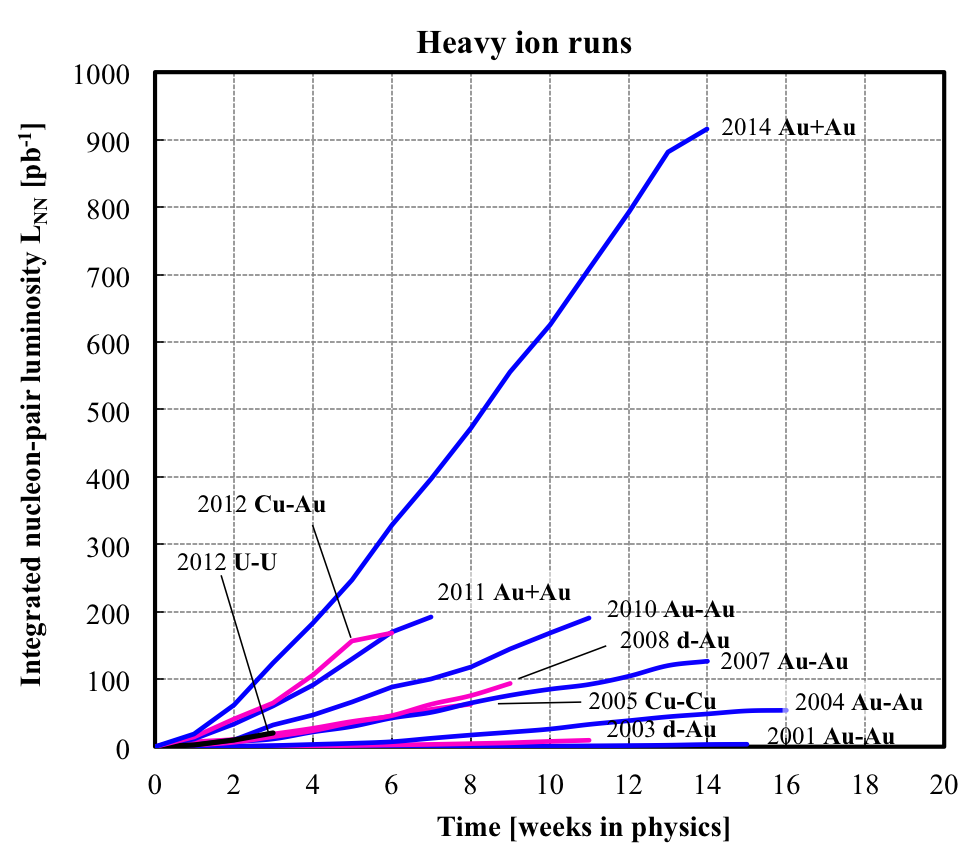}
}
\caption[RHIC luminosity, past and present]{Left: the luminosity for $\sqrt{s_{NN}}=200$~GeV \AuAu\ collisions as a function of time in the 2014 RHIC Run, compared to RHIC and RHIC II design luminosity.
Right: the integrated luminosity for all RHIC heavy ions runs.
}
\label{Fig:RHICLum}
\end{figure}

RHIC began operations in 2000 with the capability of colliding nuclei from deuterons to Au at center-of-mass energies of 
200 GeV per nucleon pair.\footnote{RHIC also can collide  polarized protons at energies up to $\sqrt{s}=510$~GeV. This unique capability forms the basis of the RHIC Spin program\cite{Aschenauer:2015eha} discussed in the Cold QCD Town Meeting White Paper.}
Recommendation IV of the 2007 Long Range Plan endorsed the ``RHIC II'' luminosity upgrade, envisioned as a 10-20 fold increase above the design luminosity of $2 \times 10^{26}\ \mathrm{cm^{-2} s^{-1}}$ for full energy \AuAu\ collisions. 
While this upgrade was a high priority of the nuclear science community, the then-estimated cost of $\sim$\$150M drove a proposed funding profile that would have provided this capability no earlier than 2016. Remarkably, breakthroughs in both transverse and longitudinal stochastic cooling by the BNL Collider-Accelerator Department scientists have made it possible to achieve luminosities well above the RHIC II specification 3 years earlier at roughly 15\% of the cost projected in 2007. The new RHIC luminosity performance is shown in the left panel of Figure~\ref{Fig:RHICLum},
demonstrating that the \AuAu\ luminosity from the 2014 RHIC run routinely exceeded RHIC II luminosities. As a result, the integrated luminosity for \AuAu\ collisions acquired in Run 14 significantly exceeds the sum of {\em all} previous RHIC runs\cite{RHICPerformance}, as shown in the right panel of Figure~\ref{Fig:RHICLum}.

RHIC's capabilities were extended in 2011 with the commissioning 
of an Electron Beam Ion Source\cite{Alessi:2010zz} 
(EBIS)\footnote{EBIS was partially funded by NASA for the study of space radiation effects on extended human space missions
at NASA's Space Radiation Laboratory located at BNL.}, 
which replaced the aging Tandem Accelerator as a source for injection into the AGS/RHIC accelerator chain. The new physics enabled by EBIS was demonstrated in definitive fashion in 2012, when RHIC became the first collider to produce \UU\ collisions. 
The intrinsic deformation of the ${}^{238}$U nucleus provides a valuable tool 
in constraining models of the initial state energy deposition 
and separating the effects from overlap geometry and quantum fluctuations in the initial state on the hydrodynamic flow in the final state.
Similarly, in 2014 RHIC provided the first \HeAu\ collisions to study the development of odd-harmonic hydrodynamic flow resulting from the three nucleons in ${}^3$He. Both of these developments are described in more detail in Section~\ref{Sec:Flow}.

Another demonstration of the flexibility of the RHIC is the Beam Energy Scan (BES) in search of the QCD critical point.  Phase I of this campaign was conducted in 2010, 2011 and 2014. In addition to full energy \AuAu\ collisions at $\sqrt{s_{NN}} = 200$~GeV, data were taken for \AuAu\ collisions at 
$\sqrt{s_{NN}}$ 62.4 GeV, 39 GeV, 27 GeV, 19.6 GeV, 14.5 GeV, 11.5 GeV and 7.7 GeV. As discussed in more detail in Section~\ref{Sec:BES} this range of energies explores the region in the QCD phase diagram (as a function of baryon chemical potential $\mu_B$ and temperature) in which it is predicted that the smooth cross-over transition observed at the highest RHIC energies transforms into a first-order phase transition. The capability to explore this regime is a unique feature of RHIC.

Both the PHENIX\cite{PHENIX} and the STAR\cite{STAR} experiments at RHIC have undergone significant upgrades since the last Long Range Plan. PHENIX extended its measurement capabilities at forward angles with a Muon Piston Calorimeter\cite{Chiu:2007zy}, which is now being upgraded with a pre-shower detector\cite{Campbell:2013zw}. 
STAR increased its data acquisition and triggering capabilities via the DAQ1000\cite{STAR:DAQ1000} and High Level Trigger\cite{STAR:HLT}
projects. The Time of Flight\cite{STAR:TOF} detector greatly extended their particle identification capabilities
and together with the muon telescope  detector\cite{Ruan:2009ug} significantly improved its capabilities to study physics in the di-lepton channel.
Both experiments installed silicon-based tracking systems for vertexing and heavy-flavor detection, 
the VTX\cite{Nouicer:2008pf,Taketani:2010zz} and FVTX\cite{Aidala:2013vna} for PHENIX and the HFT\cite{Kapitan:2008kk,Qiu:2014dha} for STAR. These new capabilities, together with the greatly increased RHIC luminosity, are the realization of the RHIC II upgrade envisioned in the 2007 Long Range Plan.

\begin{figure}[!htb]
\begin{center}
\includegraphics[width=0.8\textwidth]{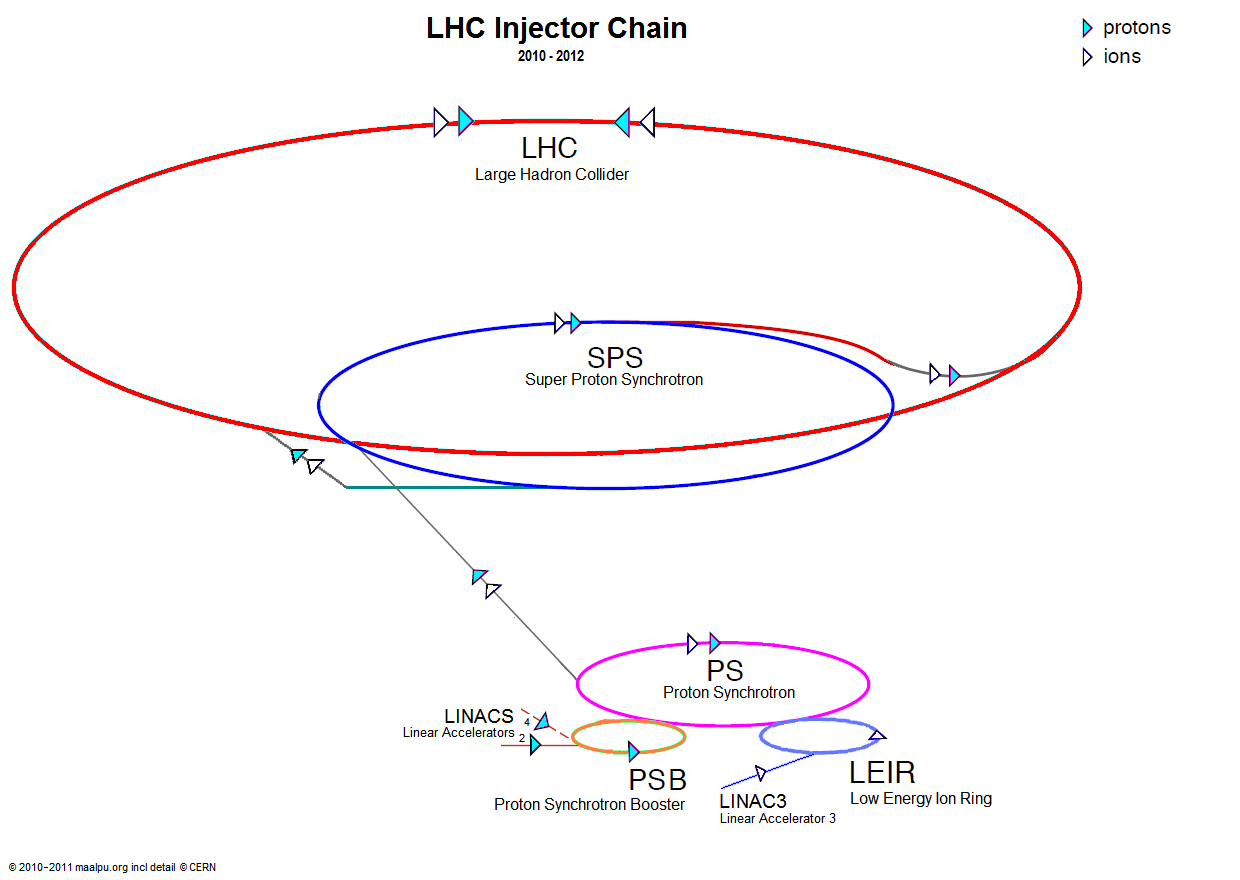}
\caption{Schematic view of the LHC accelerator complex}
\label{fig:lhc_complex}
\end{center}
\end{figure}
 
Late 2009 marked the beginning of CERN LHC operations, with a pilot run providing
first proton+proton collisions with collision energies of 0.9 and 2.36~TeV. 
The first year of LHC physics running began with p+p running at 7~TeV in April 2010 and 
ended with first Pb+Pb collisions at $\rootsNN = 2.76$~TeV in November and 
December, increasing the center-of-mass energy explored in heavy ion collisions by a factor 
of 14 compared to RHIC. The heavy ion program, which was foreseen in LHC
planning from the beginning, uses most of the elements of the LHC 
accelerator chain shown in in Figure~\ref{fig:lhc_complex}, beginning
from the electron cyclotron resonance ion source and linear accelerator, 
which provide $^{208}$Pb ions stripped to Pb$^{29+}$ at 4.2 MeV/n.
After further carbon foil stripping, bunch shaping and 
electron cooling in the CERN Low Energy Ion Ring (LEIR),  
Pb$^{54+}$ ion bunches are sent to the CERN PS, accelerated and fully stripped,
yielding Pb$^{82+}$.  After acceleration in the SPS to to 177 GeV/n,
injection and acceleration in the LHC are the final steps. 
Using 120 colliding bunches for each beam, a peak luminosity of 
$3 \times 10^{25} \mathrm{cm}^{-2} \mathrm{s}^{-1}$ was achieved 
in the 2010 LHC run.  This corresponds to an 
integrated luminosity of 7~$\mu \mathrm{b}^{-1}$ delivered to each
of three interaction regions for the ALICE, ATLAS and CMS experiments
participating in heavy-ion data taking. 

\begin{figure}[ht]
\centerline{
\includegraphics[width=0.55\textwidth]{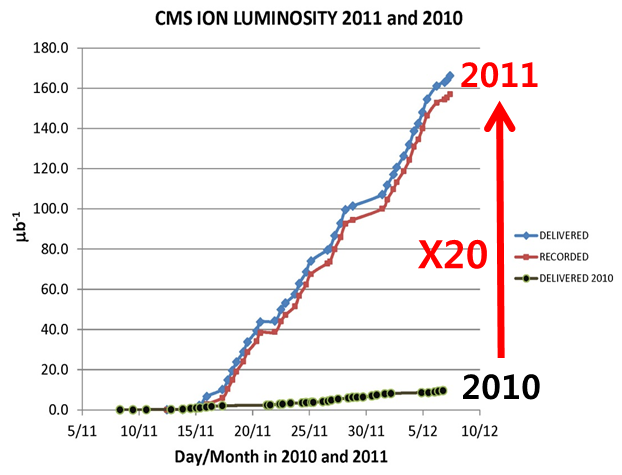}
\includegraphics[width=0.45\textwidth]{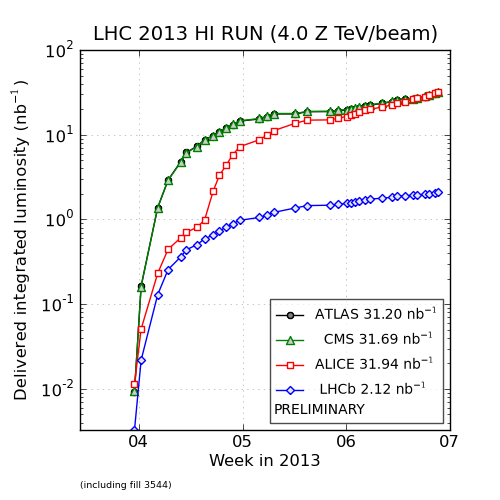}
}
\caption[LHC luminosity]{Left: Delivered and recorded integrated luminosities at 
CMS for $\sqrt{s_{NN}}=2.76$~GeV \PbPb\ collisions as a function of 
time for the 2010 and 2011 LHC Pb+Pb runs.
Right: Integrated luminosity for the $\sqrt{s_{NN}}=5.02$~TeV p+Pb 2013 LHC Run.
}
\label{Fig:LHCLum}
\end{figure}
For the November--December 2011 run, an increase in the number of 
colliding bunches to 360 per ring, as well as improved focusing, allowed 
an increase in peak luminosity by a factor of 15--20, reaching close to 
design collision rates. The total delivered luminosity per interaction region 
was about 150~$\mu \mathrm{b}^{-1}$, with delivered and 
recorded luminosity in CMS shown in Figure~\ref{Fig:LHCLum} (left).

The third heavy-ion data taking period in early 2013 provided proton+lead and lead+proton
collisions at 5.02~TeV, following a pilot run in fall 2012.
Although this mode of operation was not foreseen in the baseline design of the LHC,
beams were commissioned in 10 days. The physics requirements of all experiments 
were met in three weeks of physics running, resulting in an integrated luminosity of up to 
35~$n \mathrm{b}^{-1}$ and record intensity levels, Figure~\ref{Fig:LHCLum} (right).  
A p+p reference data set at 2.76~TeV was also recorded. 

All four major LHC detectors have participated in heavy-ion data taking
in the 2009--2013 period (Run I), with ALICE, ATLAS and CMS taking p+p, p+Pb and
Pb+Pb data, and LHCb taking p+p and p+Pb data. ALICE\cite{ALICE} has been optimized
for heavy-ion operations, with large kinematic coverage, in particular at low 
transverse momenta, and charged particle identification over a wide 
momentum range using a variety of techniques. ATLAS\cite{ATLAS} and CMS\cite{CMS} are general purpose
collider detectors, with a particular focus on high data taking rates, full 
azimuthal coverage over a wide rapidity range and high resolution at very 
high transverse momenta. Although designed for discovery physics in p+p 
collisions, the high granularity of ATLAS and CMS makes them suitable
for heavy-ion collisions as well, in particular for high momentum 
probes where they complement the strengths of the ALICE detector. The heavy-ion 
related program of LHCb\cite{LHCb} was focussed on quarkonia measurements in p+Pb 
collisions at forward rapidities.

%% file: tex/Flow.tex
\subsection{Hydrodynamics and Collective Flow}
\label{Sec:Flow}

The initial discovery of strong elliptic flow at RHIC~\cite{Ackermann:2000tr} 
and the characteristic hydrodynamic signature of mass ordering 
in the medium response~\cite{Adler:2003kt,Kolb:2003dz} focussed a great deal 
of attention on improving the relativistic hydrodynamic description 
of the quark-gluon plasma. (See Ref.~\cite{Gale:2013da} for a recent review.)
One of the most important recent discoveries made in heavy ion collisions since the last Long-Range Plan
is the persistence of density fluctuations from the initial state. 
Recent work~\cite{Mishra:2007tw,Voloshin:2003ud,Takahashi:2009na,Sorensen:2010zq,Alver:2010gr,Qiu:2011iv,ALICE:2011ab,Adare:2011tg,ATLAS:2012at,Adamczyk:2013waa}
demonstrates that these fluctuations survive through the expansion of the fireball and appear as correlations between produced particles.
Most previous approaches had approximated the incoming nuclei as smooth spheres and the initial overlap region as an ellipse. The survival of density and geometry fluctuations was first hinted at in measurements of cumulants related to the shape of the elliptic flow distribution~\cite{Adler:2002pu,Miller:2003kd}. The picture started to become more clear after measurements were made in Cu+Cu collisions where the relative fluctuations were more prominent in the smaller system~\cite{Alver:2008zza}.  Ultimately, a new paradigm emerged as the structure of the initial state was found to play a central role 
in determining the azimuthal anisotropies with respect to the event plane angles $\Phi_n$, parameterized in terms of transverse momentum $p_T$ and azimuthal angle $\phi$ as
\begin{equation}
\label{Eq:vnDef}
\frac{d^2n}{p_T dp_T d\phi}
\sim
1 
+ 2 v_2(p_T) \cos 2 (\phi - \Phi_2)
+ 2 v_3(p_T) \cos 3 (\phi - \Phi_3)
+ 2 v_4(p_T) \cos 4 (\phi - \Phi_4)
+ \dots
\end{equation}
Previous measurements that were focused almost exclusively on the dominant $v_2$ were generalized to $v_n$, a spectrum carrying information about both the initial densities in the collision and the dissipative properties of the subsequent plasma phase~\cite{Heinz:2013th}. The survival of the initial state fluctuations is related to the earlier finding that the QGP discovered at RHIC is the most perfect fluid known~\cite{Teaney:2003kp,Romatschke:2007mq,Song:2010mg} with a viscosity to entropy ratio near the string theory limit~\cite{Kovtun:2004de}. 
The low viscosity plasma phase acts as a lens (albeit of strongly non-linear character), faithfully transferring the geometric structure of the initial density distributions, with its associated distribution of pressure gradients which act as a hydrodynamic force, into the final state. There it shows up most prominently as correlations between produced particles. Quantum fluctuations in the initial state cause these correlations to fluctuate from event to event.

\begin{figure}[ht]
\includegraphics[width=1.\textwidth]{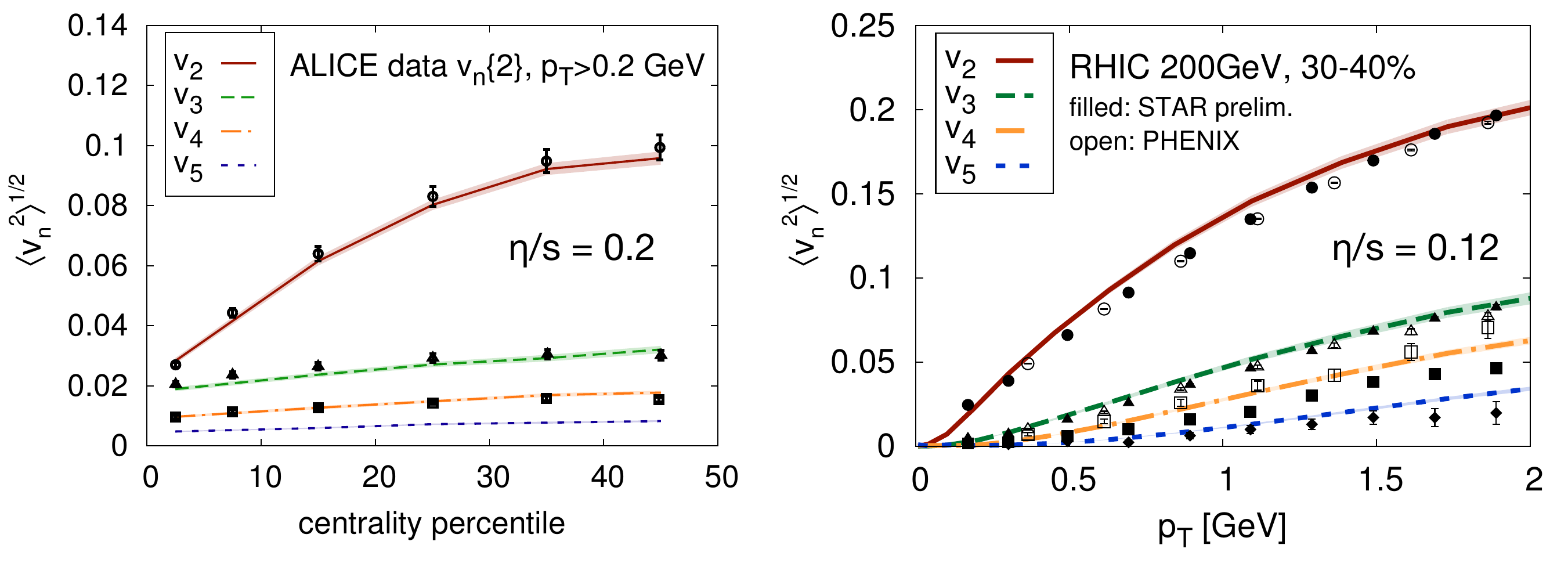}
\caption[Elliptic flow $v_2$ compared to a hydrodynamic model]{Model calculations compared to measurements of the harmonic decomposition of azimuthal correlations produced in heavy ion collisions~\cite{Gale:2013da}. The left panel shows model calculations and data for $v_n$ vs. collision centrality in Pb+Pb collisions at $\sqrt{s_{NN}}=2.76$ TeV. The right panel shows similar studies for the $p_T$ dependence of $v_n$ in 200 GeV Au+Au collisions. The comparison of the two energies provides insight on the temperature dependence of $\eta/s$. }
\label{fig:vn}
\end{figure}

Descriptions of these new phenomena have required the development of a new dynamical framework for heavy-ion collisions. It includes i) modeling of initial-state quantum fluctuations of nucleon positions and sub-nucleonic color charges and the color fields generated by them, ii) a description of the pre-equilibrium dynamics that evolves the initial energy-momentum tensor by solving either the (2+1)-dimensional Yang-Mills equations for the gluon fields (weakly-coupled approach) or Einstein's equations of motion in five-dimensional anti-deSitter space (strongly-coupled approach), followed by iii) the rapid transition, event-by-event, to second-order viscous relativistic fluid dynamics, and iv) a late-stage hadron phase described by microscopic transport calculations. 
While there is widespread agreement on the general structure of such a standardized dynamical approach, it has not yet reached the level of uniqueness that would justify calling it the ``Little Bang Standard Model'' \cite{Heinz:2013wva}. 
Model comparisons with experimental data that illustrate the state of the art in dynamical modeling can be found in 
Refs.~\cite{Schenke:2010rr,Song:2010mg,Song:2010aq,Schenke:2011bn,Schenke:2012wb,Gale:2012rq,Song:2013tpa,Song:2013qma,vanderSchee:2013pia,Habich:2014jna}. With the existence of a reliable equation of state from lattice 
QCD calculations~\cite{Bazavov:2009zn,Borsanyi:2010cj,Borsanyi:2013bia,Bazavov:2014pvz} a crucial degree of uncertainty in hydrodynamic modeling could be eliminated, enabling the development of a complete hydrodynamic space-time model. With this full space-time picture in hand, the comparisons of model calculations to 
harmonic decompositions of correlation functions ($\sqrt{v_{n}^{2}}$) at RHIC and the LHC (shown in Figure~\ref{fig:vn}) have reduced the uncertainty on $\eta/s$ by a factor of 10~\cite{Gale:2013da}. With this newfound precision, studies suggest that $\eta/s$ is smaller for RHIC collisions (right panel of Figure~\ref{fig:vn}) than it is at the LHC (left panel), consistent with a temperature dependent $\eta/s$ with a minimum near the critical temperature. In the next phase of study we seek to 1) accurately determine the temperature dependence of $\eta/s$ (aided by the Beam Energy Scan Program at RHIC described
in Sections~\ref{Sec:BES},~\ref{Sec:FacilitiesFuture} and~\ref{Sec:CP}) and 2) develop a clearer picture of the high density gluon fields discussed in Section~\ref{Sec:Saturation} that form the precursor of the plasma phase (aided by the p+A program 
and ultimately by an Electron Ion Collider).

\begin{figure}[hbt]
\begin{center}
\centerline{  \includegraphics[width=1.0\textwidth]{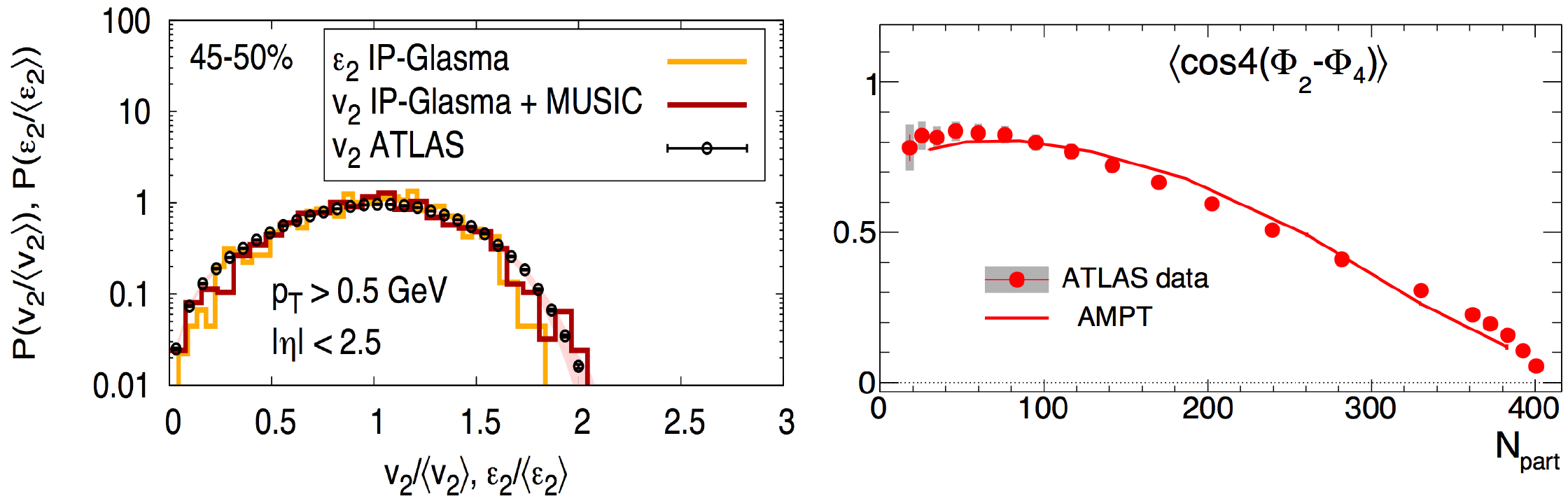}}
\caption[Hydrodynamic and transport models compared to flow observables]{Comparision of the $p(v_2)$ (left panel) and correlation between $\Phi_2$
  and $\Phi_4$ (right panel) measured for Pb+Pb collisions at
  $\sqrt{s_{NN}}=2.76$ TeV with hydrodynamic model~\cite{Gale:2012rq}
  or transport model~\cite{Bhalerao:2013ina} calculations.}
  \label{Fig:ebye}
\end{center}
\end{figure}

What is needed to turn this standard dynamical framework into the ``Little Bang Standard Model''? One fundamental challenge along the way is the need to determine {\em simultaneously} the space-time picture of the collective expansion and the medium properties that drive this expansion~\cite{Heinz:2013wva}.
A unique and reliable determination of these two unknowns
will be informed by measurements of multiple flow observables sensitive to
medium properties in different stages of the evolution~\cite{Bhalerao:2011yg,Heinz:2013th,Jia:2014jca}. Due to the
large event-by-event fluctuations in the initial state collision
geometry, the matter created in each collision follows a different
collective expansion with its own set of flow harmonics (magnitude
$v_n$ and phases $\Phi_n$). Experimental observables describing
harmonic flow can be generally given by the joint probability
distribution of the magnitude $v_n$ and phases $\Phi_n$ of flow
harmonics:
\begin{equation}
\label{eq:flow1}
p(v_n,v_m,..., \Phi_n, \Phi_m, ...)=\frac{1}{N_{\mathrm{evts}}}\frac{dN_{\mathrm{evts}}}{dv_ndv_m...d\Phi_{n}d\Phi_{m}É}.
\end{equation}
Specific examples include the probability distribution of individual
harmonics $p(v_n)$, correlations of amplitudes or phases between different
harmonics ($p(v_n,v_m)$ or $p(\Phi_n,\Phi_m)$), and flow de-correlations in transverse and longitudinal
directions. These observables can be accessed through measurements of correlations with three or more
particles. 
The joint probability distribution (\ref{eq:flow1}) can be fully characterized experimentally by measuring the complete set of moments recently identified in Ref.~\cite{Bhalerao:2014xra}. 
With the added detail provided by these measurements,
hydrodynamic models can be fine-tuned and over-constrained, thereby
refining our understanding of the space-time picture and medium
properties of the QGP created in heavy ion collisions. Initial measurements of some
of these observables~\cite{Aad:2013xma,Aad:2014fla,GranierdeCassagnac:2014jha} and 
comparison to hydrodynamic and transport models~\cite{Gale:2012rq,Heinz:2013bua,Qiu:2012uy,Bhalerao:2013ina} 
have already provided unprecedented insights into the nature of the initial density
fluctuations and dynamics of the collective evolution, as seen in Figure~\ref{Fig:ebye}. 

\begin{figure}[hbt]
\begin{center}
\centerline{  \includegraphics[width=.95\textwidth]{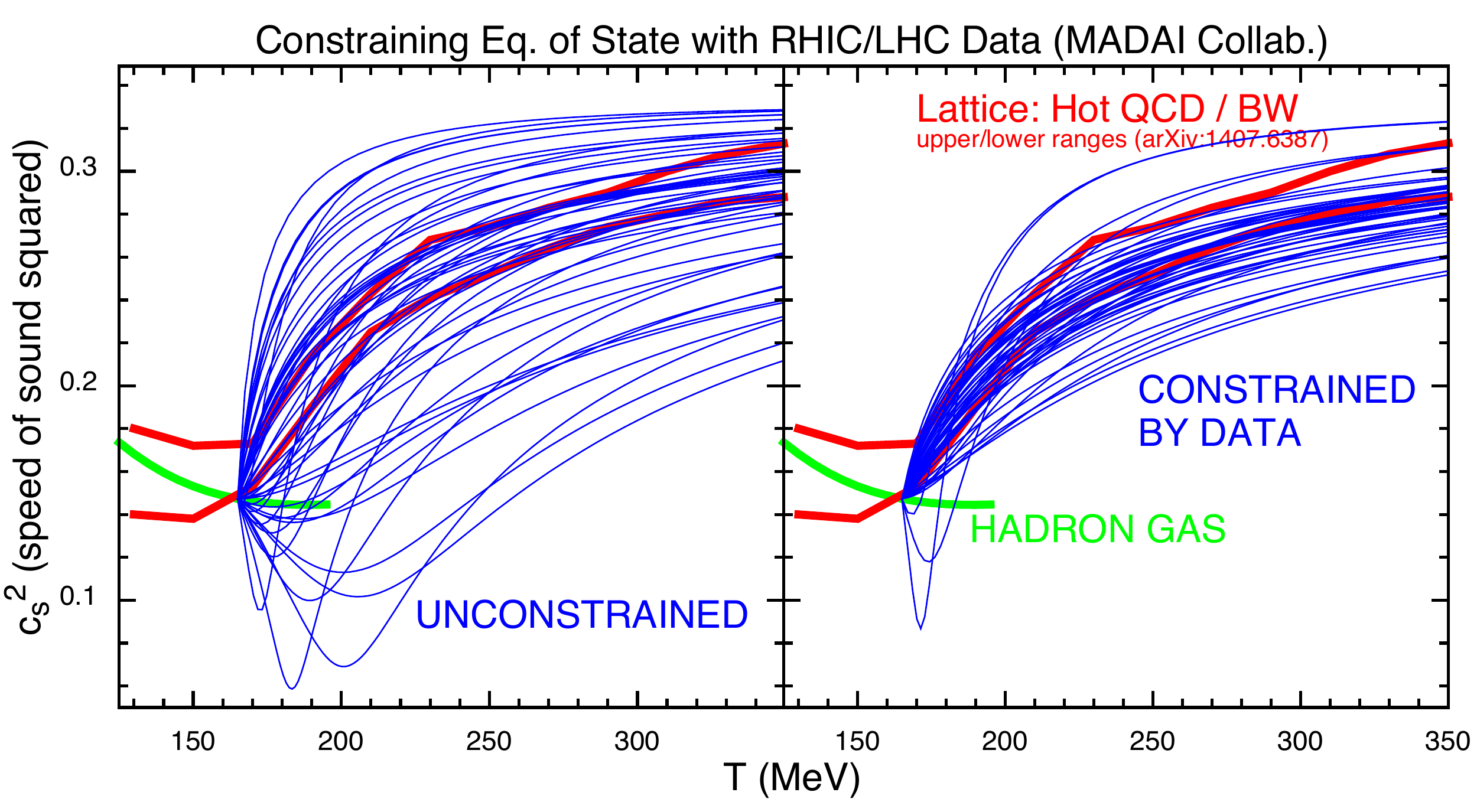}}
\caption[Constraints on the equation of state from RHIC and LHC data]{ Studies of the QCD equation of state from Lattice QCD calculations and from models constrained by data from RHIC and the LHC~\cite{Novak:2013bqa}. The right panel shows that data prefer an equation of state consistent with lattice QCD demonstrating that our model of the collision dynamics is good enough to allow us to study the emergent properties of QCD. }
  \label{fig:EOS}
\end{center}
\end{figure}

\begin{figure}[hbt]
\begin{center}
\centerline{  \includegraphics[width=.95\textwidth]{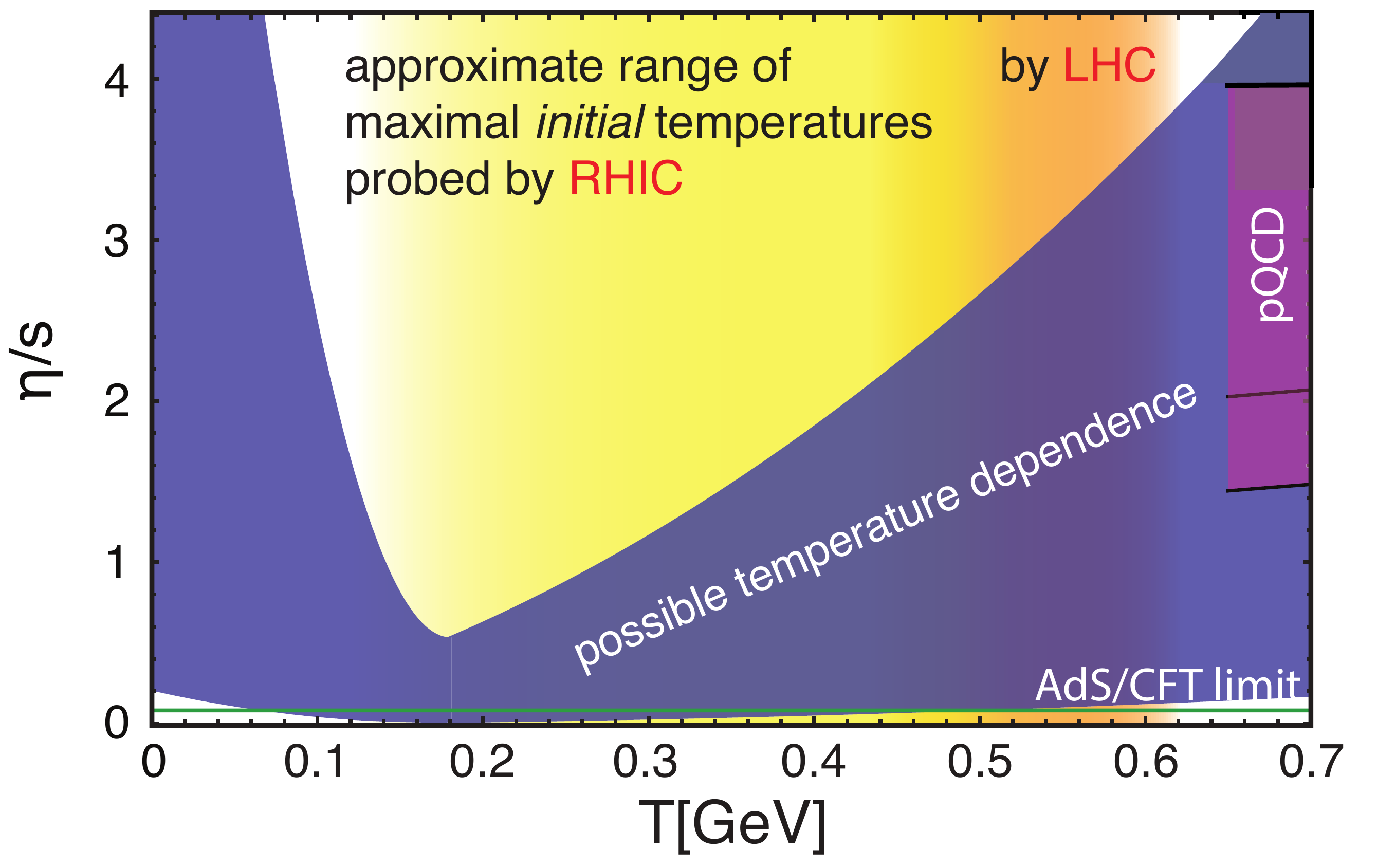}}
\caption[Temperature dependence of the viscosity to entropy density $\eta/s$]{The temperature dependence of the viscosity to entropy density $\eta/s$. The blue band represents the range allowed by our current understanding based on models compared to data with a minimum at the transition temperature. pQCD calculations and the string theory limit are also shown. The shaded vertical regions represent the ranges of initial temperatures probed by RHIC and the LHC.}
  \label{fig:visc}
\end{center}
\end{figure}

The agreement between the models and the data shown in
Figure~\ref{fig:vn} and Figure~\ref{Fig:ebye} suggests that the
essential features of the dynamic evolution of heavy ion collisions
are well described by our current models.  However, these model calculations
depend on the values assumed for many parameters, so reliable determination of the QGP
properties requires a systematic examination of the full parameter
space. An example of such an exploration~\cite{Novak:2013bqa} is shown
in Figure~\ref{fig:EOS} where the shape of the QCD equation of state (EOS) is treated as a
free parameter. 
he left panel shows a random sample of the thousands of possible Equations of State, constrained only by results on the velocity of sound obtained by perturbative QCD at asymptotically high temperature and by lattice QCD at the crossover transition temperature. They are compared to the EOS determined from lattice QCD \cite{Bazavov:2014pvz}.The right panel shows a sample of the Equations of State allowed by experimental data. The results of this study suggest that data at RHIC and the LHC require an EOS consistent with that expected from QCD. This demonstrates that our model of heavy-ion collisions describes the dynamics of the collisions well enough that we can extract information on the emergent properties of finite temperature QCD from the experimental traces left by the tiny droplet of QGP created in the collisions. These state-of-the-art models can therefore be used to both determine properties of finite temperature QCD currently inaccessible to lattice calculations and to provide an accurate space-time profile needed for modeling other processes like jet quenching. 
Figure~\ref{fig:visc}
shows a schematic representation of our current uncertainty on the
temperature dependence of $\eta/s$ in QCD matter. 
While many of the existing measurements are accurate enough, as seen in Figure~\ref{fig:vn}, to determine $\eta/s$ with much greater precision {\it if all other model parameters were already known}, the non-linear simultaneous dependence of the observables on multiple parameters does not yet allow one to translate the high quality of these experimental data into a more precise estimate of $\eta/s$.
The studies shown in
Figures~\ref{fig:vn}, \ref{Fig:ebye} and \ref{fig:EOS} 
suggest, however, that a more complete set of measurements of the moments of the joint probability distribution (\ref{eq:flow1}) at the LHC and RHIC (particularly in the Beam Energy Scan), coupled with extensive quantitative modeling, will provide the desired access to $(\eta/s)(T)$ in and around the transition temperature where hadrons melt into quark-gluon plasma, and strongly reduce the width of the blue uncertainty band in Figure~\ref{fig:visc}.

While the paradigm of collective flow phenomena in a strongly coupled,
opaque QGP fluid has been firmly established in sufficiently central
collisions of heavy nuclei, it was generally expected that the
magnitude of collectivity would diminish as the system size
decreases. As the mean-free-path of the matter approaches the
characteristic size of the system, the effects of viscous damping
become more important and the validity of a hydrodynamic description
becomes more suspect. Evidence for this trend has been observed in
peripheral A+A collisions. As such, no collective flow was anticipated
in \pp\ and \pA\ collisions. Surprisingly, 
correlations that are
long-range in rapidity and similar to those measured in A+A collisions
have now also been observed at the LHC in rare high-multiplicity
p+p collisions~\cite{Khachatryan:2010gv} (corresponding to high
gluon-density initial states). 
In several ways, these resemble the
correlations in central A+A collisions which have been widely accepted as
evidence of collective flow~\cite{Ollitrault:1992bk,Voloshin:1994mz}.
Subsequent measurements revealed similar phenomena in high
multiplicity \pPb\ and \dAu\ at both the
LHC~\cite{CMS:2012qk,Abelev:2012ola,Aad:2012gla} and
RHIC~\cite{Adare:2013piz}. The dependence of the correlations in
\pPb\ or \dAu\ on $p_T$,
multiplicity~\cite{Aad:2013fja,Chatrchyan:2013nka,Abelev:2014mda,Aad:2014lta},
pseudorapidity~\cite{CMS-PAS-HIN-14-008}, and particle
species~\cite{ABELEV:2013wsa,Khachatryan:2014jra} reveal similarities
to those observed in A+A collisions. In particular, the mass ordering
of $v_n(p_T)$ is reminiscent of the effect from a common radial flow
boost in A+A collisions~\cite{ABELEV:2013wsa,Adare:2014keg,Khachatryan:2014jra}, and
multi-particle correlations show unambiguously that the novel correlations in
high-multiplicity \pPb\ collisions are collective in
nature~\cite{GranierdeCassagnac:2014jha}.

The origin of collectivity in these small systems is a topic of
debate. While hydrodynamic models with strong final-state interactions
may provide a natural interpretation for many of the observed features
in the
data~\cite{Bozek:2009dt,Bozek:2011if,Bozek:2012gr,Schenke:2014zha,Werner:2013ipa},
their apparent applicability in such small systems, along with the required assumption of
rapid thermalization, challenges our understanding~\cite{Niemi:2014wta}. Meanwhile, other novel
mechanisms, mainly related to the initial-state quark and gluon
correlations, have also been proposed as alternative interpretations of the
observed long-range correlations in \pA\ and \dA\ collisions, 
and they have even provided qualitatively successful descriptions for \pp\
collisions~\cite{Dumitru:2010iy,Dusling:2012cg,Dusling:2013oia,Gyulassy:2014cfa,Dumitru:2014yza}. Disentangling
initial- and final-state effects to distinguish between these various approaches poses a theoretical and experimental
challenge. Recent data from $^3$He+Au collision may shed some light on
the question, as will improved correlation measurements at forward
rapidity, where the presence of the correlation structures is
most surprising. 
Further insights are expected as
comprehensive studies of the system-size and geometric dependence
become available in \pA\, \dA\ and $^3$He+A
collisions~\cite{Nagle:2013lja} where the relative contributions of
initial- and final-state correlations are expected to vary. 
This program will allow us to explore the boundary of perfect
fluidity in QCD matter at the smallest scales ever
achieved~\cite{Shuryak:2013ke}. 

Addressing these open questions on the possible role of hydrodynamics
in the smallest hadronic systems will play an
important role not only in completing our standard model of a strongly
coupled QGP matter, but also in providing new opportunities to probe
the structure of protons. 
If indeed
final-state effects described by hydrodynamic flows are proven to be
the dominant source of correlations, the presence of a tiny low viscosity
fluid enables the study of protons and sub-nucleonic scale
fluctuations at very short time
scales~\cite{Coleman-Smith:2013rla,Schenke:2014zha,Schlichting:2014ipa}. 
The high-density gluon state inside a
proton is of fundamental interest as the equations of QCD are expected
to become classical~\cite{McLerran:1993ni,Gelis:2010nm}. This
transition has the potential to reveal how a classical system can emerge from
QCD. In light of recent exciting observations, this topic should be
studied in future \pA\ programs covering the wide kinematic range
provided by RHIC and the LHC and ultimately in an EIC which is the
highest priority for new construction in our community.

%% file: tex/Jets.tex
\subsection{Parton Energy Loss and Jet Modification}
\label{Sec:Jets}

\subsubsection{Introduction}
\label{Sec:JetIntro}

In processes involving a hard scattering, highly virtual partons
are produced; these then undergo successive branchings resulting in a parton
shower~\cite{Field:1976ve}. The  ensemble of produced particles is highly collimated about
the direction of the initial parton and contains a range of different
momentum scales. The properties of these objects, known as jets~\cite{Sterman:1977wj}, and
how they emerge from perturbative QCD (pQCD) calculations have been extensively studied in high-energy
physics~\cite{Feynman:1978dt,Field:1977fa}.

One of the successes of the early portion of the RHIC program
was the discovery of jet quenching\cite{Adcox:2001jp,Adler:2002xw}: the phenomenon in which jet showers are modified by interactions with the
medium~\cite{Bjorken:1982tu}. The most noticeable outcome of this scattering in the medium is an enhanced rate of bremsstrahlung leading to a 
loss of energy and momentum by the most energetic (leading) partons in the shower, which results in a depletion of high momentum hadrons. 
The first measurements of the suppression of high-\pT\ hadrons
and di-hadron correlations established that the medium was highly
opaque to colored probes. This indicated that the medium contains a
high density of unscreened color charges which lead to considerable modification of hard jets. 
Later measurements of single and dihadron observables at both RHIC and the LHC have significantly restricted
the variety of viable theoretical approaches to jet modification in a dense medium. 

The virtuality of a hard parton within a jet represents its intrinsic scale, which is also the scale at which it resolves fluctuations in the medium. 
As partons in the jet cascade down to lower virtualities, they probe the medium over a multitude of length scales. 
As long as the virtuality (and the related resolution scale) of a parton is much larger than $\Lambda_{QCD}$, it will be weakly coupled to the medium and 
pQCD  describes its propagation, via both scattering and radiation in the medium.
The largest virtualities reside, on average, with the leading or highest energy parton, with lower energy partons decaying more rapidly to lower virtualities. 
It is this expectation that led to the formulation of the earlier pQCD based leading parton energy loss calculations~\cite{Gyulassy:1993hr,Wang:1991xy,Baier:1996kr,Baier:1996sk,Zakharov:1996fv,Zakharov:1997uu}. 
While scattering in the medium slows down the rate at which
the virtuality (and the related resolving power) decreases,
several partons within a jet still may lose sufficient amounts of energy and/or virtuality in the medium and encounter 
non-perturbatively strong coupling~\cite{Chesler:2008uy,Chesler:2008wd,Friess:2006aw,CasalderreySolana:2006rq}. 
While the fate of such partons is under some debate, the evolution of the leading hard partons with virtualities $Q^{2} \gg \Lambda_{QCD}^{2}$ has been successfully  
described using pQCD and factorized transport coefficients. This formalism and the experimental results related to it are described Section~\ref{q-hat-e-loss}. 

With the tremendous improvement in experimental abilities at the LHC, it has now become possible to reconstruct and isolate an entire jet from the dense fluctuating medium.
Experimental issues related to this development are described in Section~\ref{Sec:jet-reconstruct}. The modification of entire jets in a medium involves dynamical processes at energy scales 
ranging from the perturbative hard scale down to the non-perturbative soft scale of the medium. Recent empirical observations along with new theoretical insight dealing with phenomena at intermediate scales are discussed is 
subsection~(\ref{jet-mod}).
As jets provide probes at a
variety of scales, a program of systematic jet measurements can be
used to study the emergence of strongly-coupled behavior that has been
observed at long wavelengths in flow measurements, as well as its interplay with short distance fluctuations which affect the jet core. 

\subsubsection{Leading hadron suppression and jet transport coefficients}~\label{q-hat-e-loss}

At the time of this writing, jets have been measured at the LHC with energies up to several hundred GeV, while jets at RHIC have been measured with energies up to 50 GeV.
These exhibit non-trivial interaction with the medium over a wide range of scales.  
The collimated shower of partons contain a central, \emph{hard core} that consists of the leading partons, which carry a majority of the jet's momentum.
These partons are the dominant contributors to leading hadron analyses such as the single hadron inclusive nuclear modification factor, or leading hadron triggered near and away side correlations. 
Calculations that focus solely on this hard core~\cite{Gyulassy:2000er,Arnold:2002ja,Qin:2007rn,Wang:2001ifa,Majumder:2009ge,Majumder:2011uk} 
have been quite successful at describing several of these single particle observables. 

When produced in vacuum, jets tend to shower to several partons with progressively lower virtualities. 
In the case where the jet is produced in a medium, those partons in the jet with 
virtualities that considerably exceed $\Lambda_{QCD}$ are weakly coupled to the medium. Therefore, the 
radiation from and scattering of these partons are calculable in pQCD. 
The scatterings induce shifts both in the momentum of the propagating partons and in their virtuality. As a result, these scatterings change the radiation pattern of the shower by inducing longitudinal drag (and associated longitudinal diffusion), transverse diffusion, and enhanced splitting of the propagating partons. The transport coefficients 
$\hat{q}$~\cite{Baier:2002tc} and $\hat{e}$~\cite{Majumder:2008zg} quantify the transverse diffusion and longitudinal drag experienced by a single hard parton, and are the leading transport coefficients that modify the propagation and in-medium splitting of hard jets.

The transport coefficient $\hat{q}$ characterizes the mean-squared momentum per unit length acquired by a parton transversing the medium.
Formally, it is the Fourier transform of the Lorentz-force-Lorentz-force correlator, 
which for a near on-shell parton traveling in the negative light cone direction with momentum $q^{-}$ in a gauge with $A^{-}=0$ gauge is
\begin{eqnarray}
\hat{q} (x^{-}) = \frac{8\pi^{2} \alpha_{S} C_{R}}{ N_{c}^{2} - 1} \int \frac{d^{2} y_{\perp} dy^{-} d^{2} k_{\perp} }{(2 \pi)^{3}} e^{-i \frac{k_{\perp}^{2} }{2q^{-}} y^{-} + i \vec{k}_{\perp} \cdot \vec{y}_{\perp} }  \langle F^{+ \mu}  (x^{-} + y^{-}, \vec{y}_{\perp}) F^{+}_{\mu}(x^{-},0) \rangle.
\label{Eq:qhat}
\end{eqnarray}
In this expression, $\langle \cdots \rangle$ implies an expectation over the state (or states) of the medium. The field strength tensor $F^{\mu +} \equiv F^{a \mu +} t^{a}$ represents 
the soft color field of the medium off which the hard parton with color Casimir $C_{R}$ scatters (trace over color indices is implied). 
These soft matrix elements are 
factorized from the hard processes of parton propagation and splitting~\cite{Kang:2013raa}. In simplified scenarios, such as in a static thermal medium, 
they may be calculated from first principles assuming that the medium is 
weakly coupled~\cite{CaronHuot:2008ni}, strongly coupled~\cite{Liu:2006ug}, on the lattice~\cite{Majumder:2012sh}, or using a combination of weak coupling and 
lattice techniques~\cite{Panero:2013pla}. However, for the dynamical rapidly expanding medium created in relativistic heavy ion collisions, the only recourse is to extract averaged soft matrix elements by comparing experimental results to calculations that involve 
detailed treatments of hard parton production, shadowing, and final state parton propagation in media simulated with viscous fluid dynamics. 
In such calculations, $\hat{q}$ is either recalculated assuming a weakly coupled medium once the coupling constant in the medium is fit to data (as in Ref.~\cite{Gyulassy:2000er,Arnold:2002ja,Qin:2007rn}), or it is scaled with an intrinsic quantity in the hydrodynamic simulation, with the overall normalization fit to data (as in Ref.~\cite{Majumder:2011uk}).  

\begin{figure}[t]
\centerline{
\includegraphics[width=1.03\textwidth]{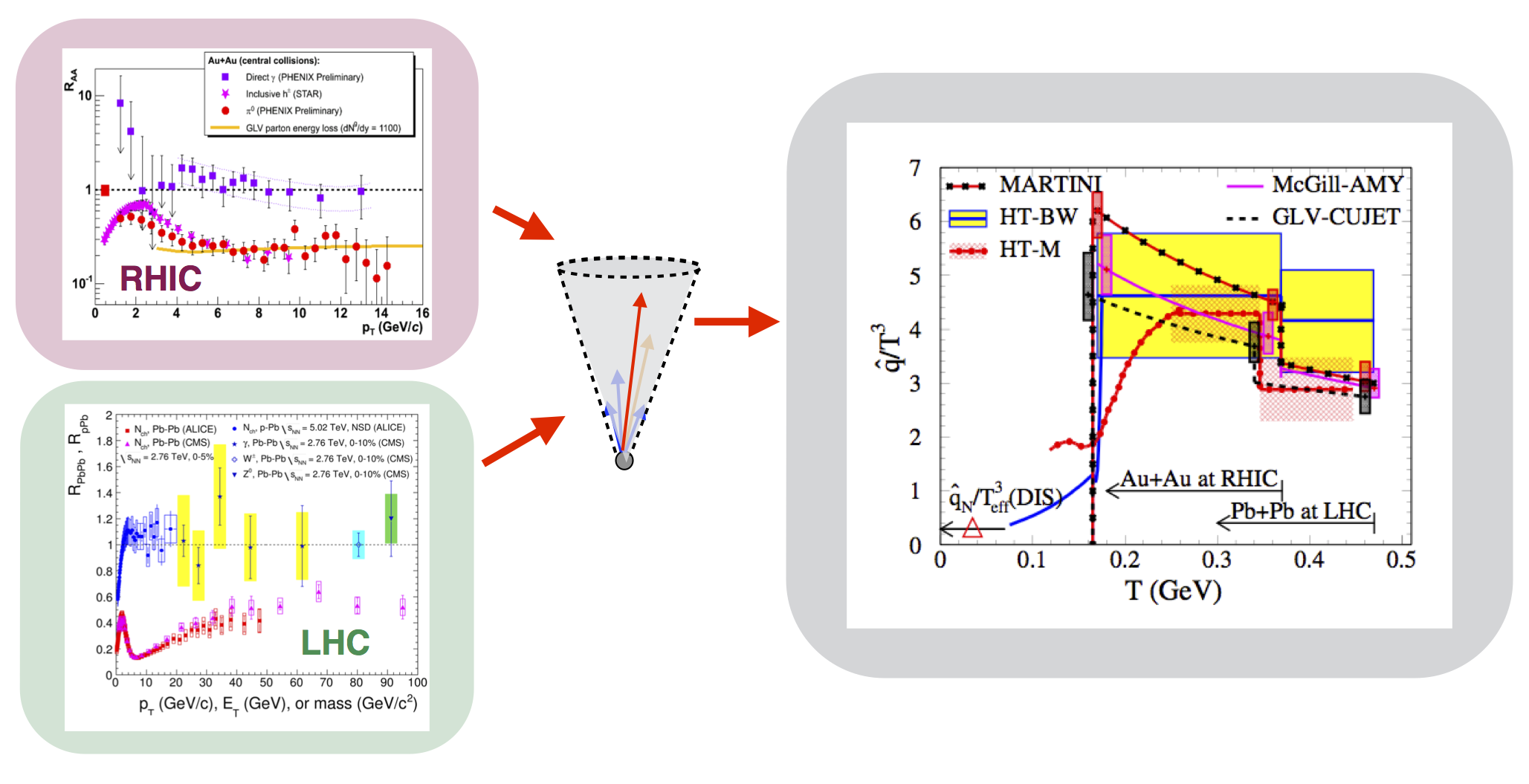}}
\caption[RHIC and LHC data compared to different pQCD energy loss energy loss schemes]{A comparison of several different pQCD based energy loss energy loss schemes to the measured leading hadron suppression in central events at RHIC and the LHC, and the extracted transport coefficient $\hat{q}$ along with its dependence on temperature.}
\label{fig:JetProgressFig1}
\end{figure}

In either of these approaches, one obtains a very good description of the experimental results on the suppression of leading hadrons from both RHIC and the LHC. Some of these fits are shown in the left panel of Figure~\ref{fig:JetProgressFig1}.
The extracted values of the leading transport parameter $\hat{q}$ and its temperature dependence are plotted in the right panel of Figure~\ref{fig:JetProgressFig1}. Using identical 
initial states and hydro simulations, 
the JET Collaboration\footnote{The JET Collaboration is one of the Topical Collaborations in nuclear theory established by the DOE Office of Nuclear Physics in response to a recommendation in the 2007 Long-Range Plan for Nuclear Physics.}\cite{JET}
has carried out the analyses of comparing the spread in values of $\hat{q}$ due to systematic differences in the pQCD based 
energy loss schemes~\cite{Burke:2013yra}. In sharp contrast to similar comparisons carried out earlier in Ref.~\cite{Bass:2008rv}, where extracted  $\hat{q}$ values differed by almost an order of magnitude, current calculations reported in Ref.~\cite{Burke:2013yra}, differ at most by a factor of 2, 
indicating the very considerable progress in our understanding of hard processes in the QGP since the time of the last Long Range Plan.

Beyond the success of applying pQCD based energy loss techniques to the in-medium modification of leading partons in a hard jet, 
this five-fold reduction in the uncertainty of the 
extracted values of transport coefficients is greatly significant, as it allows, for the first time, 
to discern a possible non-monotonic dependence of $\hat{q}$ on the temperature of the medium. Hence, such theoretical analyses have reached a level of 
sophistication where qualitative and quantitative properties of the QGP may be extracted, with quantified systematic uncertainties, using jet modification. 
%
These successes, enabled by improvements on the theoretical and experimental side, provide an anchor in our study of the modification of full jets in a medium, which involves an interplay of several different scales, as one moves from leading partons to the softer segments of the jet and their interaction with the medium. The incorporation of such theoretical improvements, and the ensuing measurements 
over the next decade
will allow jets to be used as calibrated and controlled precision probes of the microscopic structure of the quark-gluon plasma over a wide range of scales.

\subsubsection{Full jet reconstruction}
\label{Sec:jet-reconstruct}
Fully reconstructed jets have been a crucial tool used in high energy
physics, both to provide precision tests of pQCD and to understand the
topology of the hard-scattering event. Recently these techniques have
been adapted to the heavy ion environment resulting in a new set of
observables sensitive to jet quenching. Reconstructed jets are
expected to be related to initial parton kinematics and thus manifest
the kinematics of the hard scattering in the absence of medium
effects. Some of the first LHC heavy-ion results included the
observation of highly asymmetric dijet events, which provided a
striking visual demonstration of the energy
loss\cite{Aad:2010bu,Chatrchyan:2011sx}. Since these initial
measurements experimental control over the measured jet energies and
the understanding of the role of underlying event fluctuations has
improved substantially, resulting in precise measurements of jet
suppression and the properties of quenched jets. Color-neutral objects
such as a photons are not expected to experience quenching, and
measurements of direct photon production rates at
RHIC\cite{Adler:2005ig} were important in the interpretation of the
high-\pT\ hadron suppression. Recently, these measurements have been
extended to much higher \pT\
\cite{Aamodt:2010jd,Abelev:2012hxa,CMS:2012aa}, and additional probes
such as the $W$ and $Z$ \cite{Chatrchyan:2012nt,Aad:2013sla,
Chatrchyan:2011ua,Aad:2012ew} are accessible at the LHC. In events
where these objects recoil against a jet, they serve as relatively
clean probes of the jet kinematics before energy loss. The first
photon-jet measurements at the LHC \cite{Chatrchyan:2012vq} have shown
that the energy of these jets is significantly degraded and that in
many cases no recoil jet is distinguishable from the bulk medium.

\begin{figure}[t]
\centerline{
\includegraphics[width=0.99\textwidth]{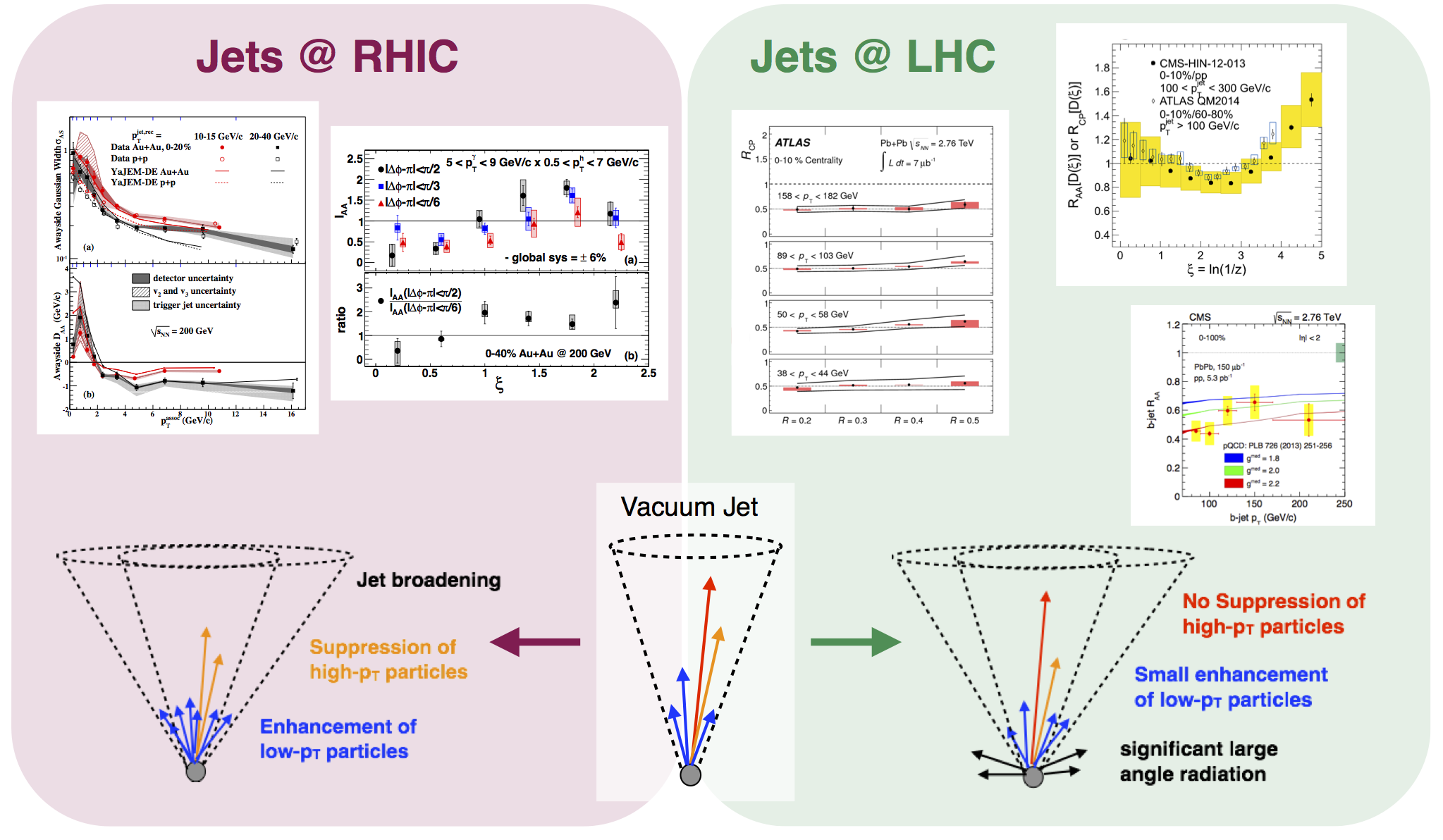}}
\caption[Qualitative summary of current jet/photon-hadron correlation measurements]{Qualitative summary of current jet/photon-hadron correlation measurements at RHIC (left panel) and selected full jet, $b$-jet $R_{AA}$ and jet fragmentation function measurements at the LHC (right panel). }
\label{fig:JetProgressFig2}
\end{figure}

\subsubsection{Jet modifications by the medium}~\label{jet-mod}
  
  While the experiments at RHIC have discovered that the medium created in heavy-ion collisions is opaque to energetic partons and initiated ground breaking attempts to measure the parton energy loss with jets, it was at the LHC where experimentalists have been able to leverage high kinematic reach of the collisions and unambiguously provide the evidence for jet momentum modifications. Utilizing the inclusive rates of high-energy jets and the momentum balance of back-to-back jets, photon-jet and hadron-jet coincidences, well described by pQCD in elementary/pp collisions, the experimental findings at the LHC strongly advanced the understanding of the depletion of the energetic jets in heavy-ion collisions. From these measurements two fundamental properties of jet-medium interactions have been established:
  \begin{itemize}
  \item the energy that is deposited into the medium is not recovered/contained within the jet cone as the rates of high-energy fully reconstructed jets in heavy-ion collisions are much less (about a factor of two) as compared to the expectation from proton-proton collisions (Figure\ \ref{fig:JetProgressFig2}, right panel);
  \item the interaction with the medium does not alter the direction of the propagating jet: although the distribution of the di-jet energy balance is strongly modified there is no sign of medium induced accoplanarity of the jet pairs.
  \end{itemize}
  Moreover, studies of the internal jet structure revealed that the recoiling jets that are 
selected to lose a significant portion of their energy on average show moderate or only small modifications of their fragmentation pattern as compared to jets fragmenting in vacuum (Figure\ \ref{fig:JetProgressFig2}, right panel). On the other hand, the subsequent search for the radiated energy has provided evidence that it is redistributed over large angles away from the jet axis and hadronizes into soft (of the order of GeV) particles. In contrast, measurements utilizing jet/photon-hadron correlations at RHIC indicate that (on average) the fragmentation pattern is suppressed at high \pT\ and significantly enhanced at low \pT\ accompanied by only a moderate broadening of the jet structure (Figure\ \ref{fig:JetProgressFig2}, left panel). 
There now exist several model studies~\cite{CasalderreySolana:2011rq,Qin:2010mn} as well as a Monte-Carlo study~\cite{Perez-Ramos:2014mna} based on pQCD which demonstrate similar effects. It should also be pointed out that, at a qualitative level, such observations are not inconsistent with expectations for how jets
should lose energy in a strongly coupled plasma that are based upon
calculations done in model systems that can be analyzed using holographic duality.
Reconciling these findings in a coherent and quantitative theoretical parton energy loss framework will ultimately require from the experimental side a suite of similar jet measurements at RHIC to allow a direct comparison to LHC results.
  
\subsubsection{Heavy quark energy loss}

Heavy quarks (in particular $b$ and $c$ quarks) provide both a systematic test of the underlying formalism of energy loss as well as access to a slightly different set of transport coefficients in the medium. As such they constitute a vital addition to the suite of jet observables. It should be made clear that here we refer to intermediate energy heavy quarks, i.e., quarks with a momentum $p$ relative to the medium satisfying $p \gtrsim M_{B}$, where $M_{B}$ is the mass of the bottom quark. 
Heavy quarks that have a lower momentum interact strongly  with the medium and will be covered in Section~\ref{Sec:HF}, while
quarks with momenta that are orders of magnitude larger than the mass can be treated as light quarks. It is in this intermediate momentum region where a considerable suppression in the yield of heavy quarks has been observed, either via the nuclear modification factor of non-photonic electrons at RHIC or via the nuclear modification factors of heavy flavor mesons at the LHC. At high momentum one expects heavy quark energy loss to be similar to that of light flavors.

\begin{figure}[!htp]
   \centering
   \begin{subfigure}[b]{0.7\textwidth}
        \includegraphics[width=0.90\textwidth]{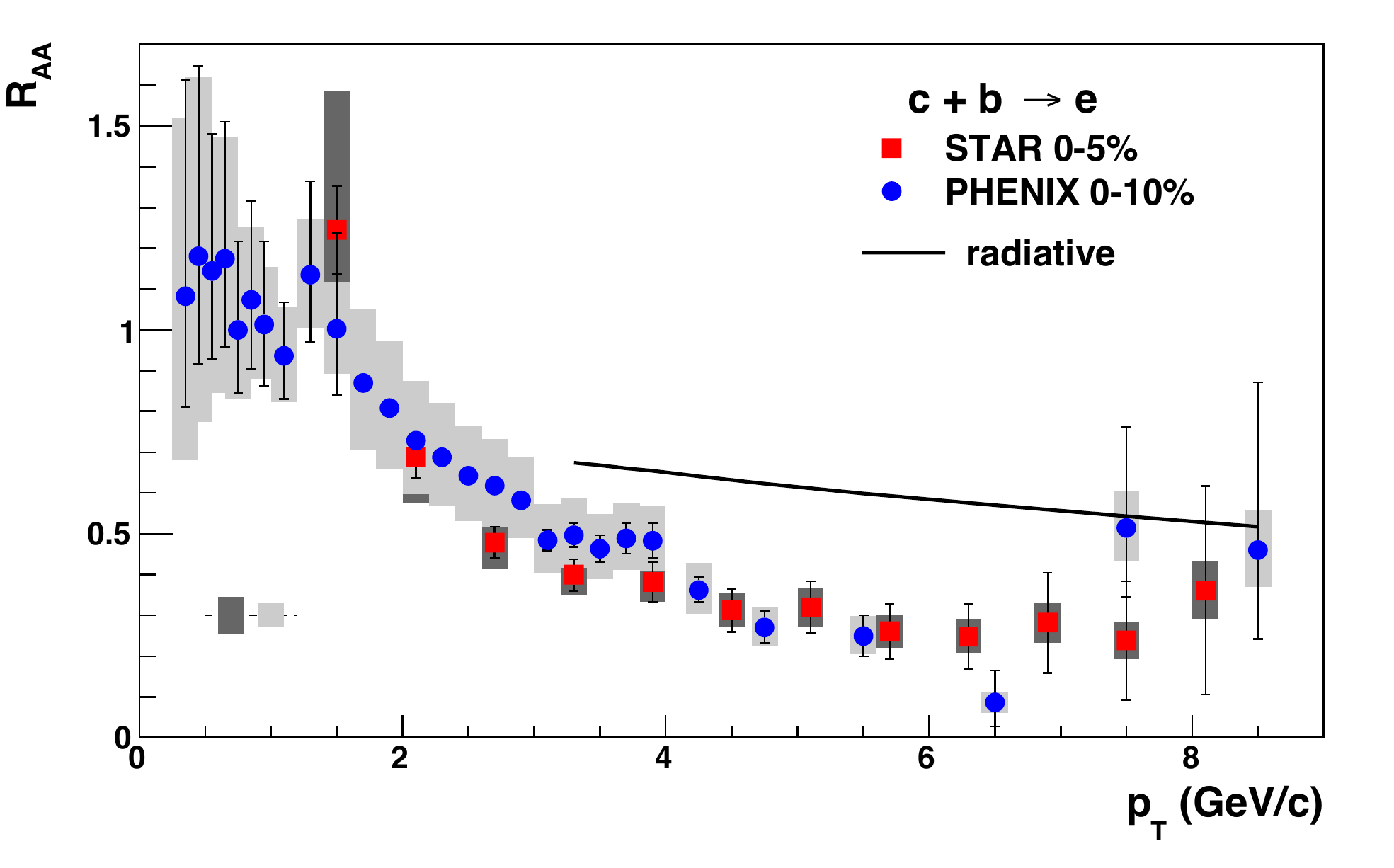}
       \caption{Calculations from ASW formalism\cite{Armesto:2003jh}.}
       \label{Fig:hqASW}
    \end{subfigure}
 
  \begin{subfigure}[b]{0.7\textwidth}
        \includegraphics[width=0.90\textwidth]{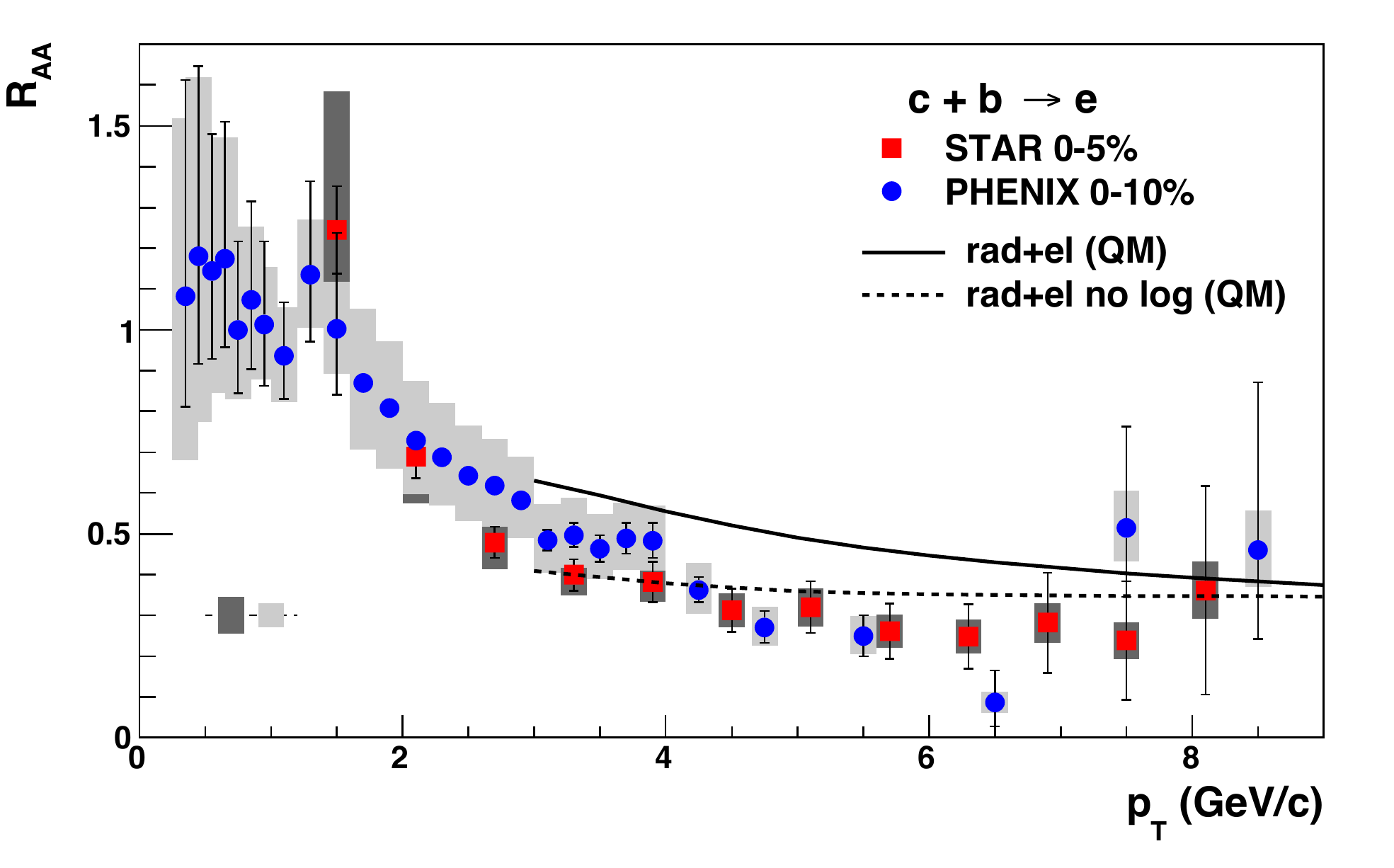}
       \caption{Calculations from Higher Twist formalism\cite{Qin:2009gw}.}
       \label{Fig:hqHT}
    \end{subfigure}
 
  \begin{subfigure}[b]{0.7\textwidth}
        \includegraphics[width=0.90\textwidth]{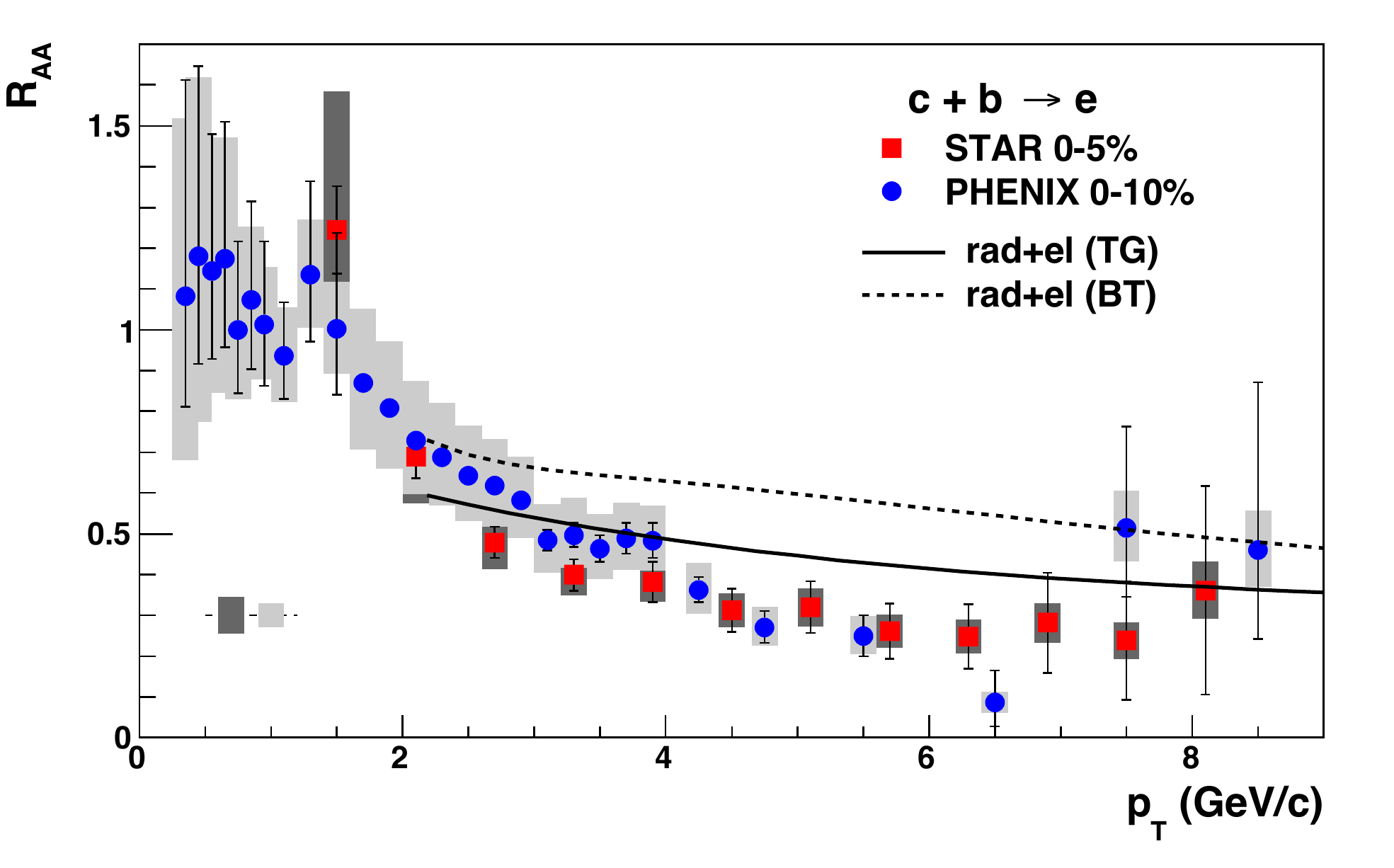}
       \caption{Calculations from WHDG scheme\cite{Wicks:2005gt}.}
       \label{Fig:hqWHDG}
    \end{subfigure}    
    
    \caption[Nuclear modification factor for semi-leptonic decay electrons from $B$ and $D$ mesons]{Nuclear modification factor for semi-leptonic decay electrons from $B$ and $D$ mesons, 
    as measured by the PHENIX\cite{Adare:2006nq} and STAR\cite{Abelev:2006db} collaborations compared to theory curves from various formalisms. The Higher Twist and WHDG contain both radiative and drag loss, the ASW calculations only contain radiative loss.}
\label{Fig:non-photonic-suppression}    
\end{figure}

The extra and unexpected suppression at intermediate momenta has been the subject of intense theoretical work over the last several years. In Figure~\ref{Fig:non-photonic-suppression}, we show the 
experimental measurements for the nuclear modification factor for non-photonic electrons (the data are identical in all three plots). 
These measurements are compared with theoretical calculations from the ASW~\cite{Armesto:2003jh}, 
Higher-Twist~\cite{Qin:2009gw}, and WHDG~\cite{Wicks:2005gt} schemes. Both the WHDG and Higher-Twist calculations contain drag loss in addition to radiative loss and fit the data much better than the ASW curve which 
only contains radiative loss. 


Such measurements and associated theory calculations showed already at RHIC that heavy quarks are much more sensitive to transport coefficients such as $\hat{e}$ than the quenching of light flavors.
Statistics accumulated at the LHC allowed for much more improved comparisons by observing the $B$ and $D$ contributions separately. Indeed, at the jet energies significantly higher than the quark mass the experimental measurements of high-\pT\  $b$-quark jets show a similar suppression pattern to that of inclusive jets. In contrast, comparison of $D$-meson production and non-prompt \JPsi\ from $B$-meson decays (for $p_{T}$ up to 20 GeV/c) indicate the predicted mass hierarchy: heavy quarks loose significantly less energy as compared to the lighter flavors. The extension of these $B$/$D$ separated measurements to RHIC is a major focus of the program discussed in Section~\ref{Sec:FacilitiesFuture}.
  
\subsubsection{Outlook and Conclusions}

  Even though these studies have provided a qualitative milestone in understanding of the jet-medium interactions they call for further investigations and demand precision. As but one example, the higher statistics of future runs at the LHC are needed both to determine precisely the in-cone modifications of the energy flux associated to the jet as well as to map out the full kinematic range of the observed phenomena down to lowest jet energies. Moreover, the higher precision is needed for gamma-jet coincidences and b-jets measurements, as well as new channels of studies will become available, such as $Z$-jet coincidences.
With these measurements and their rather complete toolkit at the LHC it is compelling to perform similar measurements at the lower center-of-mass energy provided by RHIC to determine the temperature dependence of the observed phenomena. The required program, which relies on RHIC's greatly increased luminosity and upgrades to PHENIX and STAR, is discussed in detail in Section~\ref{Sec:HardProbes}. 

%% file: tex/Saturation.tex
\subsection{Initial State for Plasma Formation and Low-$x$ Phenomena}

\begin{figure}[h!]
\centerline{
\includegraphics[width=0.75\textwidth]{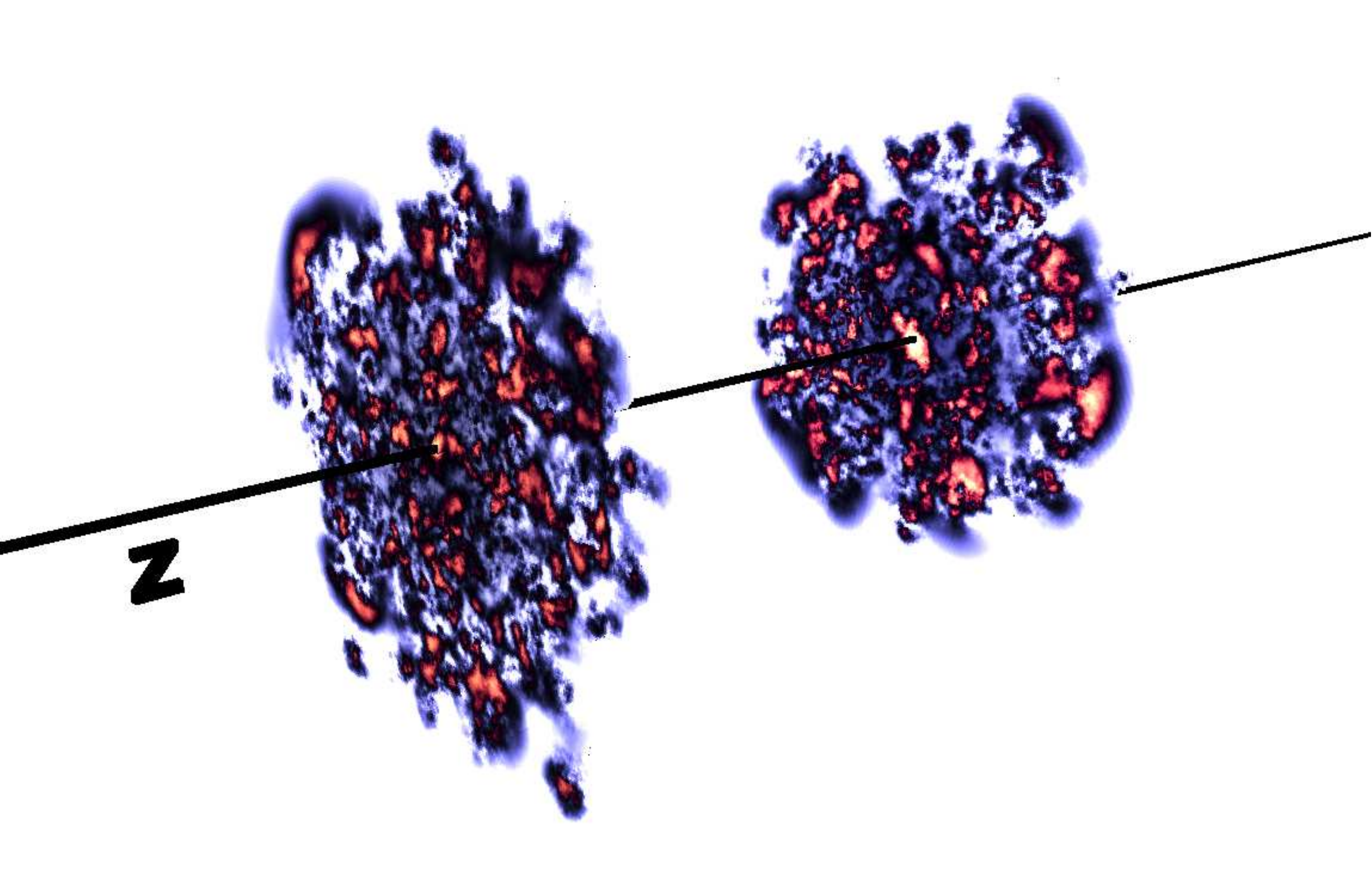}}
\caption[Color fields of the two nuclei before the collision]{Color fields of the two nuclei before the collision,
from~\cite{Schenke:2012fw}}
\label{fig:colorfield}
\end{figure}

\label{Sec:Saturation}

At the high collision energies of RHIC and the LHC, the  phase space available for radiating small-$x$ 
gluons and quark-antiquark pairs is very large. Since each emitted parton is itself a source of 
additional gluons, an exponentially growing cascade of radiation is created which would 
eventually violate unitarity. However, when the density of partons in the transverse plane
becomes large enough, softer partons begin to recombine into harder ones and the gluon 
density saturates. This limits the growth of the cascade  and preserves the 
unitarity of the $S$-matrix. 
The transverse momentum scale below which these nonlinear interactions dominate is known as 
the \emph{saturation scale} \qs . The saturation scale grows with collision energy, but also with the
size of the nucleus as $\qssqr \sim A^{1/3}$. For high enough energies \qs\ is large and
the corresponding QCD coupling weak $\as(\qs) \ll 1$. This makes it possible to calculate 
even bulk particle production using weak coupling methods, although the calculation is still  nonperturbative due to the large field strengths. Because the gluonic states have large 
occupation numbers, the fields behave 
classically. The classical field theory describing the saturation domain is
known as the ``Color Glass Condensate'' (CGC)~\cite{Gelis:2010nm}. 

The ideal probe of the CGC state are dilute-dense collisions, where a simple small
projectile collides with a large nucleus. At RHIC and the LHC proton-nucleus 
collisions provide such a tool for understanding saturation phenomena. 
Significant progress has been made in describing
the systematics of particle production as a function of transverse momentum and rapidity
in proton-proton and proton-nucleus collisions with CGC 
calculations~\cite{Albacete:2012xq,Tribedy:2011aa,Lappi:2013zma,Fujii:2013gxa,Kang:2013hta}, which are consistent with  the 
collinearly factorized perturbative QCD description~\cite{Helenius:2012wd} at high transverse momenta. 
The case of saturation effects in the multiparticle correlations as a function of azimuthal
angle and rapidity discussed in Section~\ref{Sec:Flow} remains more open. While there are contributions to these correlations 
that originate already in the nuclear wavefunctions~\cite{Dusling:2013oia}, experimental 
evidence points to strong collective behavior also in the final state of 
proton-nucleus and even proton-proton collisions. 
The versatility
of RHIC to systematically change the size of the projectile nucleus
and complement p+A with d+A, $^3$He+A etc. collisions over a wide
range of collision energies is unparalleled and a key to exploring
where these collective effects turn on.  

Yet another approach to probe the nuclear wave-function is provided by virtual photons
produced in ultra-peripheral \pA\ and \AplusA\ collisions. 
Such measurements are sensitive to the gluon structure of the nucleus at low $x$
as well as to cold nuclear matter absorption effects on produced hadrons such as the \Jpsi\ .
At RHIC studies have been of made of $\rho(1700$ production\cite{Abelev:2009aa}
and of coherent production of \JPsi\ 's and high-mass $e^+e^-$ pairs\cite{Afanasiev:2009hy}.
Much higher virtual photon fluxes are provided by the higher collision energies of the LHC, where detailed 
studies have explored the role of gluon shadowing 
in photoproduction of \JPsi\ 's in both \pPb\ \cite{TheALICE:2014dwa} and \PbPb\ collisions\cite{Abelev:2012ba,Abbas:2013oua,CMS:2014ies},
demonstrating sensitivities to Bjorken-$x$ values in both the proton and the Pb nucleus down to $x \sim 10^{-5}$.

Significant further insight into the structure of high energy nuclei
can be obtained from a polarized p+A program uniquely provided by RHIC\cite{Aschenauer:2015eha}.  
For example, it has been 
suggested~\cite{Kang:2011ni,Kovchegov:2013cva}
that comparing transverse single spin asymmetries measured in polarized p+p
and polarized p+A collisions for different nuclei and at different beam
energies could be very sensitive to the saturation scale $\qs$.
Further, as noted, proton-nucleus collisions have also provided surprises in their own right --- we now understand the resolution of these to be sensitive to the detailed spatial structure of partons in both protons and heavier nuclei. 
The importance of a polarized p+A program is therefore two-fold: (i) It will provide unique and essential information on the parton structure of proton and nuclear wave functions. (ii) The  implementation of this information in models of heavy-ion collisions will provide more sensitive  tests of and precise extraction of the parameters of the Little Bang Standard Model.
Precise and controlled access
to the high energy nucleus is needed to disentangle the effects of
strong collectivity in the initial wave functions and the final
state. An ideal complement to a polarized p+A program will be an Electron-Ion 
Collider that can measure the transverse and longitudinal 
structure of the small-$x$ gluons in nuclei.

Studying both electron-nucleus and proton-nucleus collisions will allow the direct comparison 
of a color neutral to a colored probe of the nuclear medium~\cite{Accardi:2012qut}.
In electron-nucleus collisions,
the parton kinematics are well measured. Furthermore, viewed in the appropriate
Lorentz frame,  the target is probed with a quark-antiquark dipole, which is 
theoretically well-controlled and tunable.
In proton-nucleus collisions, on the other hand, coherent multiple scattering effects
are more complicated.
The combination of data from electron-nucleus and proton-nucleus collisions
is needed to separate
the structure of the medium from the dynamics of the probe. 
The effect of the cold QCD medium on a colored probe must be consistent with theoretical
descriptions developed for partonic interactions in a hot and dense QCD medium.
Thus electron-nucleus and proton-nucleus collisions also 
provide an important control experiment for our theoretical understanding of jet quenching.

The theoretical description of the initial stage of quark-gluon plasma formation has become 
increasingly detailed. State of the art calculations~\cite{Gale:2012rq,Paatelainen:2013eea} 
now combine a 
fluctuating nuclear geometry with a microscopic QCD description of the dynamics of matter
formation that is consistent with our understanding of proton-nucleus collisions 
and deep inelastic scattering. This 
extends the description of the initial state geometry well beyond that provided
by Glauber modeling at the nucleonic level. As discussed in Section~\ref{Sec:Flow}, these initial state 
calculations, combined with detailed measurements of correlations and fluctuations in the
observed flow patterns, have 
helped 
to significantly improve the precision of the first quantitative experimental 
determinations of e.g. the viscosity/entropy ratio $\eta/s$. 


%% file: tex/HFandQuarkonia.tex
\subsection{Quarkonia and Open Heavy Flavor}
\label{Sec:HF}
 Hadrons containing heavy quarks (charm and bottom) play a special role in hot QCD matter studies.
	The heavy-quark masses, $m_{c,b}\simeq1.5,5$\,GeV, are large compared to the temperatures typically
	realized in the medium produced in heavy-ion collisions. This has important consequences.
	First, the production of heavy quarks is essentially restricted to the initial impact of the
	incoming nuclei. After that, they ``diffuse" through the produced medium, much like Brownian
	particles. The modifications of their momentum spectra relative to the initial spectra can thus
	serve as a rather direct measure of their coupling strength to the hot medium. This is so because 
        their thermal relaxation time is increased compared to that for light particles by a factor of order 
        $m_Q/T\simeq$~5-20, implying that heavy quarks in the QGP (or hadrons in hadronic matter) are 
        not expected to fully thermalize over the course of the fireball lifetime, and thus retain a 
        ``memory" of their interaction history. Moreover, at small and intermediate momenta, the heavy-quark 
        interactions become dominantly elastic and may be amenable to a potential-type prescription. 
        In the vacuum, the potential approach is well established for the description of heavy quarkonia, 
        i.e., bound states of a heavy quark and its anti-quark. The vacuum potential is characterized 
        by a color-Coulomb interaction at short distances and a linearly rising ``string" term at 
        intermediate and large distance associated with confinement. When embedded into a QGP, the 
        properties of heavy quarkonia in the QGP will thus reflect the modifications of the in-medium
        QCD potential. With increasing temperature one expects a subsequent dissolution of the 
        bound states, as the medium-induced screening penetrates to smaller distances. However, this
        simple picture is complicated by inelastic collisions with medium particles, leading to 
        a dynamical dissociation of the bound state. These processes generate in-medium widths and
        need to be accounted for in the description of quarkonium spectral functions at finite
        temperature. In heavy ion collisions, the in-medium properties of quarkonia have to be inferred
        from their production yields and momentum spectra in the final state. A thorough interpretation
        of the data thus requires systematic analyses of the centrality, energy and species dependence
        of quarkonia. This usually requires transport approaches to calculate the time evolution of 
        quarkonium distributions including both dissociation reactions and formation processes
        through recombination of diffusing heavy quarks.   
     	In the following we briefly summarize recent progress in both open and hidden heavy flavor probes of
        hot and dense QCD matter.

\input{tex/Quarkonia}

\input{tex/OpenHeavyFlavor}

%% file: tex/Quarkonia.tex
\subsubsection{Quarkonia Suppression and Enhancement}
\label{Sec:Quarkonia}

Theoretical studies of heavy quarkonia in QCD matter are carried out both from first 
principles using lattice-discretized QCD computations~\cite{Bazavov:2009us,Kaczmarek:2012ne} 
and within  effective model approaches. The latter aid in interpreting the lattice results and 
form the bridge to phenomenological implementations in heavy-ion collisions.

One of the key quantities computed in lattice QCD is the free energy, $F_{Q\bar Q}(r,T)$, 
of a static $Q\bar Q$ pair at distance $r$ in a medium at temperature $T$. In the vacuum it reduces to the well-known Cornell 
potential between the two quarks, but in the medium an additional entropy term develops, 
$F_{Q\bar Q}= U_{Q\bar Q} - T S_{Q\bar Q}$. Accurate calculations for $F_{Q\bar Q}(r,T)$, which are 
now available for full QCD with realistic light quark masses, show marked deviations from 
its vacuum form, through a progressive ``Debye screening" toward smaller distances 
as temperature increases. When utilizing the free energy as a potential in a Schr\"odinger 
equation~\cite{Digal:2001iu}, the charmonium groundstate, \Jpsi\, dissolves slightly above 
the critical temperature, while the excited states melt at or even below $T_{c}$; only the 
bottomonium groundstate, $\Upsilon$, survives up to 2-3\,$T_c$. On the other hand, if the 
potential is approximated with the internal energy, $U_{Q\bar Q}$, the \Jpsi\ survives to 
significantly higher temperatures, possibly up to 2\,$T_c$. These limiting cases are sometimes 
referred to as weak- and strong-binding scenarios, respectively. It has also become clear 
that a proper inclusion of absorptive processes in the $Q\bar Q$ system, characterizing its 
dissociation by dynamical processes, is essential to arrive at a realistic and quantitative description 
of its in-medium properties. 

Lattice computations also provide detailed information on $Q\bar Q$ correlation functions 
in euclidean space-time (where the ``imaginary-time" coordinate is related to the temperature 
of the system). Extracting the physical (real-time) $Q\bar Q$ spectral function requires an 
inverse integral transform on a limited number of ``data" points. This usually is done using 
maximum likelihood methods~\cite{Asakawa:2000tr,Aarts:2007pk}, sometimes guided by a 
parametric ansatz for the spectral function~\cite{Ding:2012sp}. In the bottomonium sector one 
can additionally utilize nonrelativistic heavy-quark effective theory~\cite{Aarts:2014cda}. 
The heavy-quark transport coefficient has been extracted from the low energy limit of the 
spectral functions in quenched QCD (no dynamical 
quarks)~\cite{Banerjee:2011ra,Ding:2012sp,Kaczmarek:2014jga}, highlighting the close connection 
between quarkonia and open heavy flavor in the QGP (cf.~Section~\ref{Sec:OpenHF}). The 
``reconstructed" charmonium spectral functions from lattice QCD have not yet led to definite
conclusions about the fate of bound states in the QGP. A promising complementary source of 
information are spatial correlation functions~\cite{Karsch:2012na,Bazavov:2014cta}; they are 
related to the 3-momentum dependence of the quarkonium spectral functions, but also encode 
information on modifications of the spectral (energy) strength in medium.   

Fruitful connections to lattice QCD results have been established with potential models, 
either through a Schr\"odinger equation~\cite{Mocsy:2005qW} or a thermodynamic $T$-matrix 
formulation~\cite{Cabrera:2006wh}. Here, the calculated  spectral functions can be 
straightforwardly integrated to obtain the euclidean correlation functions computed on the 
lattice. Both strong- and weak-binding scenarios for the $Q\bar Q$ appear to be compatible 
with the lattice data, once absorptive parts and a proper treatment of the continuum are 
included in the calculations. A better discrimination power arises when analyzing the 
interactions of open heavy flavor with the medium. This has been pursued in the $T$-matrix 
approach~\cite{Riek:2010fk}; only the strong-binding scenario produces sufficient strength 
in the heavy-quark transport coefficient to come close to open heavy-flavor observables in 
heavy-ion collisions (cf.~Section~\ref{Sec:OpenHF}).      
Recent progress in the determination of the in-medium $Q\bar Q$ potential has been made by 
utilizing the spectral functions of the so-called Wilson loop correlator, including absorptive
parts~\cite{Burnier:2014ssa}. Here, the real part of the potential turns out to be close to 
the free energy. Further theoretical work, exploiting the insights derived from both lattice QCD and 
effective models, is needed to clarify these issues and arrive at a consistent picture of 
quarkonia and open heavy flavor in the QGP.

To confront (and eventually extract) the equilibrium properties of quarkonia with (from) 
data in heavy-ion experiments, one typically adopts Boltzmann-type transport approaches to 
model the space-time evolution of quarkonium phase-space distributions, from their initial
production until the fireball freezes out. The quarkonium equilibrium properties 
are functions of 
the binding energy, in-medium heavy-quark masses (defining the continuum threshold) 
and inelastic reaction rates, e.g., $g$ + \Jpsi\ $\rightarrow$ $c$ + $\bar c$. In some 
approaches~\cite{Zhao:2010nk,Emerick:2011xu}, the in-medium quarkonium properties 
implemented into the transport equation have been checked against the euclidean correlators 
from lattice-QCD described above. Here too, current heavy ion phenomenology appears to favor a
strong-binding scenario~\cite{Zhao:2010nk,Liu:2010ej,Emerick:2011xu,Strickland:2011aa},
which would be consistent with the phenomenology for open heavy flavor.

The principle of detailed balance requires that one account for quarkonium formation reactions 
(called ``regeneration" or ``coalescence"), if the quarkonium state under 
consideration exists at the local medium temperature (relative to the ``melting temperature"). 
An accurate assessment of quarkonium formation reactions not only requires knowledge of the 
reaction rate, but also of the phase-space distributions of open heavy flavor. Again, this 
couples the problem of quarkonium production with heavy-flavor diffusion,
providing both challenges and opportunities. 

The quarkonium transport equations need to be evolved over a realistic space-time 
evolution of a given collision system and energy. Current approaches include expanding 
thermal fireball models with QGP and hadronic phase~\cite{Grandchamp:2003uw,Zhao:2010nk} 
or co-mover interactions~\cite{Capella:2007jv}, ideal hydrodynamics~\cite{Zhu:2004nw}, and
viscous hydrodynamics with local momentum anisoptropies~\cite{Strickland:2011aa}. Systematic investigations
of how sensitive the final results are to details of medium evolution models need to be 
conducted. Most of the calculations include cold nuclear matter (CNM) effects to construct 
the initial conditions of the quarkonium phase space distributions, as well as a gain 
term in the rate equation which is essential for a realistic description of charmonia 
at collider energies. For the $\Upsilon$ states, regeneration effects seem to be rather 
small, even at the LHC~\cite{Emerick:2011xu}.
Quarkonium production has also been evaluated via statistical recombination at the
QCD phase boundary~\cite{BraunMunzinger:2000px,Andronic:2003zv,Andronic:2007bi}. In essence,
this approach corresponds to the equilibrium limit of the transport approaches, but it 
allows for a more complete incorporation of open and hidden charm hadrons to account for 
their relative chemical equilibrium (at fixed charm-quark number).

An extensive program of \Jpsi\ measurements in A+A collisions has been carried out at the SPS
($\sqrt{s_{NN}}$ = 17.3 GeV), RHIC ($\sqrt{s_{NN}}$ = 200 GeV) and the LHC ($\sqrt{s_{NN}}$ = 2.76 TeV).
These measurements were motivated by the possibility of observing signals of color deconfinement through the
suppression of \Jpsi\ in the case of QGP formation\cite{Matsui:1986dk}. In fact, a strong suppression of the \Jpsi\ is observed
at all three energies, but it has become clear that the manifestation of color screening in the observed
modifications can not be uniquely determined without a good understanding of two significant competing effects.
The first of these is the modification of \Jpsi\ production cross sections in a nuclear target (known as
cold nuclear matter effects); it has been addressed at RHIC using d+Au collisions and at the SPS and LHC
using p+Pb collisions. The second effect is the recombination of charm and anticharm discussed above.

\begin{figure}[!htb]
\centerline{
\includegraphics[width=0.8\textwidth]{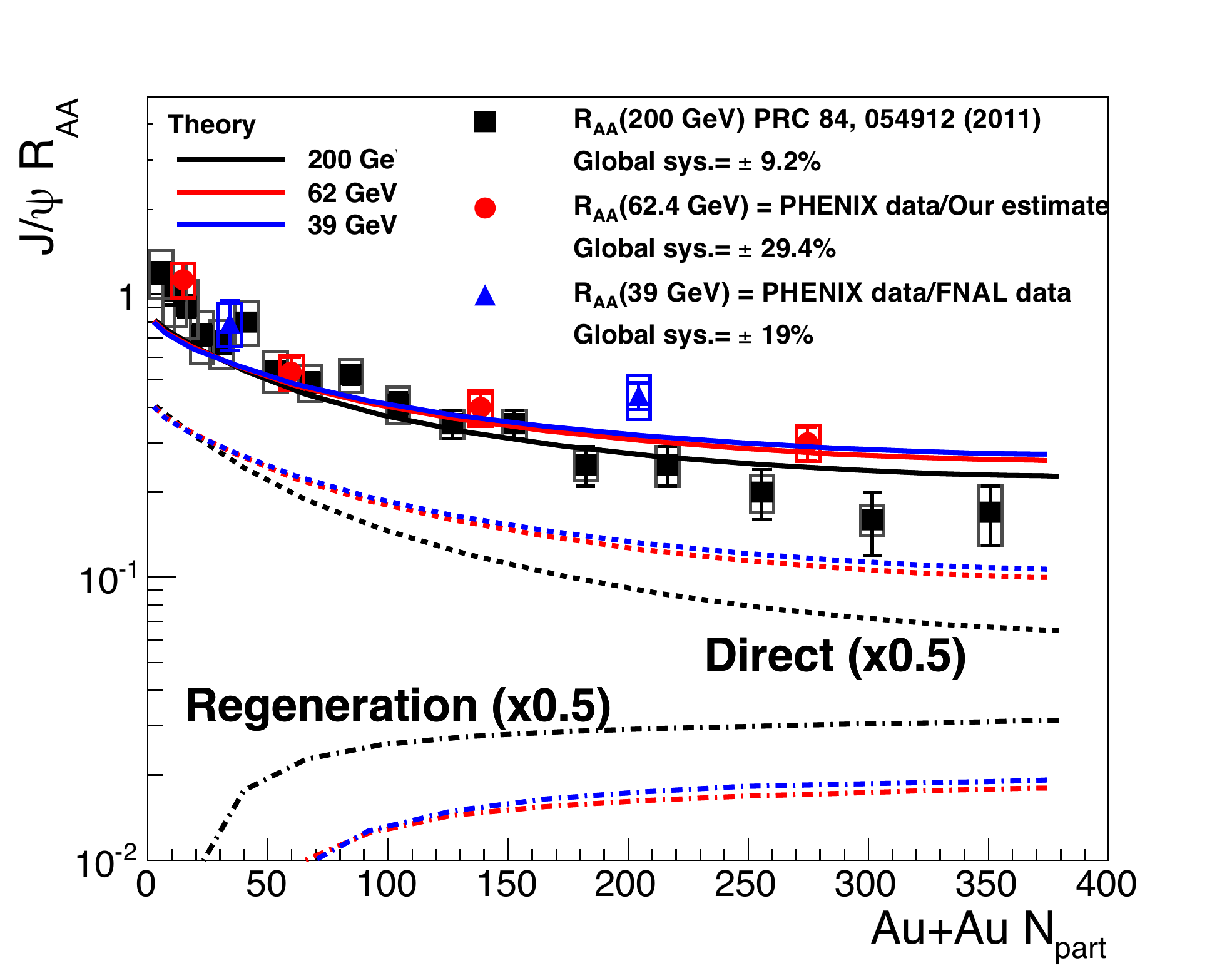}
}
\caption[PHENIX data for \Jpsi\ production compared to theory calculations]{The nuclear modification factor for \Jpsi\ production for $\sqrt{s_{NN}}$=200, 62 and 39~GeV \AuAu\ collisions from
PHENIX~\cite{Adare:2012wf}, compared with theory calculations~\cite{Zhao:2010nk} showing the contributions from direct
suppression and regeneration.
}
\label{fig:PHENIX_excitation_function}
\end{figure}

Using p+Pb and d+Au data as a baseline, and under the assumption that cold nuclear matter effects
can be factorized from hot matter effects, the suppression in central collisions due to the presence of
hot matter in the final state has been estimated to be about 25\% for Pb+Pb at the SPS~\cite{Arnaldi:2010ky},
and about 50\% for Au+Au at RHIC~\cite{Brambilla:2010cs}, both measured at midrapidity. The modification
of \Jpsi\ production in Au+Au collisions has been measured at $\sqrt{s_{NN}}$=39, 62 and 200 GeV 
by PHENIX~\cite{Adare:2012wf}
and STAR~\cite{Adamczyk:2013tvk,Zha:2014nia}.
The results for the rapidity range $1.2 < |y| < 2.2 $ is shown in Figure~\ref{fig:PHENIX_excitation_function}, 
where it is compared with a theoretical
calculation~\cite{Zhao:2010nk} that includes cold nuclear matter effects and the effects of regeneration.
The experimental observation is that the modification is similar at all three collision energies, in spite of the
differences in energy density.
Similar observations apply to data measured by STAR at mid-rapidity ($|y| < 1$)~\cite{Adamczyk:2013tvk,Zha:2014nia}.
In the model calculation, this is expected because the increased direct suppression at the 
higher energy due to stronger Debye screening is nearly compensated by the increase in the regeneration 
component~\cite{Grandchamp:2001pf}.
	
\begin{figure}[!ht]
 \centering
 \includegraphics[width=0.435\linewidth]{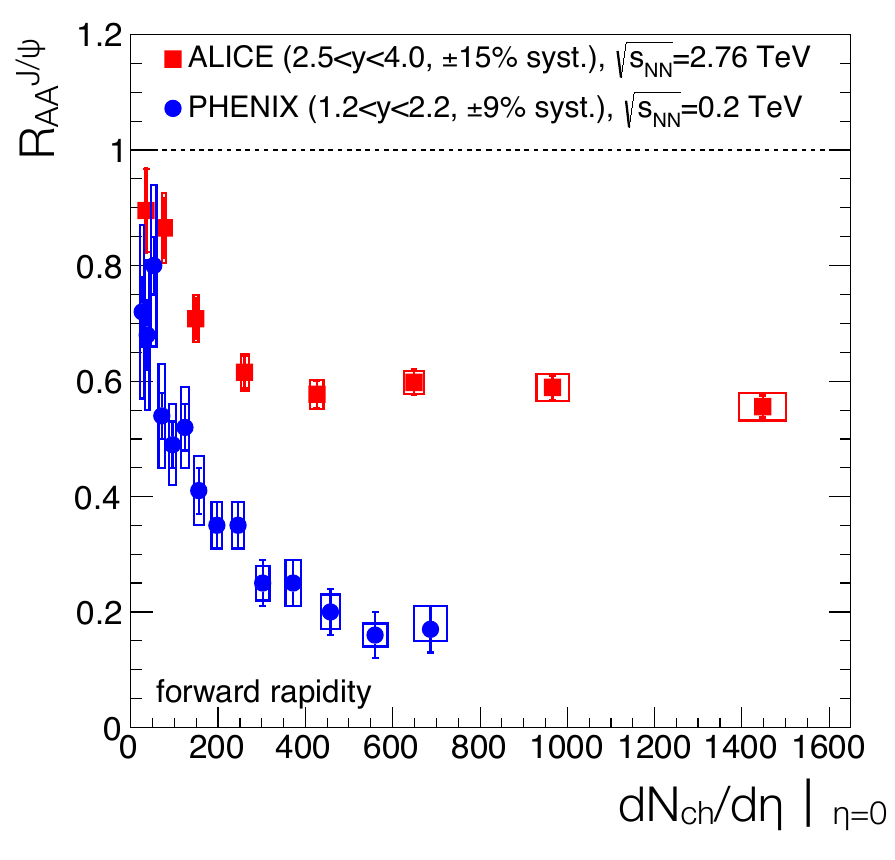}
 \includegraphics[width=0.46\linewidth]{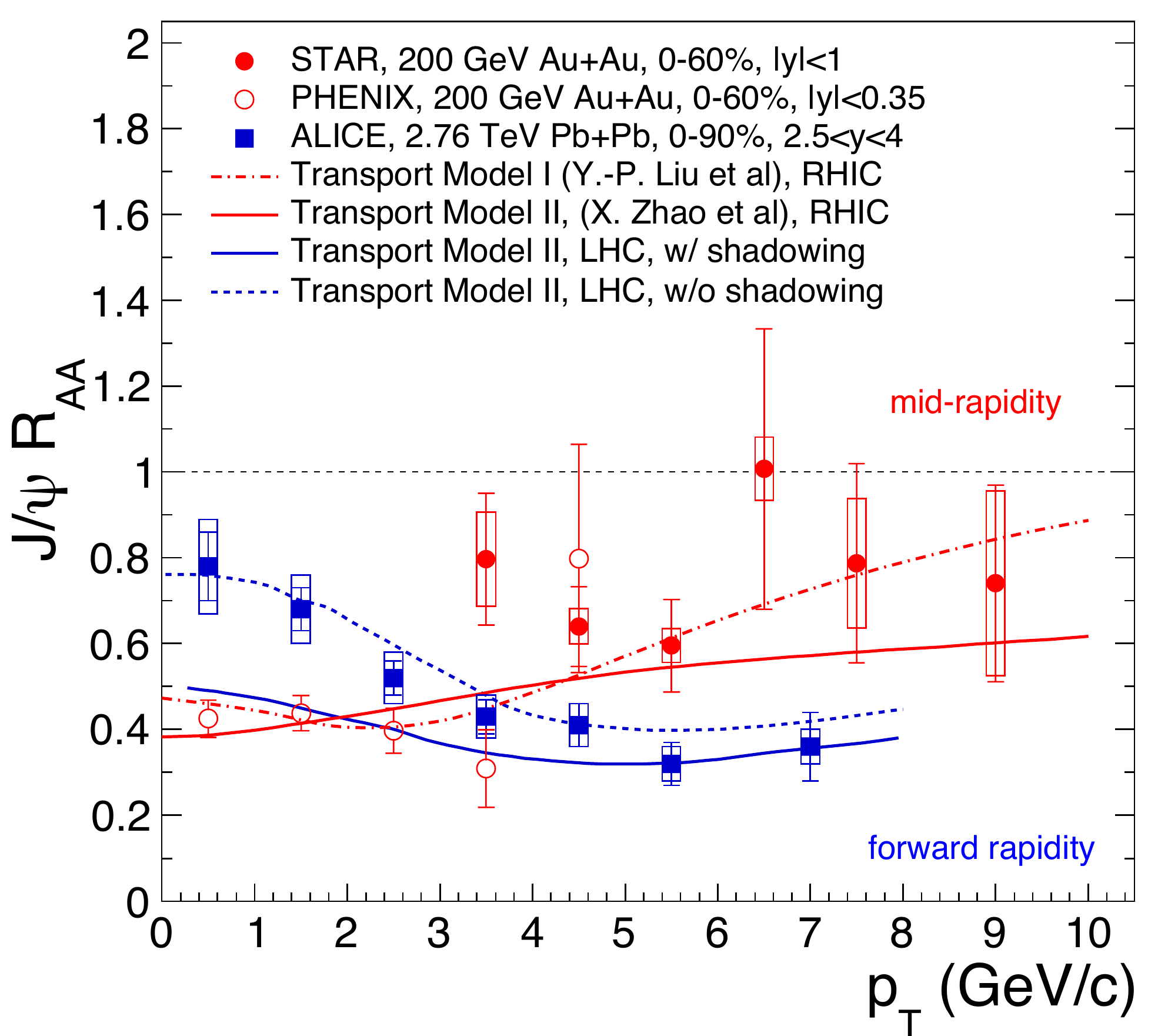}
 \caption[Comparison of RHIC and LHC data on \Jpsi\ production]{
 Left: Comparison of the nuclear modification factor for \JPsi\ production for $\sqrt{s_{NN}}=200$~GeV \AuAu\ collisions from
PHENIX and for $\sqrt{s_{NN}}=2.76$~TeV \PbPb\ from ALICE~\cite{Andronic:2014zha}. The data are plotted versus the charged
particle multiplicity at midrapidity, which is used as a rough proxy for energy density. 
Right: The transverse momentum
distributions for the same data sets,
together with higher-momentum data from STAR\cite{Adamczyk:2012ey}
showing the low momentum enhancement that would be expected from a
large coalescence contribution at the LHC energy.
The data are compared to a transport model\cite{Liu:2009nb}
which incorporates the effects of gluon scattering and coalescence.
}
\label{fig:ALICE_PHENIX_Jpsi_RAA}
\end{figure}

The first \Jpsi\
data in Pb+Pb collisions at $\sqrt{s_{NN}}$ = 2.76 TeV from ALICE~\cite{Abelev:2012rv}, measured at forward
rapidity, are shown alongside forward rapidity PHENIX data in the left panel of Figure~\ref{fig:ALICE_PHENIX_Jpsi_RAA}. 
The suppression in central collisions is found to be far greater at RHIC than at the LHC. A similar result is found
at midrapidity. This is consistent with a predicted~\cite{Zhao:2011cv} strong coalescence component due to the 
large production rate of charm and anti-charm quarks in a central collision at the LHC. This explanation is 
corroborated by the transverse-momentum spectra~\cite{Adare:2006ns,Adamczyk:2012ey,Abelev:2013ila}, 
which exhibit the expected~\cite{Zhao:2011cv,Liu:2009nb}
low-momentum enhancement generated by the coalescence contribution, 
as shown in the right panel of 
Figure~\ref{fig:ALICE_PHENIX_Jpsi_RAA}.

\begin{figure}[!htb]
\centerline{
\includegraphics[width=0.8\textwidth]{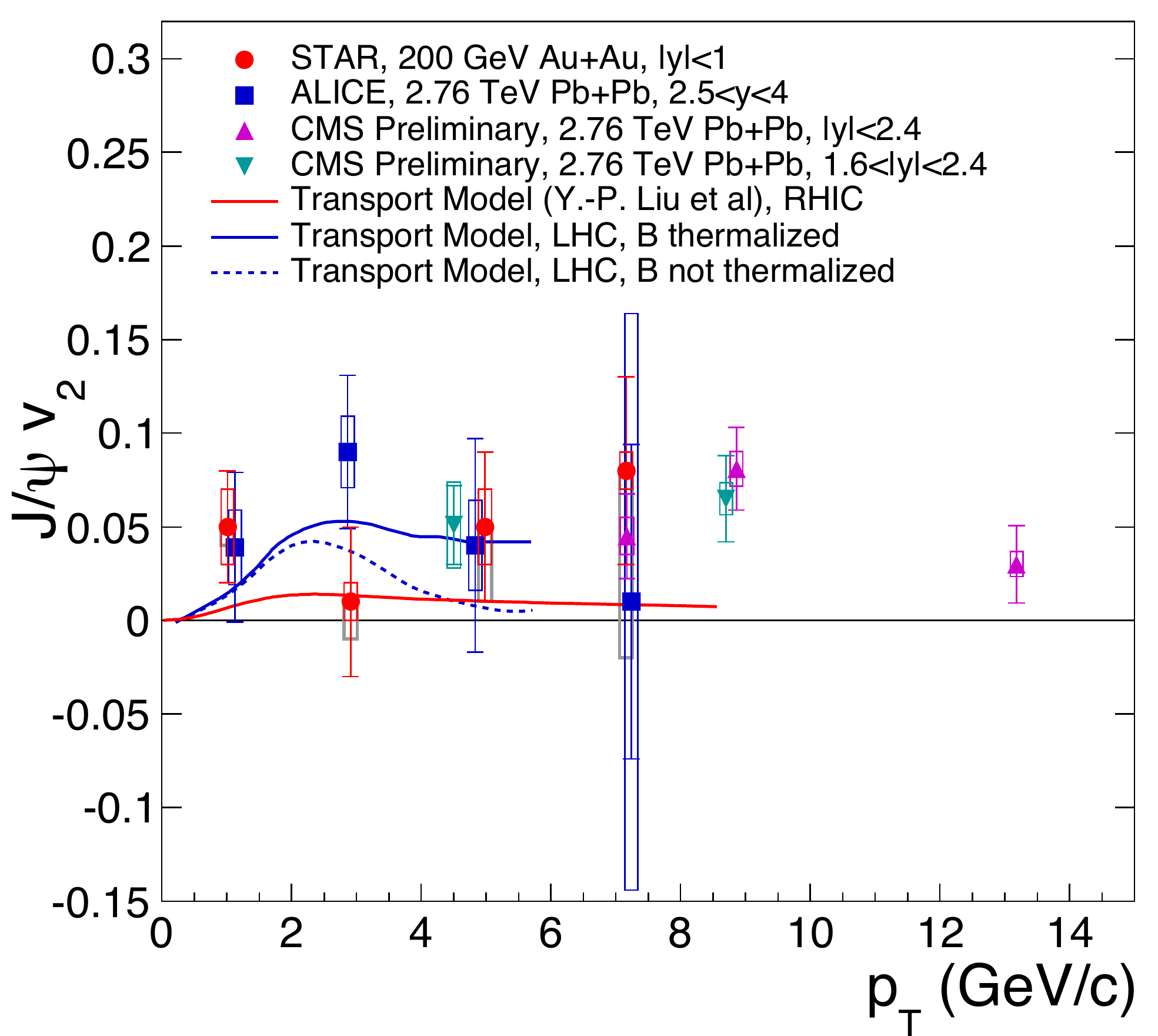}
}
\caption[\Jpsi\ elliptic flow measurements at RHIC and LHC compared to theory]{
The transverse momentum dependent elliptic flow measurements of
\Jpsi\'s from STAR\cite{Adamczyk:2012pw}, ALICE\cite{ALICE:2013xna}, and CMS\cite{Moon:2014lia} compared with the theoretical calculations\cite{Liu:2009gx}.
}
\label{Fig:JPsi_v2}
\end{figure}

One of the crucial tests of the combined effects of color screening
and coalescence in \Jpsi\ production is provided by the measurement of the \Jpsi\ elliptic flow.
All other hadrons acquire flow velocities through their interaction with 
the QGP medium and simultaneously exhibit
significant elliptic flow and strong nuclear suppresssion. 
In contrast, disassociation of \Jpsi\'s from color screening
is predicted to result in strong nuclear modification with the absence of a flow signal, while
\Jpsi\'s later regenerated from coalescence production will carry flow from the thermalized charm quarks from which they are formed. 
Figure~\ref{Fig:JPsi_v2} 
shows the transverse momentum dependent elliptic flow measurements\cite{Adamczyk:2012pw,ALICE:2013xna,Moon:2014lia} 
from RHIC and the LHC together with a corresponding theoretical calculation\cite{Liu:2009gx}.
The data are consistent (within the current large statistical errors) with a model incorporating
disassociation from color screening at both RHIC and the LHC in combination with
significant regeneration of \Jpsi\'s at the LHC from coalescence of thermalized charm and anti-charm quarks,
The larger role of the coalescence  mechanism at the LHC results from the much 
higher cross section for charm production in the initial phase of the collision at LHC energies as compared to RHIC.
There is great promise that the  future availability of data sets 
with improved statistical precision
at the widely spaced collision energies of RHIC and the LHC, 
in combination with studies constraining CNM effects with \pA\ data, 
will lead to a quantitative
understanding of the role of coalescence from nearly thermalized charm and anti-charm quarks.
However, a more direct window on the Debye screening and dissociation effects alone is expected from
a systematic analysis of bottomonium production, as we will now discuss.
		
Bottomonium production is believed to have several advantages over charmonia as a probe of deconfinement in the
QGP. First, the $\Upsilon(1S)$, $\Upsilon(2S)$ and $\Upsilon(3S)$ states can all be observed with comparable
yields via their dilepton decays. Second, bottom production in central collisions is $\sim$ 0.05 pairs at RHIC
and $\sim$ 5 pairs at LHC~\cite{Brambilla:2010cs}. At RHIC, one expects this to effectively remove any contributions
from coalescence of bottom and anti-bottom quarks (although some care has to be taken, since the ratio of bottomonium over
open bottom states in pp collisions is $\sim$0.1\% and thus a factor of 10 smaller than in the charm
sector; thus, even small regeneration, even from a single pair in the reaction, can potentially be significant).
This makes the $\Upsilon$ suppression at RHIC dependent primarily on color screening and dissociation reactions,
as well as cold nuclear matter effects. Recent theoretical calculations~\cite{Emerick:2011xu} support the
assertion that the coalescence production for $\Upsilon$'s is small at RHIC. At LHC energies, 
bottom coalescence could become comparable with charm coalesence at RHIC, i.e. at the 10's of percent level.
Since the $\Upsilon(1S)$, $\Upsilon(2S)$ and $\Upsilon(3S)$ have a broad range of radii, precise measurements
of bottomonia modifications at RHIC and LHC energies will provide information over a large range of
binding energies at two widely different initial temperatures, for a case where the modification is dominated
by Debye screening effects.
	
\begin{figure}[!htb]
\centerline{
\includegraphics[width=0.7\textwidth]{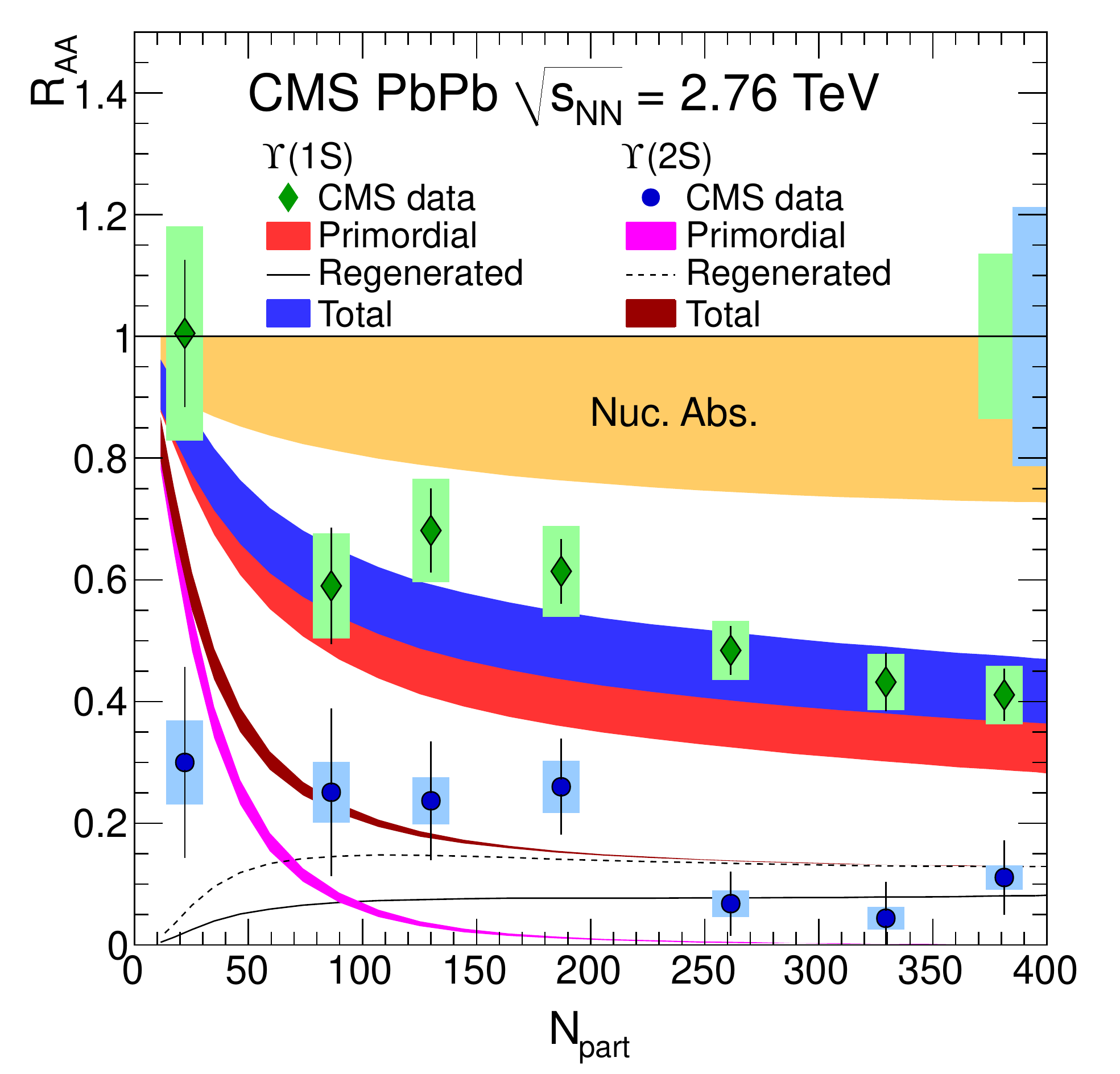}
}
\caption[CMS measurements of $\Upsilon$ production compared to theory]{The nuclear suppression factor $R_{AA}$ for the $\Upsilon(1S)$ and $\Upsilon(2S)$ states measured at $\sqrt{s_{NN}}$=2.76 TeV
by CMS~\cite{Chatrchyan:2012lxa}. The theory calculation~\cite{Emerick:2011xu} includes cold nuclear matter effects
and the contributions from the surviving primordial \Jpsi\ and those which are formed by regeneration are shown, as well
as the total.
}
\label{fig:CMS_Upsilons}
\end{figure}

The CMS experiment at the LHC has mass resolution that is sufficient to cleanly separate all three of the
$\Upsilon$ states at midrapidity using dimuon decays~\cite{Chatrchyan:2012lxa}. The data obtained in Pb+Pb collisions at
$\sqrt{s_{NN}}$=2.76 TeV for the $\Upsilon(1S)$ and $\Upsilon(2S)$ states are shown in
Figure~\ref{fig:CMS_Upsilons}, where they are compared with a model calculation~\cite{Emerick:2011xu} that includes
both cold nuclear matter effects and regeneration. The data show much stronger suppression of the $\Upsilon(2S)$
than the $\Upsilon(1S)$. The $\Upsilon(3S)$ is even more strongly suppressed, 
and as a result the yield is too small to determine accurately the nuclear suppression factor. 
The theory is in good agreement with the data, although better statistical precision is needed for strong
theoretical constraints. Future Pb+Pb data at the LHC will be measured at $\sqrt{s_{NN}}$=5.5 TeV, and will have greatly increased
statistical precision.

\begin{figure}[!htb]
\centerline{
\includegraphics[width=0.9\textwidth]{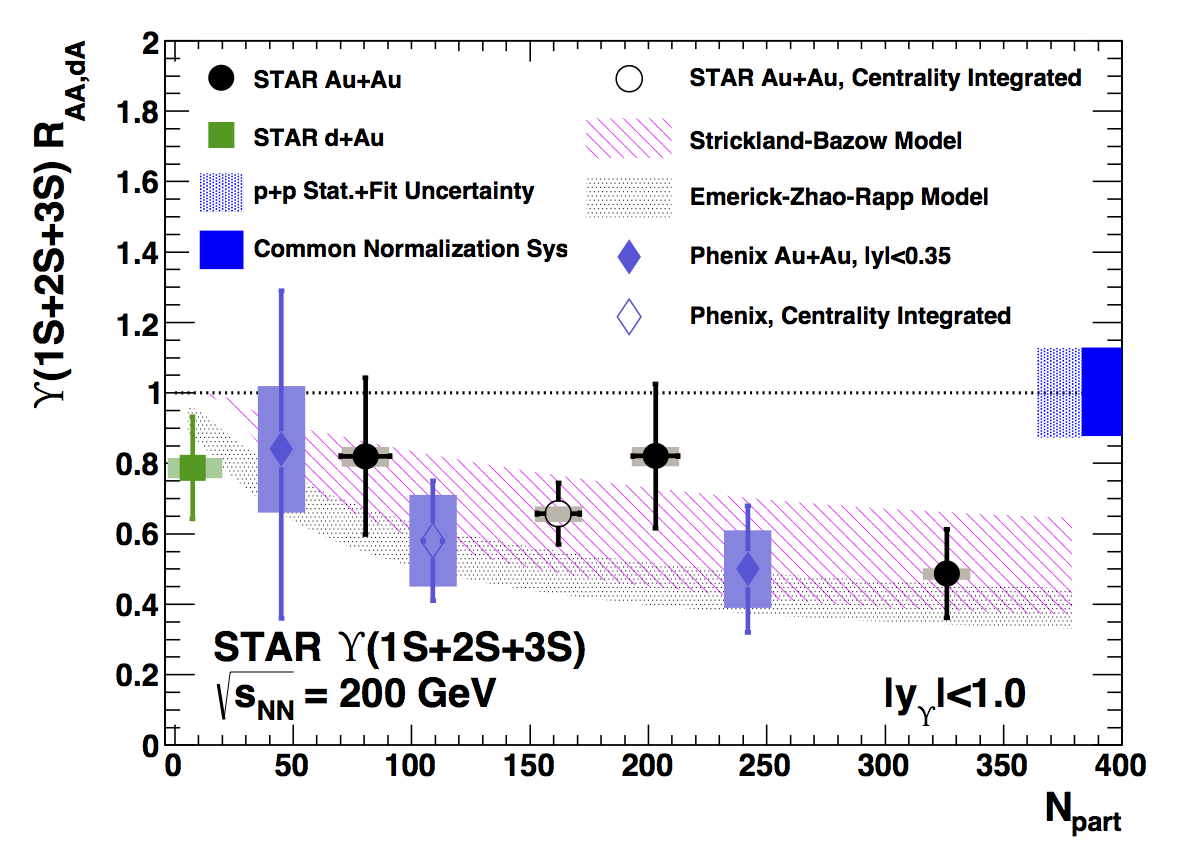}
}
\caption[RHIC measurements of $\Upsilon$ production compared to theory]{The nuclear suppression factor $R_{AA}$ for the $\Upsilon(1S+2S+3S)$ states measured at $\sqrt{s_{NN}}$=200 GeV
by PHENIX\cite{Adare:2014hje} and STAR~\cite{Adamczyk:2013poh}. The theory calculations are from~\cite{Emerick:2011xu,Strickland:2011aa},
}
\label{fig:RHIC_Upsilons}
\end{figure}

By the end of Run 3 at
the LHC (approximately 2023) CMS will have measured very precise cross sections for the three
$\Upsilon$ states in p+p, p+Pb and Pb+Pb collisions. A mass-resolved measurement of the modifications
of the three upsilon states with similar precision at RHIC energy would be extremely valuable for all of
the reasons outlined above. However $\Upsilon$ measurements at RHIC have been hampered by a
combination of low cross sections and acceptance, and insufficient momentum resolution to resolve the
three states. At RHIC there are measurements of the modification of the three states
combined in Au+Au by PHENIX~\cite{Adare:2014hje} and STAR~\cite{Adamczyk:2013poh}. These data
are shown in Figure~\ref{fig:RHIC_Upsilons}, along with two theory
calculations\cite{Emerick:2011xu,Strickland:2011aa} of the modification of the three Upsilon states combined.
The data available so far have limited statistical precision, and do not place strong constraints on models.
	
There are, however, good prospects for future $\Upsilon$ measurements at RHIC. STAR recorded data in the 2014
RHIC run with the new Muon Telescope Detector (MTD), which measures dimuons at midrapidity~\cite{Ruan:2009ug}.
The MTD has coverage of $|\eta| < 0.5$, with about 45\% effective azimuthal coverage. It will have a
muon to pion enhancement factor of $\sim$ 50, and the mass resolution will provide a clean separation
of the $\Upsilon(1S)$ from the $\Upsilon(2S+3S)$, and likely the ability to separate the 2S and 3S states by fitting.
	
On a longer time scale, the proposed sPHENIX detector~\cite{Aidala:2012nz} at RHIC 
discussed in Section~\ref{Sec:FacilitiesFuture} would begin operation in
2021. It is designed to measure $\Upsilon$'s via their dielectron decays at midrapidity. The 100 MeV mass
resolution is sufficient to cleanly separate all three $\Upsilon$ states. Pions are suppressed relative to electrons
by a factor of 90. A combination of very high luminosity, good mass resolution, good background rejection and large
acceptance (about a factor of 7 larger than the MTD) leads to a data set with precision comparable to that expected
from the CMS data by 2023, and on a similar time scale.

%% file: tex/OpenHeavyFlavor.tex
\subsubsection{Open Heavy Flavor Dynamics}
	\label{Sec:OpenHF}

 The diffusion of a heavy particle through a heat bath of light particles can be 
 quantified by the spatial diffusion coefficient $D_s$. In 
 relativistic systems, such as the QGP, it is convenient to express this transport 
 coefficient in units of the thermal wavelength of the medium, $1/2\pi T$. This renders 
 $D_s(2\pi T)$ a dimensionless quantity which characterizes the (inverse) coupling 
 strength of the diffusing particle to the medium. As such, it is expected to be 
 proportional to the ratio of shear viscosity to entropy, $\eta/s$. For example, in 
 the strong coupling limit of conformal field theories, one has 
 $D_s(2\pi T) \simeq 1$~\cite{Herzog:2006gh,CasalderreySolana:2006rq} and 
 $\eta/s=1/4\pi$ (where small values for each of these quantities indicate strong coupling). 
 
In contrast to the result for strongly coupled theories, perturbative systems
generate much large values for the heavy quark diffusion coefficient. For example,
early calculations based on perturbative methods of the diffusion coefficient for charm and bottom quarks in 
 a QGP~\cite{Svetitsky:1987gq} produced values 
 of $D_s(2\pi T) \simeq 30 \sim {\cal O}(1/\alpha_s^2)$ for a strong coupling 
 constant of $\alpha_s\simeq$~0.3-0.4,  and with a value varying only weakly with temperature. 
 We now know from a 
 phenomenological point of view that these values for $D_s$ are too large to account for 
 the open heavy-flavor (HF) observables in heavy-ion collisions at RHIC and 
 the LHC~\cite{Rapp:2009my}. Later it was found that the formal convergence of the 
 perturbative series requires much smaller values for $\alpha_s$~\cite{CaronHuot:2008uh}. 
 This calls for nonperturbative methods to assess the heavy-flavor diffusion 
 coefficient in QCD matter.  
 
 \begin{figure}[tbh]
 \centerline{ \includegraphics[width=1.00\textwidth]{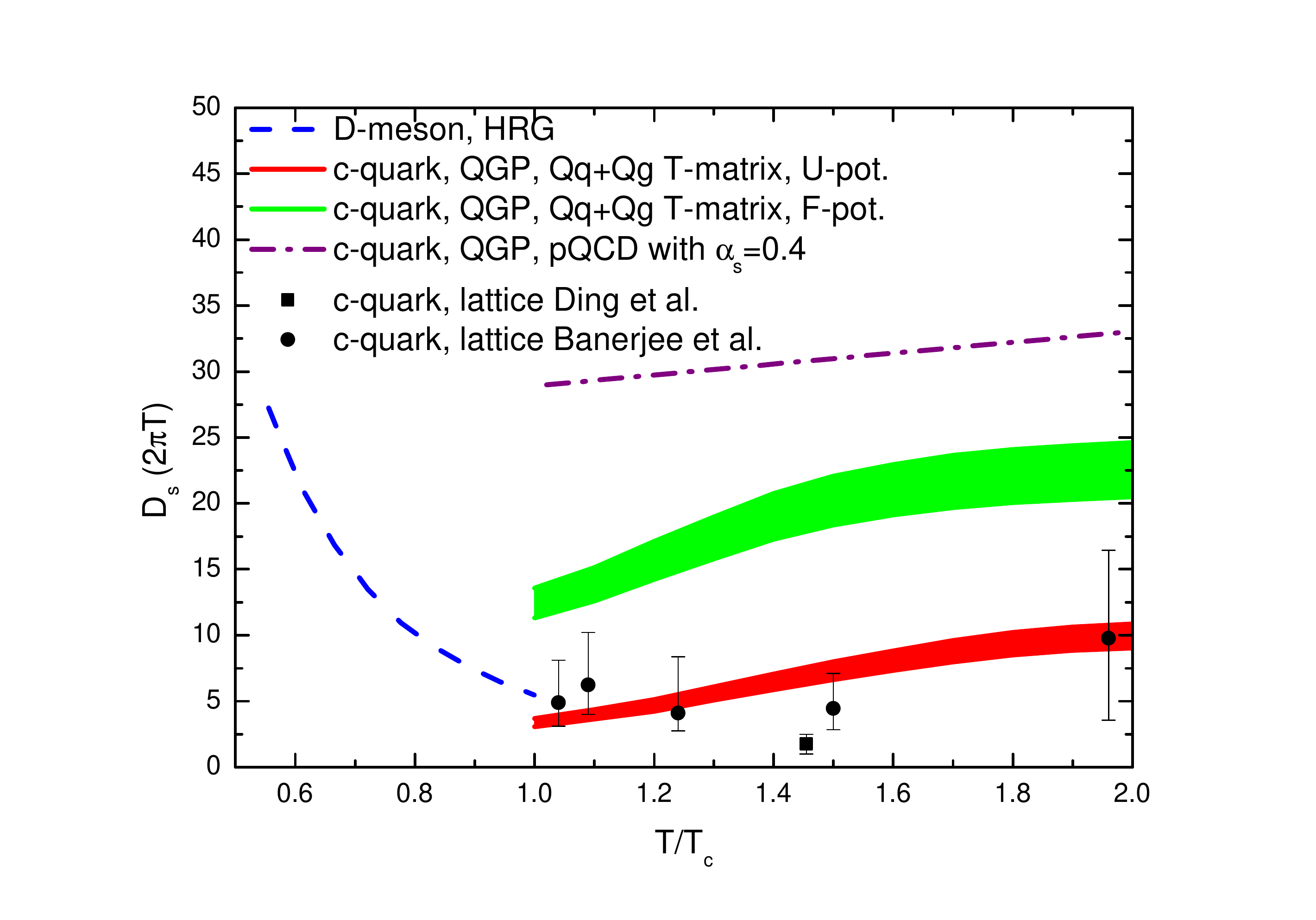} }
 \caption[Spatial diffusion coefficient for charm quarks and $D$-mesons]{Spatial diffusion coefficient for charm quarks in the QGP ($T>T_{\rm c}$)
 and $D$-mesons in hadronic matter ($T<T_{\rm c}$), in units of the thermal wavelength.
 The data points are extracted from quenched 
 lattice QCD~\cite{Banerjee:2011ra,Ding:2012sp,Kaczmarek:2014jga} while the bands are 
 obtained from potential-based $T$-matrix calculations~\cite{Riek:2010fk,Huggins:2012dj} 
 using either the free (green) or internal (red) energies from lattice QCD. The dash-dotted 
 line corresponds to leading order perturbation theory~\cite{Svetitsky:1987gq}.
 }
 \label{fig:Ds-2piT}
 \end{figure}
 Progress has been made to extract $D_s$ from first principles in thermal lattice QCD, 
 by computing euclidean heavy-quark (HQ) correlation functions and reconstructing the 
 low-energy limit of the pertinent spectral function (which defines the transport coefficient). 
 Thus far this has been done in quenched QCD (i.e., in a gluon plasma without dynamical 
 quarks)~\cite{Banerjee:2011ra,Ding:2012sp,Kaczmarek:2014jga}, resulting in a range of 
 values of $D_s(2\pi T) \simeq$~2-6 for temperatures between 1-2~$T_c$, as shown in 
 Figure~\ref{fig:Ds-2piT}. 
 To make closer contact to experiment, it will be necessary to extend these studies 
 to QCD with dynamical quarks, and to compute the 3-momentum dependence of the transport 
 coefficient. The latter can be alternatively expressed through the thermal relaxation 
 rate, $\gamma_Q=T/(m_Q D_s)$, where $m_Q$ is the HQ mass in the QGP, or heavy-meson mass 
 in hadronic matter. 
 
 Non-perturbative calculations of the HQ transport coefficients have also been carried out 
 in the thermodynamic $T$-matrix formalism~\cite{vanHees:2007me,Riek:2010fk,Huggins:2012dj}, 
 which is based on a potential approximation for HQ scattering off thermal partons. The 
 in-medium potential can, in principle, be extracted from thermal lattice QCD, see, e.g., 
 Ref.~\cite{Burnier:2014ssa}. Current uncertainties are usually bracketed by employing 
 either the HQ internal or free energies from the lattice. Using the internal energy one 
 finds values of $D_s(2\pi T)\simeq$\,3-5 at temperatures close to $T_{\rm c}$, increasing 
 to about 10 at 2\,$T_{\rm c}$, for both charm and bottom quarks, see Figure~\ref{fig:Ds-2piT}. 
 For the free energy, the $D_s$ values are about a factor of 2-4 larger. The relaxation 
 rates from the $T$-matrix formalism predict an appreciable 3-momentum dependence, decreasing 
 toward perturbative values at high momenta. Close to $T_{\rm c}$, resonant structures develop 
 in the heavy-light quark $T$-matrices, suggestive of the onset of hadronization. 
 
 Recent work has demonstrated the importance of also treating the  $D$ meson diffusion in the hadronic 
 phase. The pertinent transport coefficient has been estimated 
 in heavy-meson chiral perturbation theory~\cite{Laine:2011is} and in effective 
 hadronic theories including resonance 
 scattering~\cite{He:2011yi,Ghosh:2011bw,Abreu:2011ic,Tolos:2013kva}. The $D$-meson 
 diffusion coefficient significantly decreases as $T_{\rm c}$ is approached from below. 
 There is increasing consensus that its hadronic values~\cite{He:2011yi,Tolos:2013kva} 
 come close to the nonperturbative approaches on the QGP side. This suggests that the 
 heavy-flavor diffusion coefficient develops a minimum across the phase transition 
 region, with a near-continuous temperature dependence when passing from hadronic 
 to partonic degrees of freedom, as one would expect in a cross-over 
 transition~\cite{He:2011yi,He:2012df,Tolos:2013kva}. 
 
 \begin{figure}[tbh] 
 \centerline{ \includegraphics[width=0.90\textwidth]{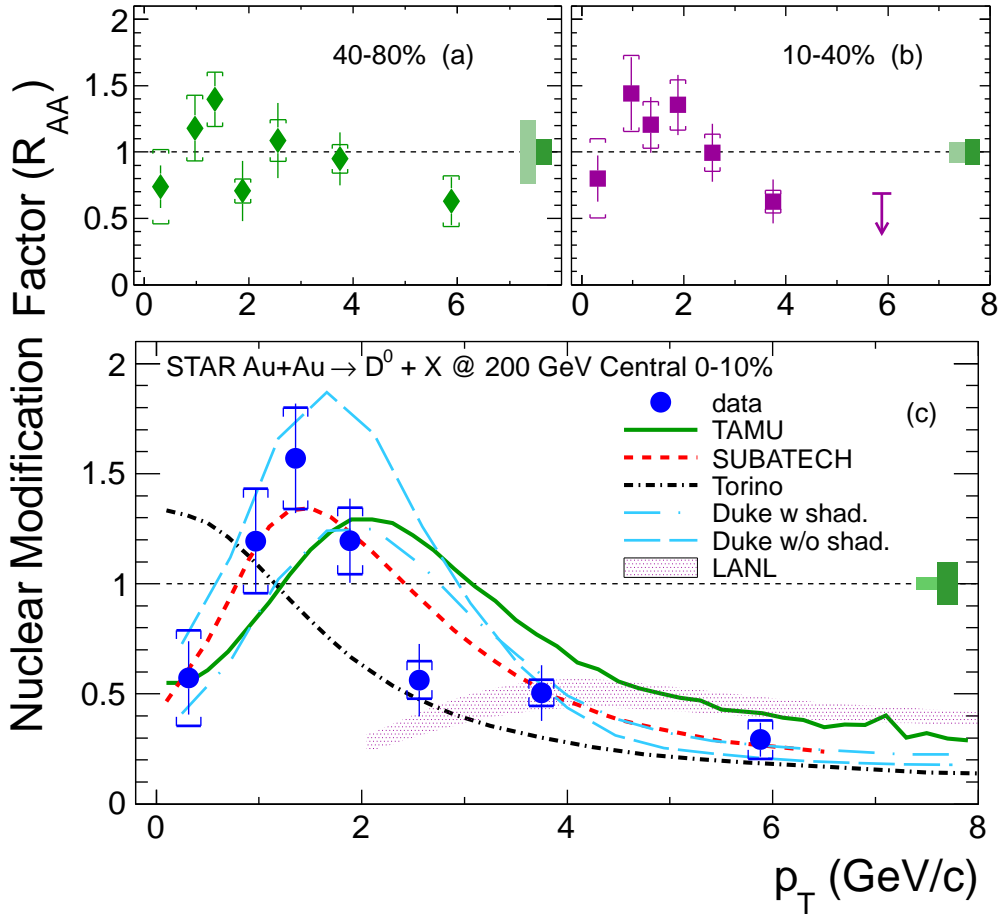} } 
 \caption[STAR measurements of $D$-meson production compared to theory]{Nuclear modification factor of $D$-mesons in 200 GeV \AuAu\ collisions 
 at various centralities as measured by STAR~\cite{Adamczyk:2014uip}, compared to 
 theoretical 
 calculations~\cite{Adil:2006ra,Gossiaux:2008jv,He:2011qa,Alberico:2013bza,Cao:2013ita} 
 } 
 \label{fig:D-RAA-star} 
 \end{figure} 
 Remarkable results for open heavy-flavor observables have recently been obtained at 
 both RHIC and the LHC. The STAR~\cite{Adamczyk:2014uip} and 
 ALICE~\cite{Alice:2012ab,Abelev:2013lca,Abelev:2014ipa} collaborations have, for the 
 first time, been able to extract the nuclear modification factor ($R_{AA}$) and 
 elliptic flow ($v_2$) of $D$ mesons. The STAR measurement of $D$ mesons in 200 GeV 
 Au-Au reaches down to rather low transverse momenta ($p_T$)~\cite{Adamczyk:2014uip}, 
 showing intriguing evidence for a maximum in the $R_{AA}(p_T)$ as seen in 
 Figure~\ref{fig:D-RAA-star}. Such a structure is a tell-tale signature for collective 
 behavior of $D$ mesons, which in turn requires a strong coupling of $c$ quarks and $D$ mesons 
 to the expanding medium. As part of the thermalization process, the heavy-flavor
 particles are dragged along in the fireball expansion and accumulate in a momentum range
 characteristic of the medium's collective flow velocity, while low- and high-momentum states are 
 depleted. This feature can be described by theoretical calculations which implement 
 (a) a sufficiently small diffusion coefficient of $ D_s (2\pi T) \leq5$ into dynamical 
 evolution models~\cite{He:2011qa,Gossiaux:2008jv,Cao:2013ita}, and 
 (b) heavy-light 
 quark coalescence in the hadronization process. The precise location of the
 flow bump turns out to be rather sensitive to the underlying bulk evolution model. 
 Systematic comparisons of the different ingredients to the theoretical models and 
 improved precision in the data are required to disentangle the different effects and 
 arrive at quantitative results for the heavy-quark transport coefficient. 
 Additionally, measurements of the $R_{AA}$ of $D_s$ mesons (containing one charm and 
 one strange anti-/quark), would be very helpful as coalescence processes of $c$ quarks 
 with the enhanced strangeness content of the QGP significantly augment the flow 
 bump~\cite{He:2012df}.  

 A critical role in the determination of the transport coefficient is played by measurements of the elliptic flow parameter
 $v_2$ of the heavy-flavor particles. Electrons and muons from semileptonic decays 
 of $D$- and $B$-meson have been found to carry a rather large $v_2$ in heavy ion
 collisions at both RHIC~\cite{Adare:2006nq,Mustafa:2012jh} and the LHC~\cite{Sakai:2013ata}. 
 Important midterm goals at RHIC are to disentangle the $B$ and $D$ contributions to the semileptonic decays, and to 
 obtain an independent measurement of the $v_2$ of directly reconstructed $D$ mesons. 
 The former will be extracted from existing RHIC 2014 Au+Au displaced vertex 
 measurements by PHENIX (using the VTX and FVTX detectors)~\cite{Nouicer:2012pr}, and by 
 STAR (using the HFT detector)~\cite{Kapitan:2008kk,Qiu:2014dha}. The $v_2$ of directly reconstructed $D$ mesons 
 will be obtained from the same data set using the STAR HFT~\cite{Qiu:2014dha}. 
 At the LHC, ALICE has 
 conducted first measurements of the $D$-meson $v_2$ in 2.76\,TeV \PbPb\ 
 collisions~\cite{Abelev:2013lca}, and found large values; CMS has been able to extract 
 a $B$-meson $R_{AA}$ through their displaced-vertex decays into $J/\psi$'s (so-called 
 non-prompt $J/\psi$'s~\cite{Chatrchyan:2012np}, which is approximately unity for small  
 momenta and turns into a suppression leveling off at $\sim$0.5 at high momenta. As 
 in the $D$-meson sector, this is consistent with collective behavior and thus indicative 
 for a strong bottom coupling to the medium, providing further valuable model constraints.

 An outstanding  issue is the determination of the temperature dependence of the transport coefficient.  
 Experimentally, one lever arm is provided by the correlation between $v_2$ and the 
 $R_{AA}$. Since the bulk medium $v_2$ takes several fm/$c$ to build up, a large $v_2$ 
 of the HF particles is indicative for a strong coupling in the later QGP phases of the 
 fireball evolution, and through hadronization. 
 A suppression in the $R_{AA}$, on the other hand, begins immediately in the early 
 high-density phases, especially at high $p_T$. Model calculations to date
 cannot easily account for the large $D$-meson $v_2$ at LHC without overestimating 
 the suppression in the $R_{AA}$. This corroborates a strong coupling in the vicinity
 of $T_{\rm c}$. The second lever arm is provided by going to lower collision 
 energies where the system starts out at smaller QGP temperatures, closer to $T_{\rm c}$. 
 First data for the heavy-flavor electron $R_{AA}$ and $v_2$ have been extracted
 from a 62\,GeV run at RHIC~\cite{Adare:2014rly,Adamczyk:2014yew}, and show evidence 
 for a non-vanishing $v_2$ and marked modifications in the $R_{AA}$. They are not 
 imcompatible with model calculations that utilize a strong heavy-flavor coupling around 
 $T_c$~\cite{He:2014epa}, but the data precision does not yet suffice for clear conclusions. 
 While varying the collision energy is a valuable tool to explore the temperature 
 dependence of these phenomena, it is necessary to account for the 
 more pronounced role of the Cronin effect at these energies, i.e., 
 a modification of the heavy-flavor spectra through cold nuclear matter effects, before 
 the QGP forms. 
Here  \pA\ collisions will be important to quantify these effects in order to provide a realistic starting 
point for assessing the hot-medium effects.



%% file: tex/Dileptons.tex
\subsection{Thermal Radiation and Low-Mass Dileptons}
        \label{Sec:EM}

Electromagnetic (EM) radiation from the fireballs created in heavy-ion collisions
has long been recognized as a valuable probe of hot and dense
matter~\cite{Feinberg:1976ua,Shuryak:1978ij}. Once produced, photons and dileptons
traverse the fireball and reach the detectors without further
re-interaction. Therefore, their measured spectra are direct signals from the hot and dense
phases of the fireball. In particular, thermal radiation from the locally equilibrated
matter contains unique information about the QCD medium.

The physics potential of EM radiation can be gleaned from the expression for its
local emission rate. At a temperature $T$ it is given by
\begin{equation}
R_{\rm EM} = {\rm const} \ f(E;T) \ \rho_{\rm EM}(M,p;T)
\label{eq:rate}
\end{equation}
where $f(E;T)\simeq \exp(-E/T)$ is the thermal distribution function and
$\rho_{\rm EM}(M,p;T)$ the EM spectral function; $M$ is the invariant mass of the
dilepton ($M$=0 for photons), $p$ its 3-momentum and $E=\sqrt{M^2+p^2}$ its energy.
The mass dependence of $\rho_{\rm EM}$ reflects the operational degrees of freedom:
in vacuum, the low-mass region ($M\le 1$~GeV) is saturated by the light vector mesons
$\rho$, $\omega$ and $\phi$, while the intermediate-mass region (1.5~$\le M/\GeV \le$~3)
is characterized by a perturbative quark-anti-quark continuum.

Calculations of thermal dilepton and photon spectra suitable for comparison to
experiment require the emission rate, Eq.~(\ref{eq:rate}), to be integrated over a
realistic space-time evolution of the fireball in heavy-ion collisions, e.g., by
using hydrodynamic models. The following properties of the QCD medium can then be
studied with EM emission spectra in heavy-ion collisions.
\begin{description}
\item[In-Medium Properties of Vector Mesons.]
In the low-mass region, dilepton radiation from the fireball monitors the
medium effects on the vector mesons as the QCD phase transition is approached and
surpassed. In the vacuum, the properties of light hadrons are governed by the
spontaneously broken chiral symmetry induced by the formation of a quark-anti-quark
condensate. The reduction of the condensate in the QCD medium~\cite{Borsanyi:2010bp}
therefore imposes marked changes on the hadron spectrum. Low-mass dileptons are a
unique observable to measure these modifications in the vector meson mass spectrum.
\item[Temperature of the Fireball.]
Dilepton spectra in the intermediate-mass region\linebreak
($1.5 \le M/\GeV \le 3$) provide a
pristine thermometer of the fireball. Since the medium modifications of the continuum
in the EM spectral function, $\rho_{\rm EM}$, are small (suppressed by the ratio
$T^2/M^2$), the spectral slope of the radiation is solely determined by the temperature
in the thermal distribution function, $f(E;T)\simeq{\rm e}^{-E/T}$. For large masses,
its exponential form strongly favors radiation from the hottest phases of the
fireball~\cite{Shuryak:1980tp,Rapp:2004zh}, so that the observed spectra are mostly
emitted from early in the evolution. Since the mass spectra are
Lorentz-{\em invariant}, they are not distorted by a Doppler shift
from the collective expansion of the exploding fireball.
\item[Lifetime of the Fireball.]
The total yields of thermal EM radiation are a measure of the total lifetime of the
emitting fireball (the expanding 3-volume is constrained by final-state hadron yields).
The optimal mass window for this measurement appears to be
$M\simeq$\,0.3-0.7\,GeV~\cite{Rapp:2014hha}, where the radiation yields turn out to
be proportional to the fireball lifetime within about 10\%,
over a large range of heavy-ion collision energies~\cite{Rapp:2014hha}.
\item[Collective Properties of the Emission Source.]
In contrast to the invariant-mass spectra, the transverse-momentum spectra of dileptons
and photons are subject to a Doppler shift: the expansion velocity of the medium
imparts additional energy on the photons and dileptons which makes their spectra appear
``hotter". The slope of the transverse-momentum spectra is therefore determined by both
the temperature and collective-flow properties of the emitting source.
Another powerful observable is the elliptic flow of the EM radiation. Since the elliptic
flow of the bulk medium takes several fm/c to build up, the elliptic flow of the EM
radiation further constrains the origin of its emission.
\end{description}

The physics potential of accurate dilepton data has been demonstrated by the NA60
collaboration at the CERN SPS in collisions of medium-sized nuclei (Indium with A=114)
at an energy of
$\sqrt{s_{NN}}$=17.3\,GeV~\cite{Arnaldi:2008fw,Arnaldi:2008er,Specht:2010xu}. The
low-mass spectra confirmed the gradual melting of the $\rho$-meson resonance into a
structureless quark-anti-quark continuum, while the total yields translate into an
average fireball lifetime of $7\pm1$\,fm/c. The inverse slope of the radiation at
intermediate masses gives an average temperature of $205\pm12$\,MeV, signaling QGP
radiation. The radial-flow pattern in the transverse-momentum spectra confirms
predominantly hadronic and QGP sources at low and intermediate masses, respectively.
Broad dilepton measurements at different energies with heavy projectiles are needed
to exploit this potential for systematic studies across the QCD phase diagram.

\begin{figure}[tbh]
\centerline{
\includegraphics[width=0.45\textwidth]{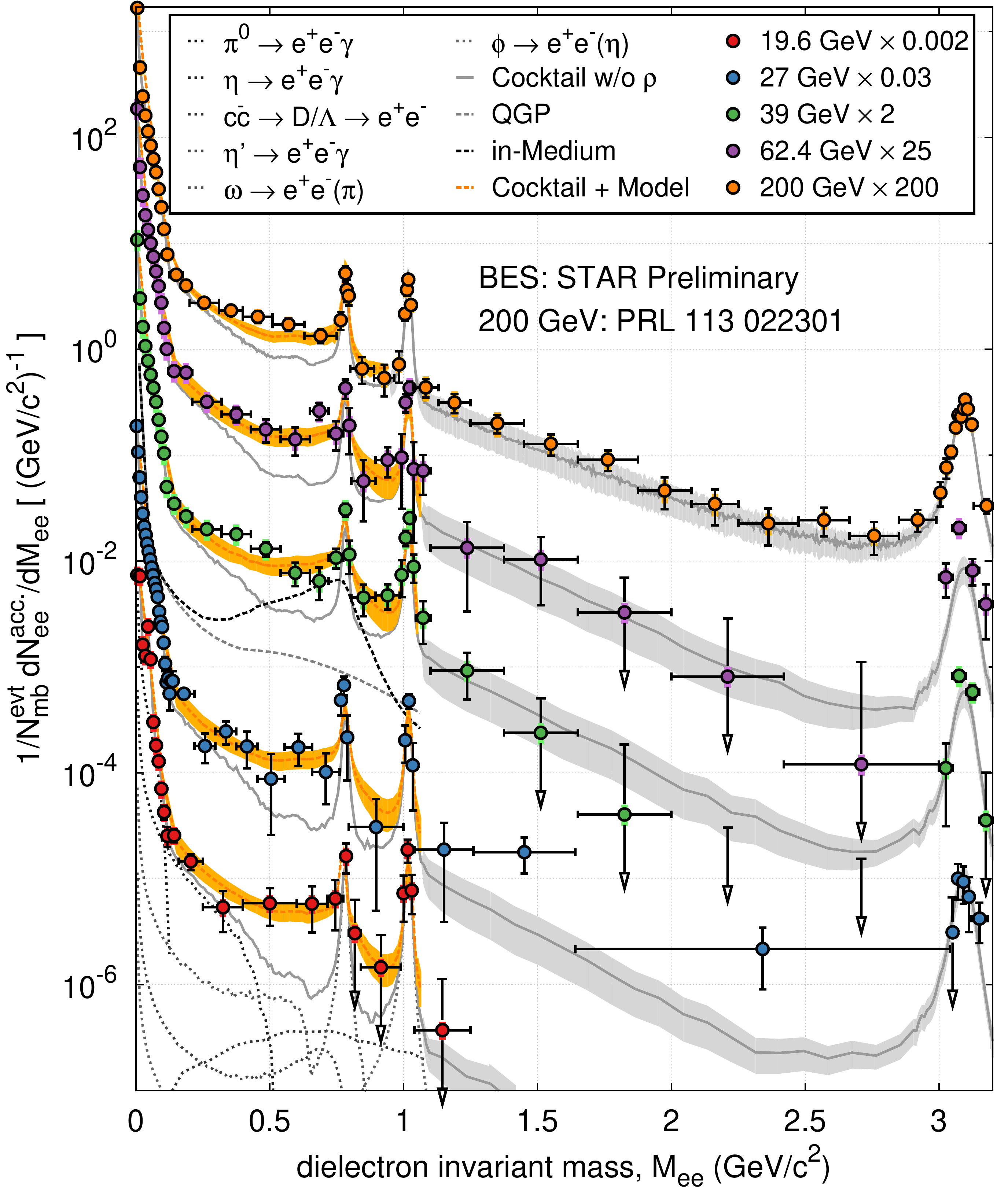}
\includegraphics[width=0.55\textwidth]{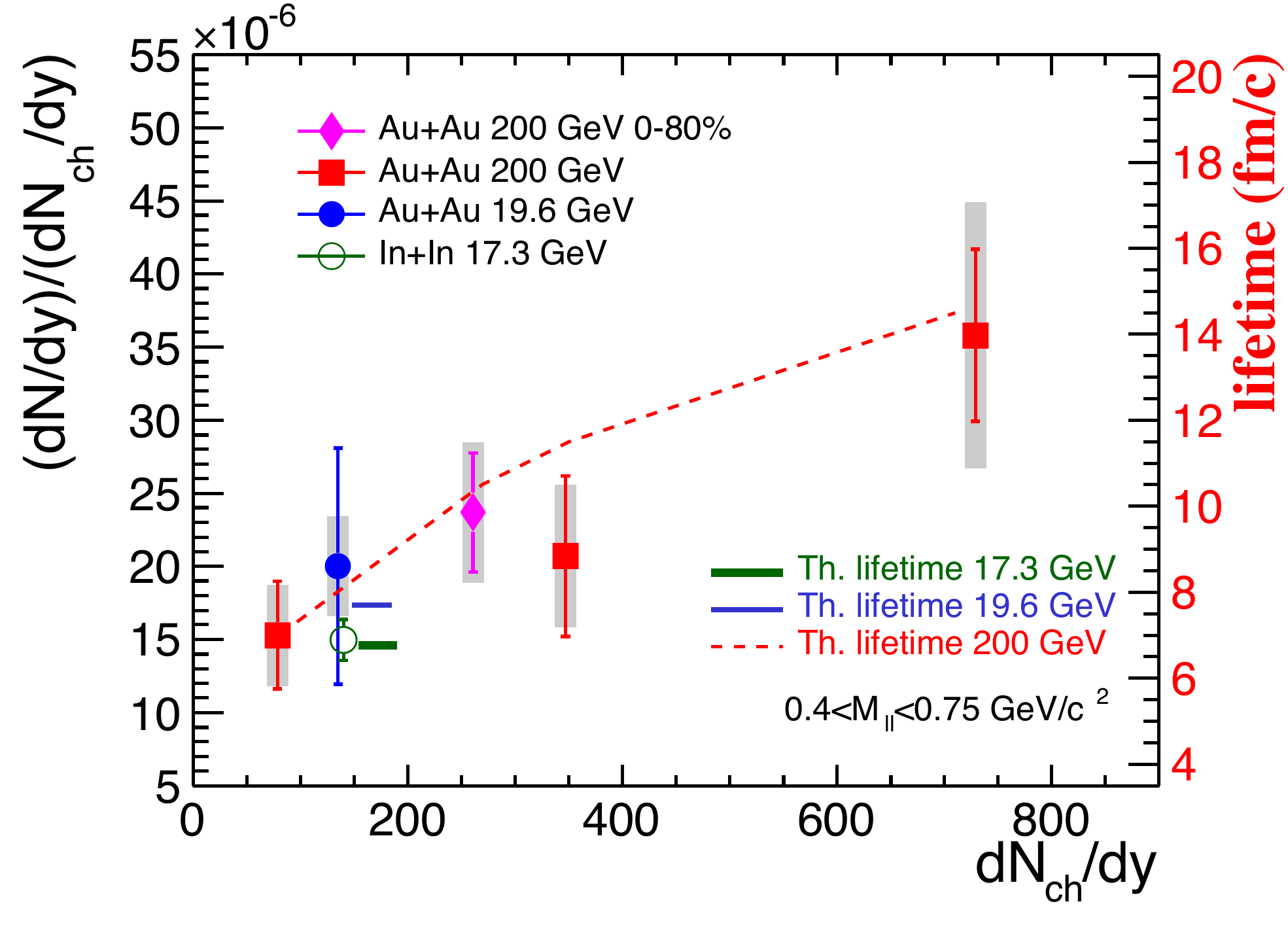}
}
\caption[STAR Beam Energy Scan results on di-electron yields]{Di-electron invariant-mass spectra as measured by
STAR~\cite{Adamczyk:2013caa,Huck:2014mfa} in \AuAu\ collisions in the first beam
energy scan at RHIC. The left panel shows the invariant-mass spectra for increasing
collision energies (bottom to top); the orange bands represent the sum of theoretically
predicted thermal radiation~\cite{Rapp:2013nxa} from QGP and hadronic matter
and final-state hadron decays (including their uncertainty). At intermediate masses
($M>1$\,GeV) the spectra are dominated by correlated heavy-flavor decays and are not yet
sensitive to thermal radiation. 
The right panel shows integrated yields of the normalized dilepton excesses 
for $0.4<M_{ll}<0.75$ GeV/$c^{2}$ as a function of $dN_{\rm ch}/dy$~\cite{Adamczyk:2015bha,Specht:2010xu}. 
The theoretical lifetimes inferred from a model calculation~\cite{Rapp:2014hha} are also shown
}
\label{fig:star-ee}
\end{figure}
A first step in this direction has recently been made by
STAR~\cite{Adamczyk:2013caa,Huck:2014mfa,Adamczyk:2015bha} in the BES program at RHIC. In \AuAu\
collisions covering energies from SPS to top RHIC ($\sqrt{s_{NN}}$=19.6, 27, 39, 62, 200\,GeV),
a sustained low-mass excess radiation was found, cf.~Figure~\ref{fig:star-ee}.
Both mass and transverse-momentum spectra are well described by thermal radiation from
hadronic matter and QGP~\cite{Rapp:2013nxa} (added to contributions from final-state
hadron decays). The data corroborate the melting of the $\rho$  as a robust mechanism of the low-mass
excess in the hadron-to-quark transition at small and moderate baryon chemical potential.
Improved measurements of the low-mass spectral shape at small chemical
potential~\cite{Adamczyk:2013caa} will be critical to discriminate theoretical
models~\cite{Rapp:2000pe,Dusling:2007su,Linnyk:2011vx,Xu:2011tz,Vujanovic:2013jpa}.
This is expected from upcoming high-luminosity \AuAu\ running at 200\,GeV. Lifetime
``measurements" via the integrated low-mass excess require less precision. 
Comparison of 
the BES-I data at $\sqrt{s_{NN}} = $ 19.6 and 200 GeV to theoretical models indicate
that the normalized excess dilepton yields in the low mass region are proportional to the calculated lifetimes of the medium, 
as shown in the right panel of Figure~\ref{fig:star-ee}.
These measurements can be performed with much improved precision in the upcoming BES-II campaign, 
providing a tool to
detect lifetime ``anomalies" possibly induced in the vicinity of a critical
point~\cite{Rapp:2014hha}.

Theoretical progress has been made in elaborating the connection of the dilepton data
to chiral symmetry restoration~\cite{Hohler:2013eba}. The broadening $\rho$-meson
spectral function used to describe the dilepton spectra has been tested  with Weinberg
and QCD sum rules which relate vector and axial-vector spectral functions to quark and
gluon condensates. With temperature-dependent condensates taken from
lattice QCD~\cite{Borsanyi:2010bp}, solutions for the axial-vector spectral function
were found which accurately satisfy the in-medium sum rules. Thus the melting of the
$\rho$-meson resonance is compatible with (the approach to) chiral symmetry restoration.
Further progress has been made in evaluating the in-medium vector correlation function,
and the associated thermal dilepton rates, in thermal lattice
QCD~\cite{Ding:2010ga,Brandt:2012jc}. The computed correlation functions in the QGP
encode a low-mass enhancement which is quite compatible with the spectral functions
that figure into the explanation of the observed dilepton spectra~\cite{Rapp:2011is}.

\begin{figure}[tbh]
\centerline{\includegraphics[width=1.0\textwidth]{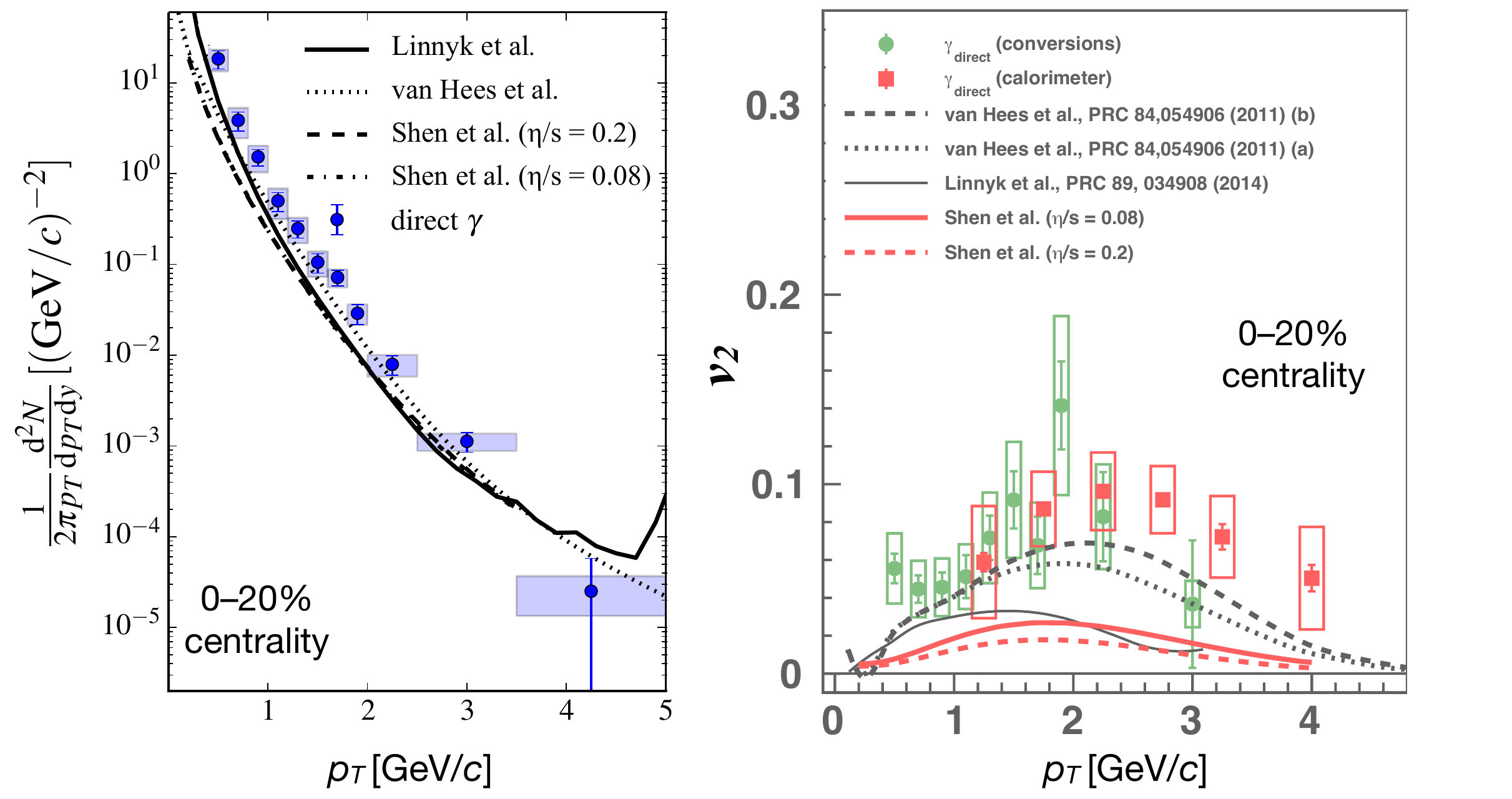} }
\caption[PHENIX results on direct photon spectra and flow compared to theory]{Direct photon spectra (left panel) and their elliptic flow (right panel) in 0-20\%
\AuAu\ collisions at 200\,GeV energy. PHENIX data~\cite{Adare:2011zr,Adare:2014fwh,Bannier:2014bja}
are compared to theoretical model calculations~\cite{vanHees:2011vb,Shen:2013cca,Linnyk:2013wma}.
}
\label{fig:phenix-gam}
\end{figure}
An excess signal of low-momentum direct photons, beyond expectations from binary
nucleon-nucleon collisions and final-state hadron decays, has been measured by
PHENIX~\cite{Adare:2008ab} in 200\,GeV \AuAu\ collisions. The transverse momentum spectra of 
the excess photons are of exponential shape with an inverse slope parameter 
$T_{\rm slope}\simeq 240\pm 30$\,MeV~\cite{Adare:2008ab,Adare:2014fwh}.
As noted above, Doppler shifts due to the collective medium expansion need to be accounted for
in extracting the temperature. Theoretical models suggest that the emission mainly originates
from the later QGP and hadronic stages of the fireball, from a rather broad window of
temperatures around $T_{\rm pc}\simeq170$\,MeV, with an average medium expansion velocity of
$\sim$0.3-0.5$c$~\cite{vanHees:2011vb,Shen:2013cca}. The experimental yields tend to be
underestimated by currently available calculations for thermal radiation, as shown in the left
panel of Figure~\ref{fig:phenix-gam}. Similar measurements are also becoming available
from STAR~\cite{Yang:2014mla}.

The direct photon excess carries a surprisingly large elliptic flow
($v_2$)~\cite{Adare:2011zr,Bannier:2014bja} as seen in the right panel of Figure~\ref{fig:phenix-gam}.
In fact, the magnitude of the flow is comparable to that of pions, which are emitted
when the fireball decouples. This is incompatible with
a large contribution to the direct photon excess from early QGP
radiation~\cite{Chatterjee:2005de,Liu:2009kta,Holopainen:2011pd,vanHees:2011vb,Dion:2011pp,Shen:2013cca,Linnyk:2013wma},
and again points to a later emission of the excess photons. The elliptic flow strength $v_2$ also is underestimated
by current theoretical calculations, but more precise data are needed to quantify the discrepancies.

The PHENIX data have triggered substantial theoretical activity. To obtain sufficiently large yields 
{\em and} a large $v_2$ from a thermal photon source the bulk medium would need to develop its final $v_2$ rather 
rapidly, probably before reaching the phase transition regime~\cite{vanHees:2011vb}. This could be driven, 
e.g., by a pre-equilibrium radial flow from a glasma evolution~\cite{Dusling:2010rm,Chen:2013ksa}.
Such favorable collective properties would also need to be accompanied by large photon rates in the phase transition
regime~\cite{vanHees:2011vb,Shen:2013vja,vanHees:2014ida} and in the hadronic
phase~\cite{Turbide:2003si}. An initially gluon-rich plasma~\cite{McLerran:2014hza}, or 
nonperturbative effects in a QGP~\cite{Gale:2014dfa}, could aid in suppressing
early electromagnetic emission when the bulk $v_2$ is still small
and thereby avoid diluting the observed strong $v_2$ pattern with the azimuthally symmetric
distribution expected for photons emitted early in the collision.

Preliminary measurements of the direct-photon triangular flow ($v_3$)~\cite{Bannier:2014bja}
support a thermal emission source~\cite{Shen:2013cca} and disfavor more exotic sources, e.g.,
from strong but short-lived primordial magnetic fields~\cite{Basar:2012bp,Bzdak:2012fr}.
The centrality dependence of the excess signal~\cite{Adare:2014fwh} is consistent
with expectations from thermal radiation, but is also compatible with scaling arguments based on
initial-state saturation effects~\cite{McLerran:2014oea}. Improved measurements of the
photon spectra and their collective properties will be critical in resolving these issues, as will 
be extending these measurements to other colliding systems. 

Closely related but independent observables are the transverse-momentum spectra and $v_2$ of
low-mass dileptons.
A first measurement of
the $v_2$ has been achieved~\cite{Adamczyk:2014lpa} but does not yet provide tangible
discrimination power. 
Again, improved and extended experimental measurements are required, as well
as additional calculations. Since photons and dileptons are intimately related in theoretical
calculations, a complete understanding of EM data and its implications for the thermal 
history of the medium will need to account for both observables
simultaneously.

%% file: tex/EnergyScan.tex
\subsection{Mapping the Phase Diagram of QCD via a Beam Energy Scan}
\label{Sec:BES}

\begin{figure}[t]
\begin{center}
\centerline{\includegraphics[width=0.65\textwidth]{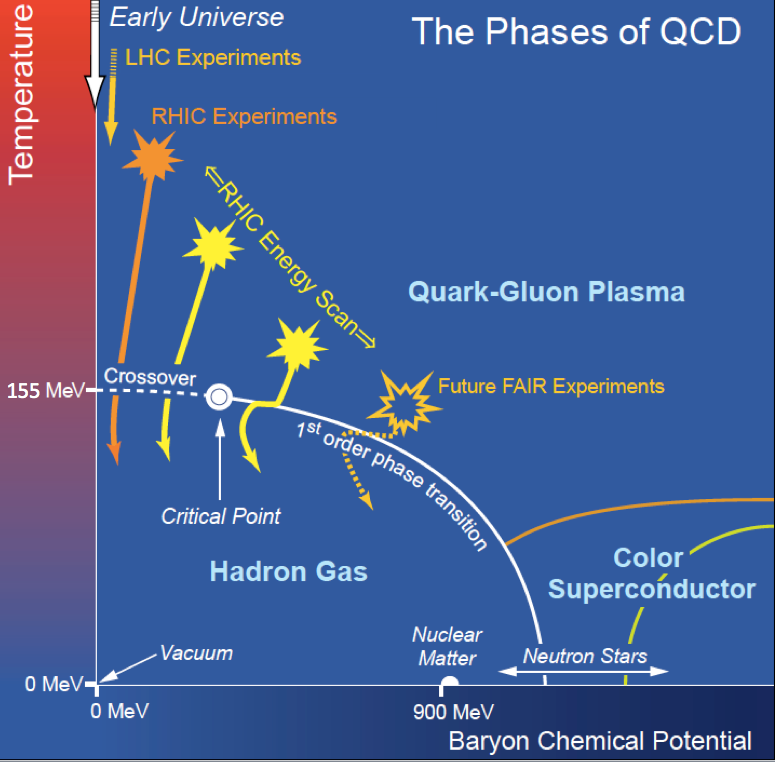}}
\caption[The QCD phase diagram]{A sketch illustrating the experimental and theoretical
  exploration of the QCD phase diagram. Although experiments at
  highest energies and smallest baryon chemical potential are known to
  change from a QGP phase to a hadron gas phase through a smooth
  crossover, lower energy collisions can access higher baryon chemical
  potentials where a first order phase transition line is thought to
  exist.}
\label{F-PD1}
\end{center}
\end{figure}

When the first protons and neutrons and pions formed in the
microseconds-old universe, and when they form in heavy-ion collisions
at the highest RHIC energies and at the LHC, they condense out of
liquid quark-gluon plasma consisting of almost as much antimatter as
matter. Lattice calculations~\cite{Aoki:2006we,Aoki:2009sc,Bazavov:2011nk} 
show that QCD predicts that, in such an environment, this condensation 
occurs smoothly as a function of decreasing temperature, with many 
thermodynamic properties changing dramatically but continuously within 
a narrow temperature range around the transition temperature 
$T_c\in [145\,\mathrm{MeV},163\,\mathrm{MeV}]$
\cite{Bazavov:2011nk,Bazavov:2014pvz}, referred to as the crossover
region of the phase diagram of QCD, see Figure~\ref{F-PD1}.
%
In contrast, quark-gluon plasma doped with a sufficient excess of
quarks over anti-quarks may instead experience a sharp first order
phase transition as it cools, with bubbles of quark-gluon plasma and
bubbles of hadrons coexisting at a well-defined critical temperature,
much as bubbles of steam and liquid water coexist in a boiling
pot. The point where the doping of matter over antimatter
(parametrized by the net baryon number chemical potential $\mu_B$)
becomes large enough to instigate a first order phase transition is
referred to as the QCD critical point. It is not yet known whether QCD
has a critical point~\cite{Stephanov:1998dy,Fodor:2004nz,Allton:2005gk,Gavai:2008zr,deForcrand:2008zi},
nor where in its phase diagram it might lie. Lattice calculations
become more difficult or more indirect or both with increasing $\mu_B$
and, although new methods introduced within the past decade have
provided some hints~\cite{Fodor:2004nz,Gavai:2008zr,Datta:2012pj}.
While these theoretical calculations are advancing through both
new techniques and advances in computational power,
at present only experimental measurements can answer these questions
definitively.

The phase diagram of QCD, with our current knowledge shown schematically
in Figure~\ref{F-PD1}, is the only phase diagram of any form of
matter in Nature that we have the opportunity of both mapping experimentally
and relating directly and quantitatively to our fundamental
description of Nature, the Standard Model. With QCD the only strongly
interacting theory in the Standard Model, mapping the transition
region of its phase diagram is a scientific goal of the highest
order. In the long term, successfully connecting a quantitative,
empirical understanding of its phases and the transitions between
phases to theoretical predictions obtained from the QCD Lagrangian
could have ramifications in how we understand phases of strongly
coupled matter in many other contexts.

{\bf RHIC's unique capability to probe the QCD phase diagram}

A major effort to use heavy ion collisions at RHIC to survey the phase
diagram of QCD is now underway. The excess of matter over antimatter
in the exploding droplet produced in a heavy ion collision can be
increased by decreasing the collision energy, which reduces the
production of matter-antimatter symmetric quark-antiquark pairs and
gluons relative to the quarks brought in by the colliding nuclei, thus
increasing $\mu_B$. Decreasing the collision energy also decreases the
maximum, {\it i.e.}~initial, temperature reached by the matter produced in
the collision. A series of heavy ion collision measurements scanning
the collision energy~\cite{BESII} can therefore explore the properties
of matter in the crossover region of the phase diagram, matter that is
neither quark-gluon plasma nor hadronic nor both at the same time, as a
function of the doping $\mu_B$. Such a program can scan the transition
region of the QCD phase diagram out to $\mu_B$ values that correspond
to collision energies below which the initial temperature no longer
reaches the transition. If the crossover region narrows to a critical
point within this experimentally accessible domain, an energy scan can
find it. RHIC completed the first phase of such an energy scan in
2014, taking data at a series of energies ($\sqrt{s_{NN}}=$ 200, 62.4,
39, 27, 19.6, 14.5, 11.5 and 7.7 GeV) corresponding to values of
$\mu_B$ that range from 20 to 400 MeV. Data from these experiments at
RHIC~\cite{Kumar:2012fb,BESII} and from previous experiments confirm that
lower-energy collisions produce matter with higher $\mu_B$, as
anticipated. 

RHIC is, and will remain, the optimal facility in the world for
mapping the phase diagram of QCD, including searching for a possible
critical point in its so far less well understood regions with larger $\mu_B$.
What makes RHIC unique is both its wide
reach in $\mu_B$ and that it is a collider, meaning that the
acceptance of detectors, and hence the systematics of making
measurements, change little as a function of collision
energy. Accelerator and detector performance has been outstanding
during the first phase of this program, referred to as Beam Energy
Scan I or BES-I. Measurements of all the important observables
targeted in the planning of this campaign have now been made in
collisions with energies varying by a factor of 25, allowing for a
first look at a large region of the phase diagram of QCD.

A selection of measurements of several observables from BES-I that exhibit interesting 
non-monotonic behavior as a function of collision energy is shown in 
Section~\ref{Sec:CP}, see Figure~\ref{F-PD2} there.  
As we will discuss in that later section, the BES-I measurements of these 
observables provide some
evidence for the softening of the equation of state in the 
crossover region of the phase diagram. 
In addition, these data may indicate
the first hints of the presence of a critical point in the phase
diagram of QCD. Above all, given the size of the present
experimental uncertainties these current measurements 
provide strong motivation
for the next phase of the BES program BES-II, described in Section~\ref{Sec:CP}, which will deliver in 2018-19
substantially greater statistics and hence substantially smaller error bars
in collisions with energies at and below $\sqrt{s_{NN}}$=19.6 GeV.
Accordingly, we defer our presentation of  highlights from the rich BES-I data set to 
Section~\ref{Sec:CP}
where we will discuss each observable in a way that synthesizes
what we have learned from data to date with what can be learned
from their measurement in BES-II in combination with anticipated advances
in theory.

%% file: tex/Exotica.tex
\subsection{Topological Fluctuations within Quark-Gluon Plasma}
\label{Sec:Exotica}

It has been known for decades that topological effects play an
important role in determining the structure of the vacuum in
non-Abelian gauge theories\cite{Belavin:1975fg}.  In QCD, fluctuations
that change the topology of the non-Abelian gauge fields are at the
same time fluctuations that create an imbalance of chirality.
Recently, considerable interest has been generated by the possibility
of exploring experimental signatures of such fluctuations in the hot
QCD matter produced in relativistic heavy ion collisions.  A key
observation is that the incredibly strong magnetic field that is
generated in off-center heavy ion collisions, together with the chiral
anomaly in QCD and the fluctuations in chirality, can produce a Chiral
Magnetic
Effect~\cite{Kharzeev:2007tn,Kharzeev:2007jp,Fukushima:2008xe,Hirono:2014oda}
in which an electric current flows along (for one sign of the
chirality fluctuation) or opposite to (for the other sign of the
chirality fluctuation) the direction of the magnetic field, resulting
in a separation of particles with opposite electric charge in a
direction perpendicular to the reaction plane of the collision, while
particles with the same electric charge show a preference for ending
up in the same hemisphere.
\begin{figure}[!htp]
\includegraphics[width=\textwidth]{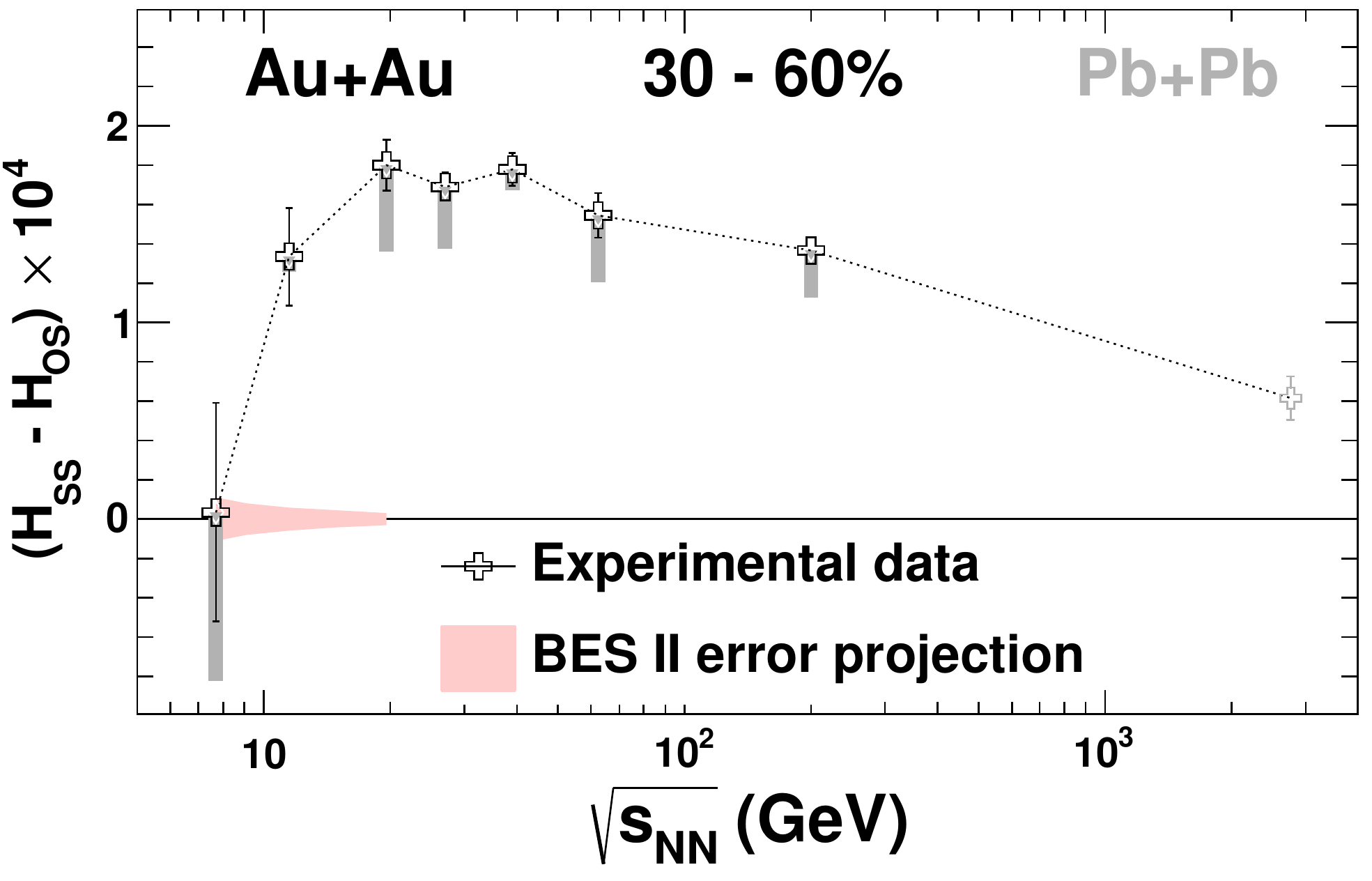}
\caption[Charge separation results from STAR and ALICE]{Collision energy dependence of the charge separation
  correlation $H$ difference between same-sign (SS) and opposite sign
  (OS) pairs in mid-central (30-60\%) \AuAu\ collisions by RHIC\cite{Adamczyk:2014mzf}
  ($\sqrt{s_{NN}}= 7.7 - 200$ GeV) and Pb+Pb collisions at LHC\cite{Abelev:2012pa}
  ($\sqrt{s_{NN}}= 2.76$~TeV). The hatched band represents the
  estimated statistical errors that will be obtained in BES II
  measurements at RHIC.}
\label{Fig:CME}
\end{figure}

Such an effect has been observed by using the three-point correlator
method~\cite{Voloshin:2004vk} to extract event-plane dependent charge
correlations in \PbPb\ collisions at the LHC~\cite{Abelev:2012pa} and
at a variety of collision energies at
RHIC\cite{Abelev:2009ac,Abelev:2009ad,Adamczyk:2014mzf}.  Conclusively
establishing that these observations result from the CME would be an
important advance in our ability to study experimentally the
topological phase structure of QCD.  In addition, the effect itself
can be used to study the properties of the medium created in the
collisions. Since the CME requires the formation of QGP, {\it e.g.}
the restoration of the chiral symmetry~\cite{Hirono:2014oda}, one
would expect that at sufficiently low energy, the QCD related
correlations should disappear. This is precisely what is observed in
$\sqrt{s_{NN}}$=11.5 GeV \AuAu\ collisions~\cite{Adamczyk:2014mzf}, as
shown in Figure~\ref{Fig:CME}.

\begin{figure}[!htp]
\includegraphics*[width=0.9\textwidth]{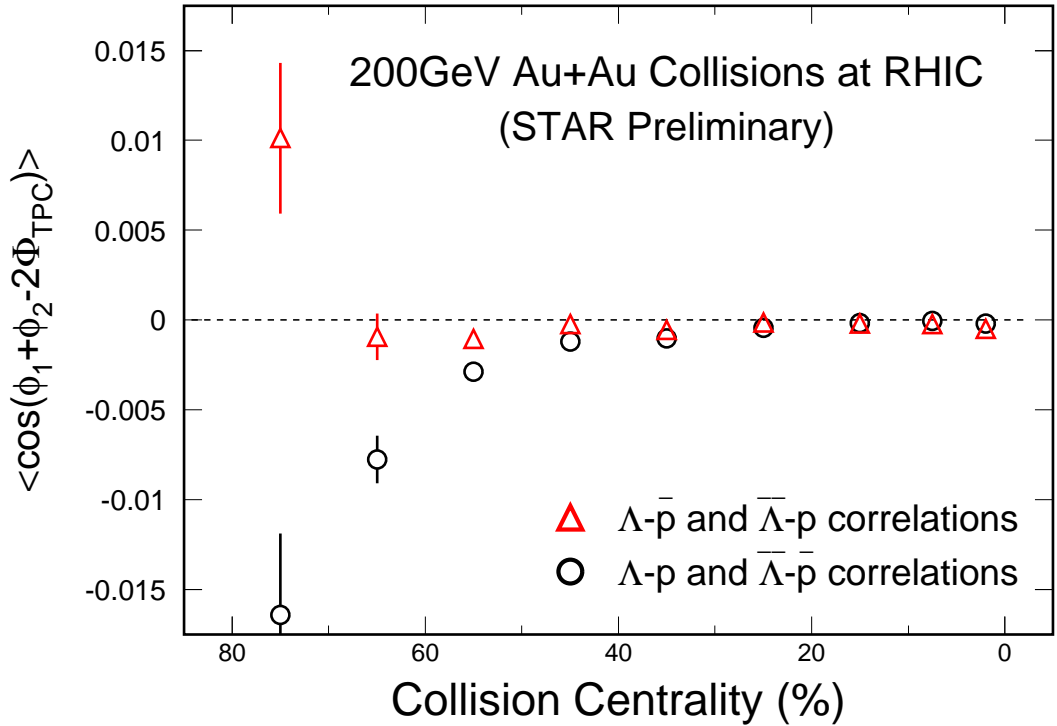}
\vspace{-1.5cm}
\caption[Baryon separation as a function of collision centrality from STAR]{Baryon separation shown as a function of collision centrality
  from $\sqrt{s_{NN}}=200$ GeV Au+Au collisions. Circles and triangles
  represent the data for the same-baryon-number and
  opposite-baryon-number correlations, respectively.  Error bars are
  statistical only. Taken from \cite{Zhao:2014aja}.}
\label{Fig:CVE}
\end{figure}
The same physical mechanism that produces the Chiral Magnetic Effect
in the presence of a magnetic field $\vec B$ can also produce a Chiral
Vortical Effect (CVE) in the presence of external angular momentum
$\vec L$~\cite{Kharzeev:2010gr}.  While high energy nuclear collisions
produce the strongest known magnetic fields $\sim 10^{18}$
gauss\footnote{The magnetic field at the surface of magnetars is on
  the order of $10^{15}$ gauss, several orders of magnitude lower than
  the field strength in high-energy nuclear collisions.}, these fields
are largest at the very beginning of the collision and decay away on a
timescale that is controlled by the electric conductivity of the
matter produced in the
collision~\cite{Tuchin:2013ie,McLerran:2013hla,Gursoy:2014aka}.
Angular momentum, on the other hand, is conserved meaning that if the
plasma is created with nonzero angular momentum this remains.  If
topological fluctuations and the associated chirality fluctuations do
have observable effects in high energy nuclear collisions, then in
addition to the charge separation along the external magnetic field in
the CME one would expect baryon number separation along the (same)
direction of the external angular momentum.  The first measurements of
observables that receive a contribution from any CVE were reported
recently for \AuAu\ collisions at 200 GeV for proton (anti-proton) and
Lambda (anti-Lambda) pairs~\cite{Zhao:2014aja}, as shown in
Figure~\ref{Fig:CVE}.  The figure indicates a preference for proton
and Lambda pairs, as well as antiproton and anti-Lambda pairs, to be
found in the same hemisphere.


The topological fluctuations that drive the CME and CVE in QCD have
analogues in other gauge theories and arise in other contexts.  For
example,
it has been argued that if an electron chirality imbalance exists in
the weakly coupled electroweak plasma at temperatures thousands of
times hotter than those achieved in heavy ion collisions, this may be
responsible for the primordial magnetic fields in the early
Universe~\cite{Joyce:1997uy}.  Furthermore, the same kinds of
topological fluctuations that make a chirality imbalance in the
quark-gluon plasma can make a baryon number imbalance in the much
hotter electroweak plasma, meaning that if these fluctuations were in
some way biased they could be responsible for the
matter-over-antimatter excess in the Universe.  In a third context,
the CME has been realized in condensed matter physics in a
(3+1)-dimensional structure called a Weyl semimetal due to the chiral
anomaly~\cite{Wan:2011aa} and in other systems as reported in
Ref.~\cite{Perks:2012aa}.
  
As can be seen from Figure~\ref{Fig:CME}, the error bars in the current
data are still large, especially in the lower energy region. As a
result, it is not yet possible to systematically investigate these
phenomena as a function of energy, nor is it possible to
determine the exact collision energy where the CME disappears. The
planned BES-II program at RHIC, discussed in Section~\ref{Sec:CP},
will provide high statistics data for CME and CVE studies in
\AuAu\ collisions at energies below 20~GeV. The estimated errors are
shown as the shaded bar in Figure~\ref{Fig:CME}.  With such
measurements, along with the planned dilepton measurements, we may
expect to gain significantly improved quantitative understanding of
the chiral properties of hot QCD matter at nonzero baryon density.


%% file: tex/CrossFertilization.tex
\subsection{Broader Impacts}
\label{Sec:Cross}
 
 The intellectual challenges posed by strongly interacting QCD matter have led to significant cross-fertilization with other fields. Theoretical tools have been both imported from and in some cases exported to fields ranging from condensed matter physics to string theory. 
 Recently results from RHIC experimentalists have been announced that have implications beyond QCD at high energy and densities. Of particular note are the ``dark photon" upper limits and the first observation of anti-$^{4}$He. 
It is worth noting that neither of these results were envisioned as part of the experimental program; 
rather they result from the very large data samples that have been acquired at RHIC in combination with exceptional experimental sensitivities\footnote{Essentially identical statements apply to the LHC detectors.}.
Also of interest is a new study in condensed matter physics of the Chiral Magnetic Effect discussed in Section~\ref{Sec:Exotica}.

\subsubsection{Dark Photons}
It has been postulated that an additional U(1) gauge boson, a``dark photon" $U$ that is weakly coupled to ordinary photons, can explain the anomalous magnetic moment of the muon $(g-2)_\mu$, which deviates from  standard model calculations by 3.6$\sigma$. By studying  $\pi^0, \eta \rightarrow e^+e^-$ decays the PHENIX experiment has extracted upper limits on U-$\gamma$ mixing at 90\% CL for decays to known particles, in the mass range 30$<m_U<$90 MeV/$c^2$\cite{Adare:2014mgk}. These results show that except for the small range 30$<m_U<$32 MeV/$c^2$ the 
 U-$\gamma$ mixing parameter space that can explain the $(g-2)_\mu$ deviation from its Standard Model value is  excluded at the 90\% confidence level.
When combined with 
experimental limits from BaBar\cite{Lees:2014xha} and and NA48/2\cite{Adlarson:2014hka}, these analyses essentially
exclude the simplest model of the dark photon as an explanation of the
$(g-2)_\mu$ anomaly.

\subsubsection{Antimatter Nuclei}
In top energy RHIC collisions matter and antimatter are formed with approximately equal rates. The rapid expansion and cooling of the system means that the antimatter decouples quickly from the matter making these types of collisions ideal for studying the formation of anti-nuclei. 
The STAR collaboration has reported detection of anti-helium 4 nuclei ${}^4\mathrm{\overline{He}}$, which is the heaviest anti-nucleus observed to date\cite{Agakishiev:2011ib}. 
The ${}^4\mathrm{\overline{He}}$ yield is consistent with expectations from thermodynamic and coalescent nucleosynthesis models, providing a suggestion
that the detection of even heavier antimatter nuclei is experimentally feasible. These measurements may serve as a benchmark for possible future observations from antimatter sources in the universe.

\subsubsection{Chiral Magnetic Effect in Condensed Matter Systems}
Recently the Chiral Magnetic Effect discussed in Section~\ref{Sec:Exotica} has been observed in a condensed matter experiment measuring magneto-transport in $\mathrm{ZrTe_5}$\cite{Li:2014bha}. 
The recent discovery of Dirac semi-metals 
with chiral quasi-particles\cite{Bor:2014aa,Neu:2014aa,Liu:2014aa} 
has created the opportunity to 
study the CME in condensed matter experiments. 
In these materials it is possible to generate a chiral charge density by the application 
of parallel electric and magnetic fields\cite{Fukushima:2008xe}.
The resulting chiral current is in turn proportional to the product of the 
chiral chemical potential and the magnetic field, leading to a quadratic dependence on the magnetic field.
This is precisely what is observed in Ref.~\cite{Li:2014bha},
where the magnetoconductance varies as the square of the applied magnetic field.
These studies can be extended to a broad range of materials, since three-dimensional 
Dirac semimetals often emerge at quantum transitions between normal and
topological insulators.
Interestingly, the qualitative features observed in \cite{Li:2014bha} 
have been reproduced in a calculation connecting chiral anomalies in 
hydrodynamics with its holographic system in the gauge/gravity duality\cite{Landsteiner:2014vua}. 

%% file: tex/FutureProspects.tex
\section{Future Prospects}
\label{Sec:Future}
The cornucopia of experimental advances and theoretical insights detailed in Section~\ref{Sec:Progress}
are but a representative sample of the advances made in the field since the last Long Range Plan.
As in all healthy scientific enterprises, our increased understanding of hot QCD matter
has engendered new questions. 
These questions are both quantitative
(``What is the precise value of $\eta/s$ at the various temperatures accessible at RHIC and the LHC?")
and qualitative
(``How does the perfect liquid behavior of quark-gluon plasma emerge from the QCD Lagrangian?")
in nature. 
In addition, opportunities exist to ask and address discovery-oriented
questions, such as ``Does the QCD phase diagram have a critical point?''
The field is poised to answer definitely such questions in the next decade,
thanks to ongoing and anticipated investments in the experimental programs 
at RHIC and the LHC, and in theoretical investigations world-wide. 
This section describes those opportunities and delineates the
progress that will follow from exploiting them.

\input{tex/FacilitiesFuture}

\input{tex/CriticalPoint}

\input{tex/HardProbes}

\input{tex/Theory}

%% file: tex/FacilitiesFuture.tex
\subsection{Facilities Evolution}
\label{Sec:FacilitiesFuture}

Planning is well underway to address the compelling physics questions presented in Section~\ref{Sec:Progress}. 
The evolution of the RHIC and LHC accelerators, accompanied by upgrades to their experiments, will provide tools uniquely suited
for determining the inner workings of thermal QCD matter. The prospect of these two facilities, operating with increased luminosity
and a range of energies spanning three orders of magnitude, coupled with greatly upgraded detector capabilities, provides an unprecedented
opportunity to resolve fundamental aspects of QCD.

\begin{figure}[ht]
\centerline{
\includegraphics[width=1.00\textwidth]{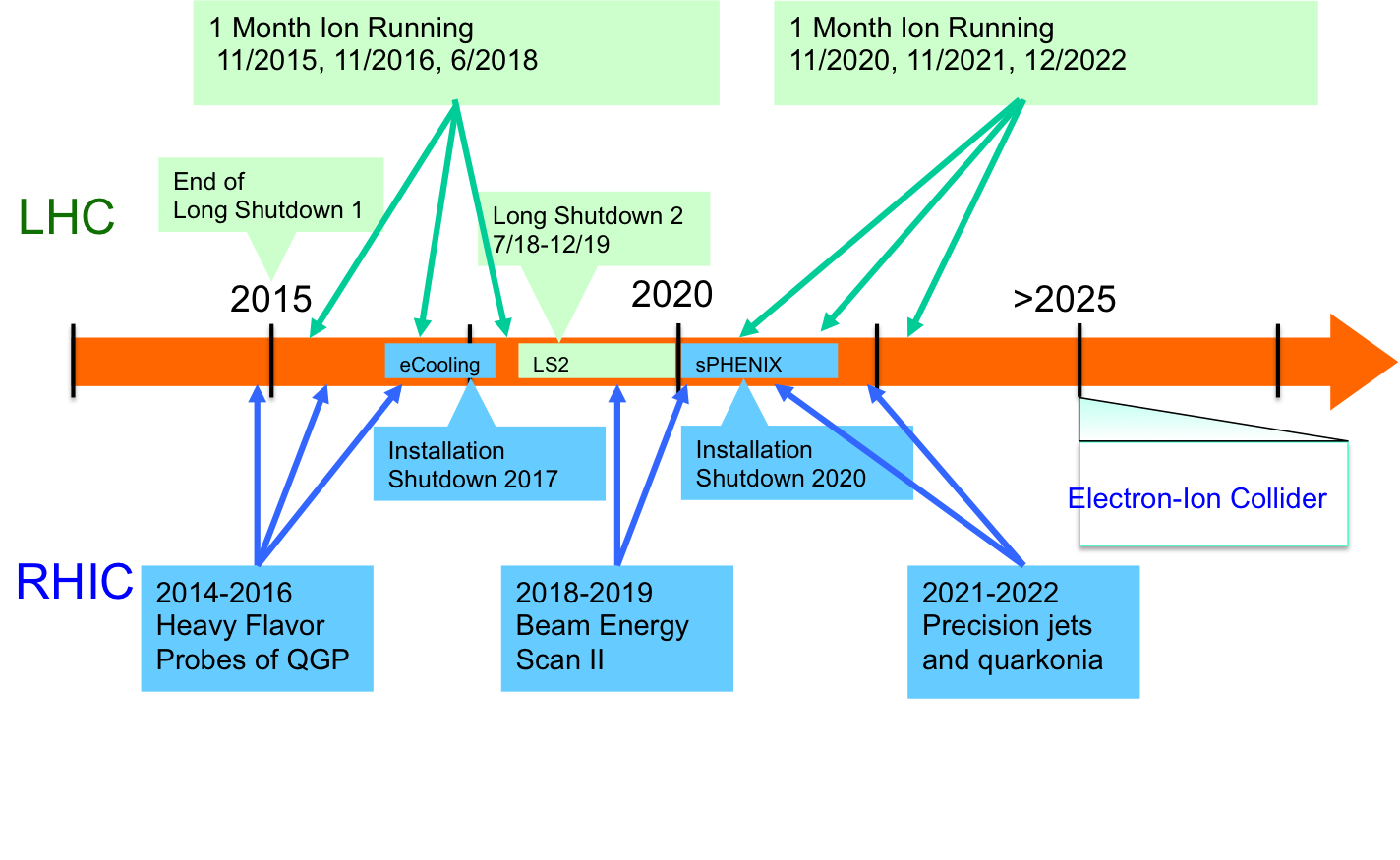}
}
\caption{The timeline for future RHIC and LHC heavy ion running.
}
\label{Fig:Timeline}
\end{figure}

\subsubsection{Facility and experiment upgrades at RHIC}
\label{Sec:RHICUpgrades}
At Brookhaven National Laboratory, the goal for the next decade is to complete the  science mission of RHIC on a schedule that permits a smooth and timely transition towards preparations for an electron-ion collider based on the RHIC complex. 
A coherent physics program has been developed to address the outstanding questions in both hot and cold QCD physics 
relevant to completing our understanding of the QGP, with a natural physical and intellectual evolution towards the 
physics addressed by an electron-ion collider.
Accordingly, future heavy ion running is divided into three campaigns: 

\begin{itemize}[leftmargin=2.0cm]

\item[\bf 2014-16:] Measure heavy flavor probes of the QGP using the newly installed silicon vertexing detectors in PHENIX and STAR.
\item[\bf 2018-19:] Conduct a fine-grained scan of the QCD phase diagram via Phase II of the RHIC Beam Energy Scan program.
\item[\bf 2021-22:] Perform precision jet quenching and quarkonia measurements following the installation of sPHENIX.

\end{itemize}

Interspersed between these three running periods are two shutdowns, as illustrated in Figure~\ref{Fig:Timeline}. 
\begin{figure}[!htp]
\centerline{
\includegraphics[width=1.00\textwidth]{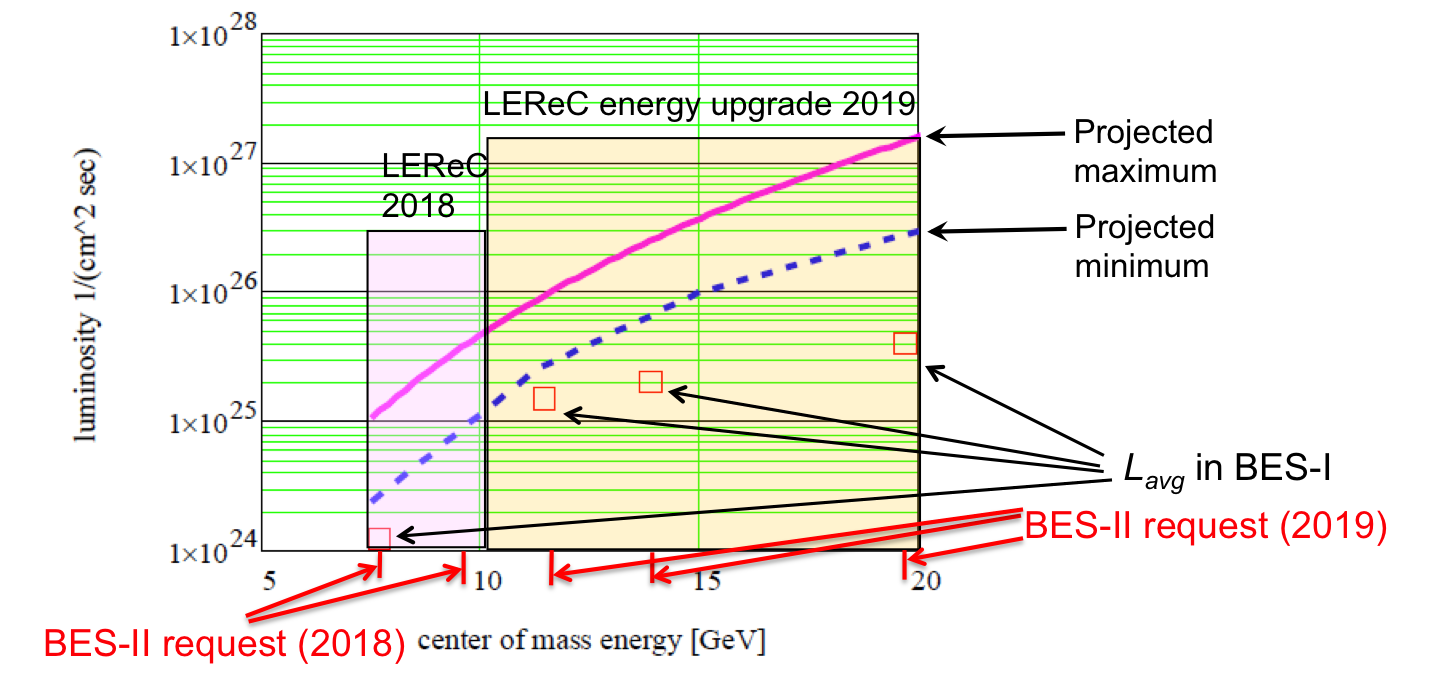}
}
\caption[Projected RHIC luminosity increase at low energies from electron cooling]{The projected RHIC luminosity increase at low energies provided by electron cooling of the ion beams. The red squares are the measured average luminosity in RHIC Beam Energy Scan~I. The blue dashed line is the minimum projection for the improvement with electron cooling; the magenta solid line is the maximum projection. Also shown are the actual luminosities achieved in RHIC BES~I. 
}
\label{Fig:eCooling}
\end{figure}
The first shutdown, in 2017,  is to allow for installation of electron cooling to increase RHIC's luminosity
at low energies\footnote{It should be noted that the current RHIC performance at low energies exceeds the original design values for luminosity by well over an order-of-magnitude. In fact, the original RHIC design was limited to energies of 20 GeV and above.}, 
as shown in Figure~\ref{Fig:eCooling}. This luminosity upgrade, in combination with targeted upgrades of the STAR detector described below, greatly increases the discovery potential of the subsequent beam energy scan in 2018-2019. The second shutdown, in 2020, will be used to install the sPHENIX detector (discussed below), prior to a period of dedicated running to explore the microscopic structure of the QGP with high energy jets and quarkonia measurements. Both the sPHENIX upgrade and the proposed STAR upgrades provide natural evolution paths towards detectors for the EIC\footnote{In addition to the upgrade paths from sPHENIX to ePHENIX and STAR to eSTAR, consideration has been given
to a new detector explicitly designed for electron-ion capabilities. Details are available in Refs.~\cite{Accardi:2012qut,Boer:2011fh}
}.

\begin{figure}[!htp]
   \centering
       \includegraphics[width=0.9\linewidth]{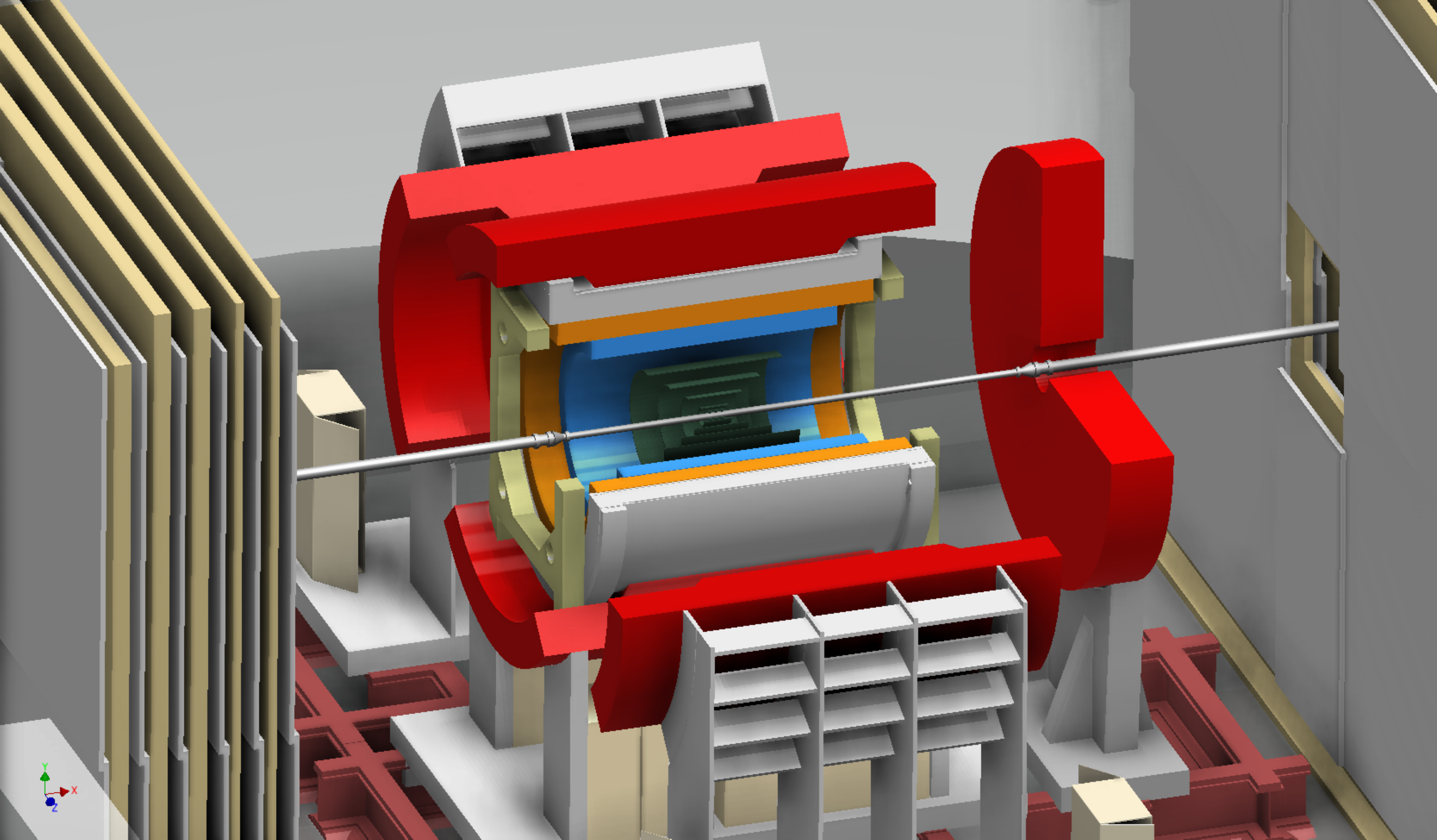}
       \vskip 4mm
       \includegraphics[width=0.9\textwidth]{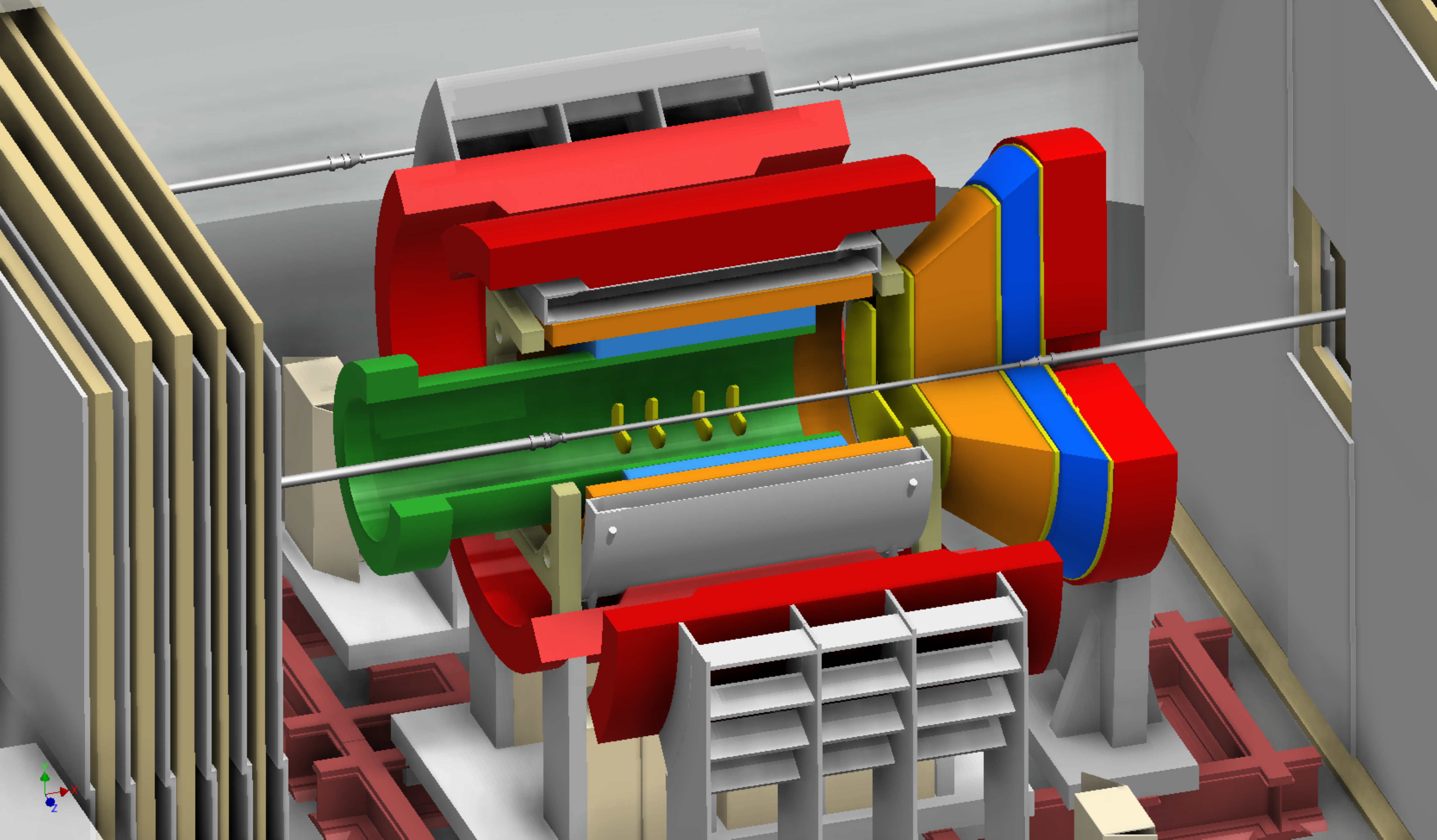}
       \caption[Proposed sPHENIX upgrade and its evolution to an EIC detector]{ (top) An engineering rendering of the proposed
         sPHENIX upgrade to the PHENIX experiment, showing the inner
         tracking system, the electromagnetic calorimeter, the BaBar
         solenoid and the hadronic calorimeter. (bottom) The evolution
         of the sPHENIX detector into a full-capability EIC detector.}
       \label{Fig:sPHENIX}
       \label{Fig:ePHENIX}
\end{figure}

{\bf sPHENIX:} The PHENIX collaboration has submitted a
proposal\cite{Aidala:2012nz} to the DOE for MIE funding (Major Item of
Equipment) to replace the PHENIX central detectors in order to provide
full hadronic and electromagnetic calorimetry along with charged
particle tracking over a pseudorapidity interval $\eta < 1$ The new
apparatus, sPHENIX, shown in Figure~\ref{Fig:sPHENIX} would
dramatically extend the range of jets measurable at RHIC and provide
precision spectroscopy of quarkonia.  With a mass resolution of better
than 100~MeV/$c^2$, sPHENIX will separately measure the $1S$, $2S$ and
$3S$ states of the upsilon, providing key information about Debye
screening in the QGP.  The full sPHENIX physics program employs
inclusive jet, dijet, $b$-tagged jet, $\gamma$$+$jet, high transverse
momentum charged hadron, jet fragmentation function, and upsilon
measurements to enable a very comprehensive and detailed investigation
of the microscopic dynamics of the QGP in the temperature range where
its coupling is at its strongest.

\begin{figure}[hbt!]
  \centering
  \includegraphics[width=0.8\textwidth]{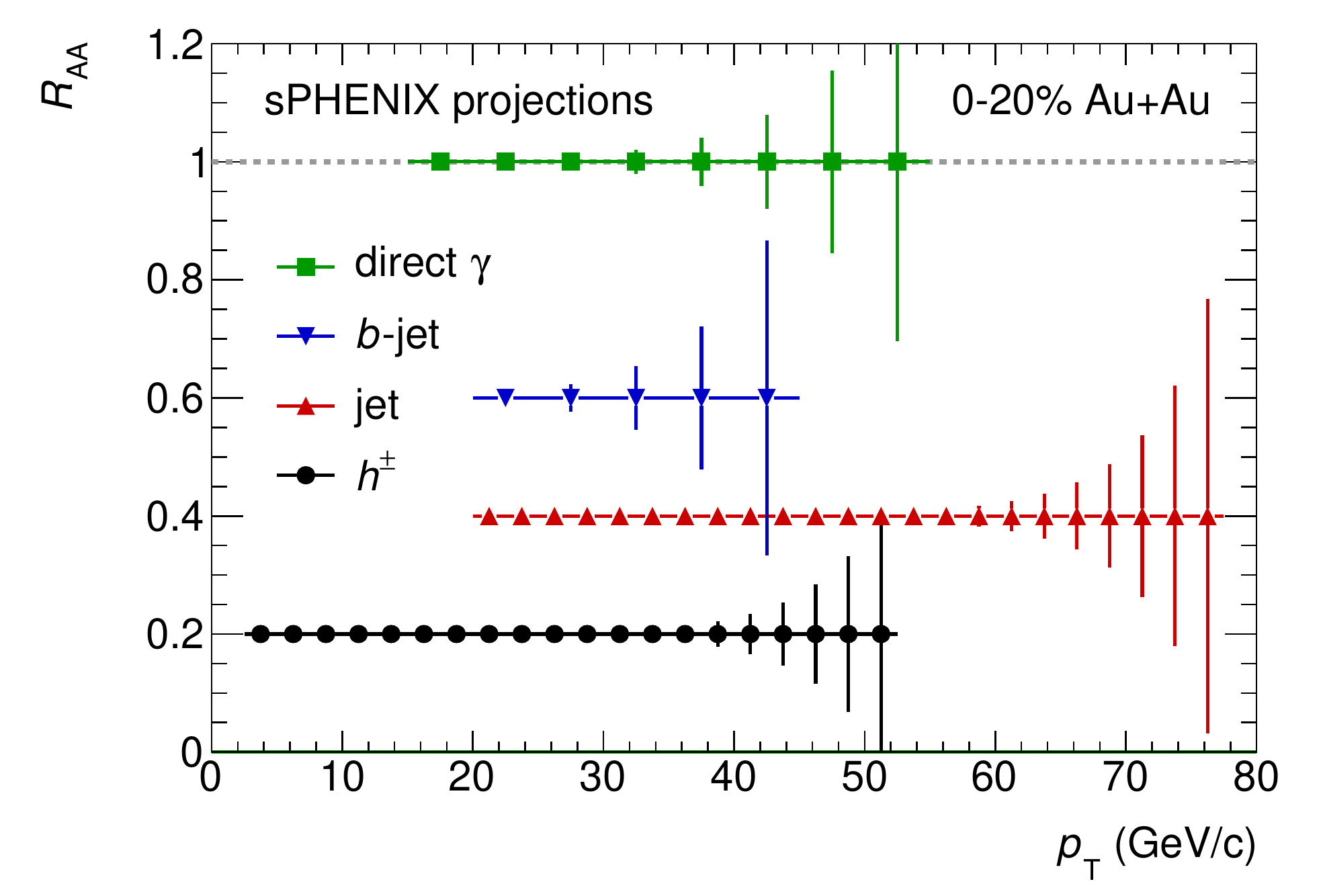}
  \caption[Projected sPHENIX statistical uncertainties on $R_\mathrm{AA}$
    for $\gamma$'s, jets, $b$-jets and $h^\pm$]{Projected statistical uncertainties on the $R_\mathrm{AA}$
    for inclusive photons (green points, assuming $R_\mathrm{AA} =
    1$), $b$-jets (blue points, assuming $R_\mathrm{AA} = 0.6$),
    inclusive jets (red points, assuming $R_\mathrm{AA} = 0.4$) and
    charged hadrons (black points, assuming $R_\mathrm{AA} =
    0.2$). These projections are made with a $b$-jet tagging
   efficiency of $50$\%, 10 weeks of $p$+$p$ and 22 weeks of Au$+$Au
    data taking.}
    \label{fig:AAphysics_projections}
\end{figure}

The very high data acquisition bandwidth of sPHENIX, combined with
RHIC~II luminosities, brings fundamentally new capabilities to the
measurement of hard probes at RHIC.  In one year, sPHENIX will record
100 billion minimum bias Au$+$Au collisions, providing an extremely
large sample of unbiased jets.  This enormous sample will be further
augmented by calorimetric triggers sampling more than 2/3 of a
trillion top-energy Au$+$Au collisions made possible by the RHIC~II
luminosity.  All told, this enables the measurements of jets in
$p$$+$$p$, $p$$+$Au and Au$+$Au beyond 70~GeV and a correspondingly
large kinematic reach for other hard probes, as shown in
Figure~\ref{fig:AAphysics_projections}.

This reach in $Q^2$, combined
with precision vertexing provides the basis for a compelling program of direct comparisons to corresponding measurements at the LHC
discussed in Sections~\ref{Sec:FutureJetCapabilities}, \ref{Sec:FutureJetProbes} and \ref{Sec:FutureQuarkonia}. The sPHENIX design takes advantage of recent technological advances in sensors and read-out electronics to minimize costs. In addition, the superconducting solenoid from the BaBar experiment at SLAC has been transferred to BNL\cite{BabarMove} for use in sPHENIX, resulting in a very considerable cost savings. Finally, the sPHENIX design provides a route for a smooth evolution to a full-capability EIC detector\cite{Adare:2014aaa}, shown in Figure~\ref{Fig:ePHENIX}. In addition, there is a an exciting extended program of polarized p+p and p+A\cite{PHENIXpppA,Aschenauer:2015eha}
if some of the EIC detector can be realized earlier.

\begin{figure}[!htp]
\includegraphics*[width=0.9\textwidth]{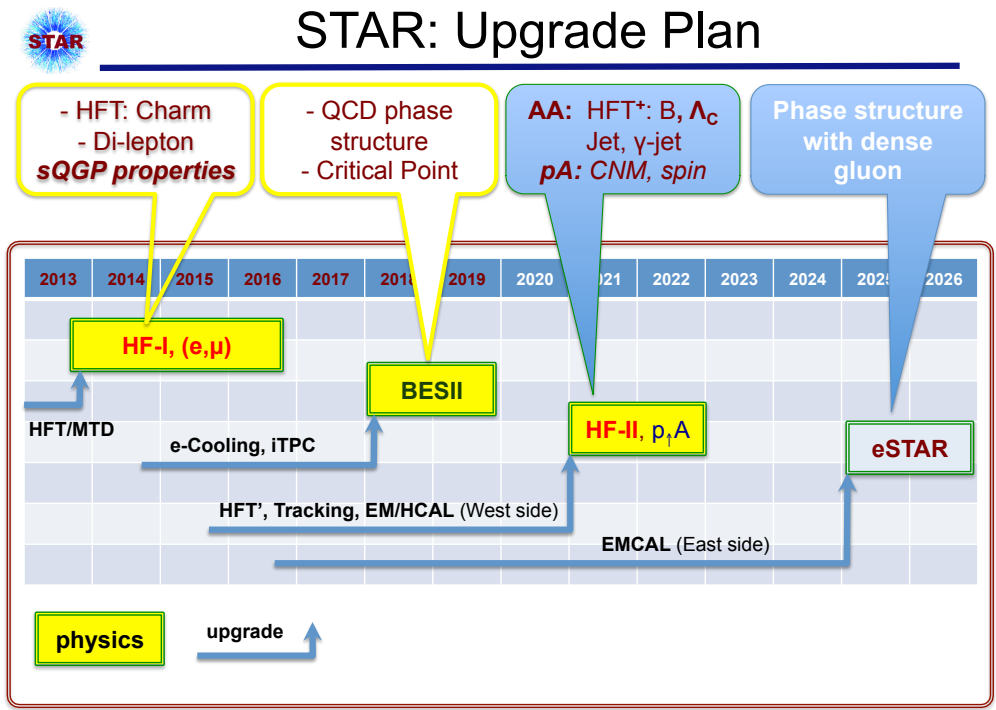}
\vspace{-0.5cm}
\caption[STAR upgrade plan towards an EIC detector]{The STAR upgrade plan,
showing the evolution from the current configuration focused on \pp\ and \AplusA\ physics at
mid-rapidity to a design emphasizing forward physics in \pp\ and \pA\ collisions\cite{STAR:FU,STAR:DecadalPlan}
and later  $e$+p and $e$+A physics at an EIC~\cite{STAR:eSTAR}.}
\label{Fig:STARplan}
\end{figure}

{\bf STAR Upgrades:} The STAR experiment has recently installed two new detector sub-systems: the Heavy Flavor Tracker discussed in Sections~\ref{Sec:Facilities} and Section~\ref{Sec:OpenHF}, and the Muon Telescope Detector described in Section~\ref{Sec:Quarkonia}.
With the completion of the HFT and the MTD,
STAR's upgrade plans next focus on  Phase II of the RHIC Bean Energy Scan.  The inner sectors of the TPC will be replaced to allow full coverage of pads increasing the pseudorapidity coverage, and extending momentum coverage to lower \pT\ . In addition STAR will  add an event plane detector which will enrich the STAR BES program by allowing for an independent measurement of the event plane  and significantly improving its resolution .  Both these and longer term upgrades improve STAR's forward capabilities 
in preparation for an eSTAR configuration in the EIC era\cite{STAR:eSTAR} as outlined in Figure~\ref{Fig:STARplan}.  

The STAR near-term proposed upgrades relevant to BES-II include:
\begin{itemize}
\item[] {\bf iTPC upgrade:} The STAR collaboration has proposed to upgrade the inner sectors of the read-out plane of the Time-Projection-Chamber (iTPC)~\cite{STAR:BESII}  in order to increase the segmentation on the inner pad plane and to renew the inner-sector wires. This upgrade will improve the resolution on both momentum and $dE/dx$ resolution, increase the track reconstruction efficiency and extend the  acceptance in 
pseudorapidity  from $|\eta| \le 1.1$ to $|\eta| \le 1.7$.
The enhanced performance made possible by the iTPC will not only benefit the BES-II physics program but will also be crucial for STARÕs future program with \pp\ / \pA\ and $e$+p/$e$+A collisions at the forward high-rapidity regions.

\item[] {\bf EPD:} The proposed Event Plane Detector (EPD) is a dedicated event-plane and centrality detector placed in the forward rapidity region 2 $\le | \eta| \le$ 4. With segmentation in both radial and azimuthal directions, the detector will provide precise measurements of the collision centrality and the event plane, as well as serving as a trigger detector for collisions at lower beam energies.
The EPD will be crucial for the physics measurements of collectivity as well as correlations in a much wider rapidity region. 
\end{itemize}

In the longer term STAR has proposed a series of mid-rapidity and forward upgrades that are complementary to the sPHENIX
physics program described above. The planned upgrades include:
\begin{itemize}
\item {\bf HFT$^+$:} The current HFT pixel integration time of $\sim 200~\mu$sec is much longer 
than the $\le 40~\mu$sec of the STAR TPC . 
In order to synchronize the TPC and HFT read-out for maximum rate capability for measuring bottom quark production at RHIC, the STAR collaboration has started design studies on the HFT$^+$, a faster version of the HFT with the state of art pixel technology. 
New technological developments permit read-out times less than
$20~\mu$sec without any increase of power consumption, so that the HFT support infrastructure
for cooling and power can be re-used for the HFT$^+$, making it a very cost-effective upgrade. 
The physics enabled by the HFT$^+$ physics is complementary to sPHENIX's jet program as well as ALICE's upgraded heavy flavor program
(to begin in 2019) at the LHC. In addition, the faster HFT$^+$ will be helpful for the heavy flavor physics in the spin program at RHIC.

{\bf Forward Upgrades:} The STAR collaboration has developed plans~\cite{STAR:FU,STAR:DecadalPlan} for measurements with forward photons, $J/\Psi$'s, Drell-Yan pairs, and di-jet and hadron/jet correlation probes, as well as $W$ and $Z$ bosons at top RHIC energy.
Measuring these probes in \pA\ collisions will further our understanding of cold nuclear matter effects in QCD processes in cold nuclear matter by studying the dynamics of partons at very small and very large momentum fractions $x$ in nuclei, and at high gluon density to investigate the existence of nonlinear evolution effects. STAR's forward upgrade plan is centered around the unique capabilities afforded with the existing STAR detector, complemented with detector upgrades including the forward calorimetric system (FCS) and forward tracking system (FTS), which are required to carry out the proposed physics program at forward rapidities. The proposed FCS and FTS upgrades were first envisioned in the STAR Decadal Plan~\cite{STAR:DecadalPlan} and represent a natural evolution of the growth of the STAR scientific program. These upgrades will be an integral part of the eSTAR configuration at eRHIC outlined in the eSTAR letter of intent~\cite{STAR:eSTAR}, see Figure~\ref{Fig:eSTAR}.
\begin{figure}[!htp]
\includegraphics*[width=0.9\textwidth]{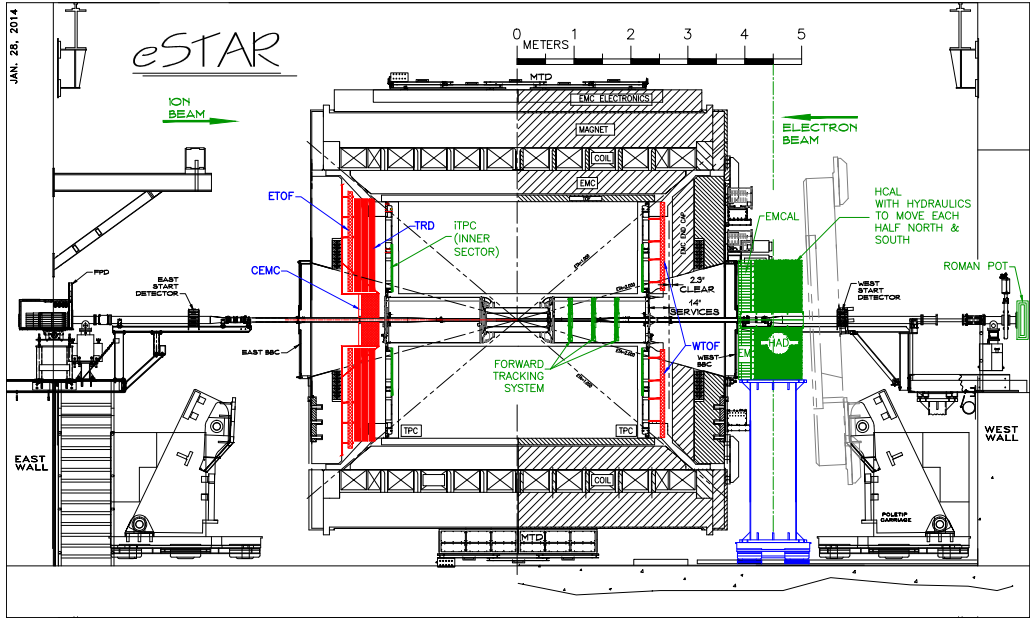}
\vspace{-1.5cm}
\caption[eSTAR layout with proposed upgrades]{eSTAR layout with the proposed upgrades of the iTPC, Forward Calorimetry System (FCS), Forward Tracking System (FTS), Endcap TOF (E/W TOF), BSO Crystal Calorimeter (CEMC), and a GEM-based transition radiation detector. In this configuration, the electron beam is from right to left while hadron beam from left to right~\cite{STAR:eSTAR}. }
\label{Fig:eSTAR}
\end{figure}
\end{itemize}

\subsubsection{Facility and experiment upgrades at the LHC}

Following the successful Run I p+p, Pb+Pb and p+Pb data taking periods, the LHC is now preparing
for Run II, forseen to include p+p, p+Pb and heavy ion data taking from 2015 to 2018. Run II
will be followed by a shutdown from 2018 to 2020 (LS2) and Run III from 2020 to 2023.
For both the p+p and Pb+Pb data taking, the LHC upgrades during the current shutdown should
allow for collisions at close to the design energy, i.e., $\sim$5~TeV for Pb+Pb. In addition, a
large increase in the Pb+Pb instantaneous luminosity is projected, with collision rates expected
to exceed those achieved in Run I by up to one order of magnitude. Combining Run II and III,
the LHC goal is to deliver about 10~nb$^{-1}$  of Pb+Pb collisions to each of ALICE, ATLAS and CMS. In combination,
the increased collision energy and luminosity will increase statistics for rare high \pT\
probes by about a factor of 200.

To exploit the improved accelerator performance, ALICE, ATLAS and CMS are undergoing
significant upgrades during the current shutdown and in the future LS2. For ATLAS and CMS, these upgrades
are mostly driven by the needs of the p+p program. Further large luminosity increases for
\pp\ will produce a larger number of collisions per bunch crossing (``pileup''), eventually
reaching multiplicities per event that are within a factor of 2 of average heavy-ion
collisions. This will require extension (e.g., adding a fourth pixel tracker layer)
and eventually replacement of the silicon inner tracker detectors of the two experiments to
cope with the increased particle densities. In addition, the rejection power of the
trigger systems is being improved by increasing the trigger granularity at the
hardware (L1) level. Both the inner tracker and trigger upgrades, as well as other
developments, are well matched to the needs of the heavy ion program in Run II and III. Of
particular importance is the improved trigger selectivity for jets in central events at L1,
which is essential to fully sample the expected collision data for jet-related probes.

ALICE is preparing for Run II with an expansion of the calorimetric coverage (EMCAL) which will allow for dijet studies and improved jet triggering. During LS2 the experiment's data taking capabilities will be significantly enhanced with major upgrades to detector readout and data acquisition systems  to allow the collection of data at the full collision rate. This includes in particular a replacement of the TPC readout with faster detectors and electronics. In
addition, a replacement of the silicon inner tracking system which would improve precision, acceptance and readout speed has been proposed as well as a proposal to add a silicon telescope in front of the current forward muon detector to improve the low \pT\ momentum resolution of the reconstructed muons. While the main physics driver for the upgrades is precision measurements of the low \pT\ open heavy flavor
program studying e.g.\ charm production and equilibration, these upgrades also benefit the
intermediate and high \pT\ jet quenching program.

%% file: tex/CriticalPoint.tex
\subsection{Mapping the Crossover and Searching for the QCD Critical Point}
\label{Sec:CP}

%
\begin{figure}[!thp]
\vspace{-0.3in}
\begin{center}
\centerline{  \includegraphics[width=0.7\textwidth]{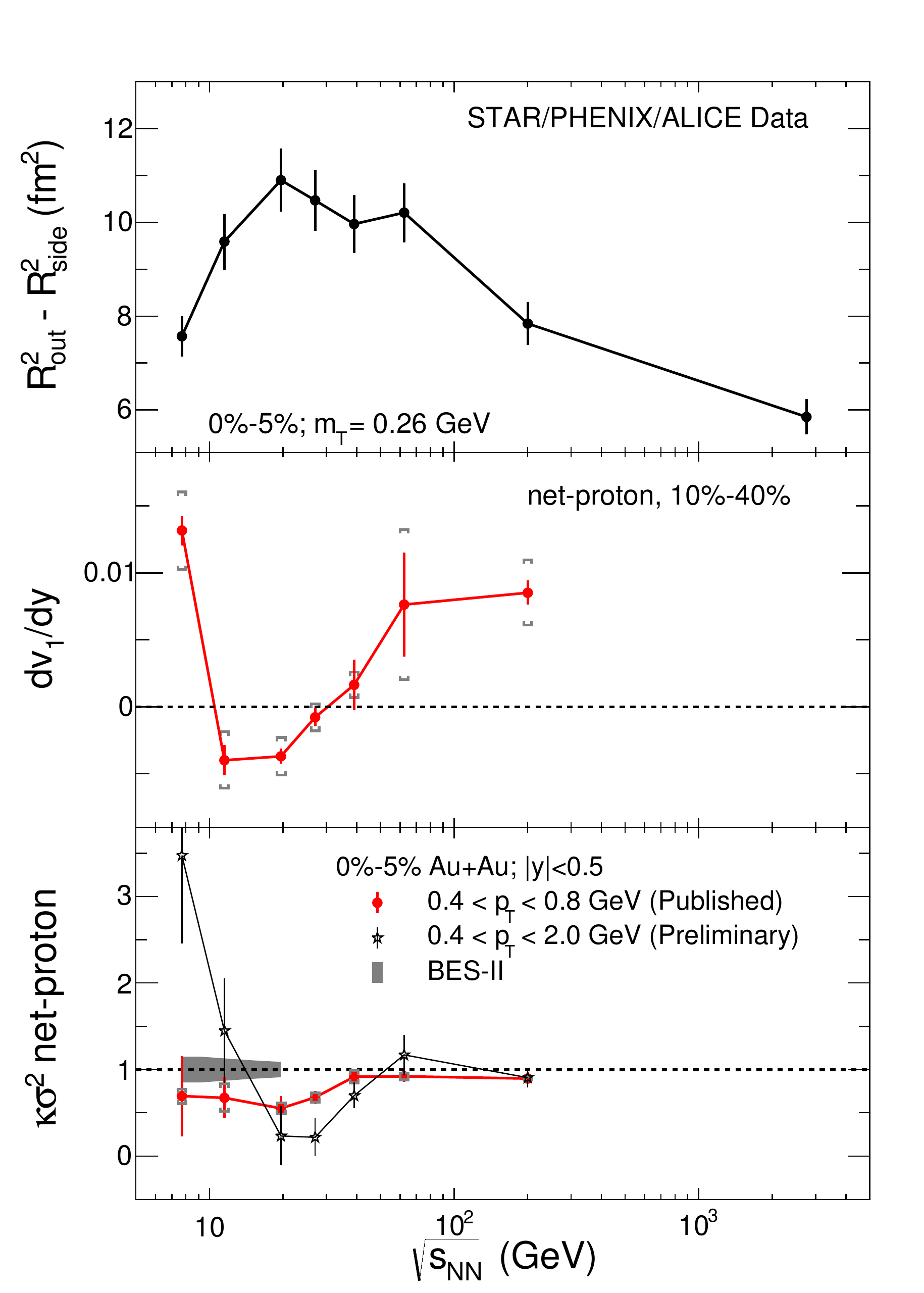}}
\caption[Observables showing non-monotonic behavior as a function of $\sqrt{s_\mathrm{NN}}$]{Three selected observables
  that all show interesting non-monotonic behavior as functions of collision energy around  $\sqrt{s_\mathrm{NN}}{\,\sim\,}15{-}40$\,GeV.  
    {\bf Top panel:}
  The difference 
  $R_{out}^{2}-R_{side}^{2}$
  between the squared radii
  in the outward and sidewards directions
  measured via two-pion interferometry
  vs. $\sqrt{s_{NN}}$ using STAR~\cite{Adamczyk:2014mxp}, PHENIX~\cite{Adare:2014qvs}, and 
  ALICE~\cite{Aamodt:2011mr}
  data. $R_{out}^{2}-R_{side}^{2}$, 
  reflects the lifetime of the
  collision fireball and was predicted~\cite{Rischke:1996em} to reach
  a maximum for collisions in which a hydrodynamic fluid forms at
  temperatures where the equation of state is softest.  
  {\bf Middle panel:} The rapidity-slope of the net proton directed flow $v_1$,
  $dv_1/dy$. This quantity is sensitive to early pressure gradients in
  the medium.  
  {\bf Bottom panel:} The kurtosis of the event-by-event
  distribution of the net proton (i.e. proton minus antiproton) number
  per unit of rapidity, normalized such that Poisson fluctuations give
  a value of $1$. In central collisions, published results in a
  limited kinematic range~\cite{Adamczyk:2013dal} show a drop below
  the Poisson baseline around $\sqrt{s_{NN}}=$27 and 19.6 GeV. New
  preliminary data over a larger $p_T$ range~\cite{CPODKurtosis},
  although at present still with substantial error bars, hint that the
  normalized kurtosis may, in fact, rise above 1 at lower
  $\sqrt{s_{NN}}$, as expected from critical fluctuations~\cite{Stephanov:2011pb}. 
  The grey band shows the much reduced
  uncertainties anticipated from BES-II in 2018-2019, for the 0-5\%
  most central collisions.}
\label{F-PD2}
\vspace{-0.1in}
\end{center}
\end{figure}
%

{\bf The need for BES-II and accompanying advances in theory}

Several observables from the
first phase of the RHIC Beam Energy Scan program (BES-I, introduced
in Section~\ref{Sec:BES})
exhibit interesting 
non-monotonic behavior as a function of collision energy,
and hence as a function of baryon number chemical potential $\mu_B$.
A selection of such measurements is shown in Figure~\ref{F-PD2}.  
At present, we await with considerable interest
the results from the final run in
the BES-I program at $\sqrt{s_{NN}}=14.5$\,GeV, where the data were taken only
a few months ago, since for a number of important observables the
measurements made previously at and below $\sqrt{s_{NN}}=19.6$\,GeV 
have quite
limited statistics. Nevertheless, it is already possible to see trends
and features in the data that provide compelling motivation 
for both a strong and concerted theoretical response aimed at quantitative
precision as well as for experimental measurements with much higher statistical precision at 
and below $\sqrt{s_{NN}}=19.6$\,GeV 
(i.e. at the largest achievable values of $\mu_B$)
that will be provided by the second phase of the Beam Energy Scan
program (BES-II) in 2018 and 2019. The goal of BES-II, as described
in more detail below, is to follow through in order to turn currently observed trends and
features into definitive scientific conclusions. As discussed in Section~\ref{Sec:RHICUpgrades}, 
the accelerator physicists at RHIC are planning a machine upgrade to
provide electron cooling to increase the beam luminosity at these
energies by about a factor of 10 \cite{BESII}. Targeted new detector
capabilities will also increase the sensitivity to the signals
described below in the BES-II campaign \cite{BESII}.

Experimental discovery of a first-order phase transition or a critical
point on the QCD phase diagram would be a landmark achievement. The
first goals of the BES program, however, have focused on obtaining a quantitative 
understanding of the properties of matter in the crossover region of the 
phase diagram as it changes with increasing $\mu_B$. Available tools
developed over the last few years now make a quantitative comparison
between theory and experiment tractable in the $\mu_B$-range
below any QCD critical point. Success in this effort, in and of itself, would
establish another major and lasting impact of the RHIC program. Questions
that can be addressed in this regime include quantitative study of the
onset of various signatures associated with the presence of
quark-gluon plasma and of the onset of chiral symmetry restoration as
one traverses the crossover region. Data now in hand from BES-I
provide key inputs and impetus toward this goal. Here we give four
examples, intended to be illustrative, of areas where a coherent
experimental and theoretical effort is expected to have substantial
impact on our understanding of QCD. In each case we note the
substantial impact expected from the additional measurements
anticipated during the BES-II:

{\bf 1.} The directed flow observable $dv_1/dy$ for net protons has been 
found to feature a dip as a function of collision energy (see middle panel 
in Figure~\ref{F-PD2}), with a minimum at energies somewhere between 
$\sqrt{s_{NN}}=11.5$ and 19.6\,GeV \cite{Adamczyk:2014ipa}. This has 
long been predicted in qualitative terms as a consequence of the 
softening of the equation of state in the transition region of the phase 
diagram \cite{Brachmann:1999mp,Stoecker:2004qu}. Several theoretical 
groups around the world have now begun hydrodynamic calculations with 
nonzero baryon density, deploying all the sophistication that has been
developed very recently in the analysis of higher energy collisions,
including initial fluctuations and a hadronic afterburner, in
applications to these lower energy collisions. These
hydrodynamic+hadronic cascade calculations will be used to compare the
$dv_1/dy$ data with equations of state in the crossover region of the
phase diagram obtained from lattice calculations via Taylor expansion
in $\mu_B/T$ \cite{Huovinen:2014woa}. This is a program where a
quantitative comparison, successful or not, will be of great interest,
since failure to describe the data could signal the presence of a
first-order phase transition. The precision of a comparison like this
will be substantially improved in 2018-19 when BES-II data will allow
$dv_1/dy$ to be measured for the first time with tightly specified
centrality; the statistics available in the BES-I data sets limit
present measurements to averages over collisions with widely varying
impact parameters \cite{Adamczyk:2014ipa}.

{\bf 2.} A second goal of the hydrodynamic calculations referred to
above will be to use identified particle BES-I $v_2$ data to map, in
quantitative terms, where and how hydrodynamics starts to break down
at lower collision energies, and where, to an increasing extent, $v_2$
develops during the hadron gas phase when viscosities are not small,
{\it i.e.}~where the contribution of the partonic phase to observed measures
of collectivity decreases in importance. A key future experimental
input to this program is the measurement of the elliptic flow $v_2$ of
the $\phi$-meson, which will be obtained with substantially greater
precision in the BES-II program. The first measurements of $v_2$ of
$\Omega$ baryons at these collision energies, also anticipated in
BES-II, will represent a further, substantial advance. Seeing $\phi$
mesons flowing like lighter mesons and $\Omega$ baryons flowing like
lighter baryons in collisions at a given energy would indicate that
the dominant contribution to the collective flow in those collisions
was generated during the partonic phase \cite{Abelev:2007rw}.

This component of the BES program, together with the following one,
will yield guidance as to what the lowest collision energies are at
which temperatures in the transition region of the phase diagram can
be explored. That is, they will tell us the largest value of  $\mu_B$ 
for which it will be
possible to use heavy-ion collisions, anywhere, to study matter in the
crossover region and search for a possible critical point.

{\bf 3.} Heavy-ion collisions at top RHIC energies and at the LHC have
now seen several experimental phenomena
\cite{Abelev:2009ac,Abelev:2012pa,Adamczyk:2013kcb} that may be
related to the chiral magnetic effect (CME
\cite{Fukushima:2008xe,Kharzeev:2010gr}, see Section~\ref{Sec:Exotica}). In
each case, alternative explanations are also being considered
\cite{Bzdak:2009fc,Pratt:2010zn}. One of the intriguing BES-I results
is that the three-particle correlations that are related to charge
separation across the reaction plane, possibly induced by the CME, are
clearly observable over most of the BES range but then seem to turn
off at $\sqrt{s_{NN}}=$ 7.7 GeV \cite{Adamczyk:2014mzf}, where the
elliptic flow $v_2$ is still robust. This is an indication that
$v_2$-induced backgrounds alone do not explain the observed
correlations. The observation that these three-particle correlations
disappear at the lowest energy could prove crucial to understanding
their origin and how they are related to the formation of QGP. On the
theoretical side, lattice QCD calculations probing the response of the
equation of state and transition temperature to the presence of
external magnetic fields~\cite{DElia:2010nq,Bali:2011qj,Bali:2014kia} are needed
to understand these signals.
Also necessary are hydrodynamic calculations incorporating magnetic fields and
chiral effects; these are being pursued by several groups, with
first results starting to appear~\cite{Hirono:2014oda}. On
the experimental side, higher statistics BES-II data will make it
possible to determine with much greater precision the $\sqrt{s_{NN}}$
at which this effect turns off and will also make it possible to
measure the (related but theoretically more robust) chiral magnetic
wave phenomenon~\cite{Kharzeev:2010gd,Burnier:2011bf}, which has also
been seen at top RHIC energy and at the LHC~\cite{Wang:2012qs,Belmont:2014lta}, 
and which should turn off below the
same $\sqrt{s_{NN}}$ where the CME-related observables
turn off, if these interpretations are correct.

{\bf 4.} Theoretical developments over the past decade have identified
specific event-by-event fluctuation observables most likely to be
enhanced in collisions that cool in the vicinity of the critical point
\cite{Stephanov:2008qz,Athanasiou:2010kw}. Higher moments of the
event-by-event distribution of the number of protons, or the net
proton number, are particularly sensitive 
\cite{Ejiri:2005wq,Athanasiou:2010kw,Karsch:2010ck}. STAR has now 
measured the first four moments (mean, variance, skewness and kurtosis) 
of the event-by-event distribution of net proton number and net charge at 
the BES-I energies \cite{Adamczyk:2013dal,Adamczyk:2014fia}. At the 
lowest collision energies, although the statistics are at present rather
limiting, there are interesting trends, including for example the drop in the
normalized kurtosis of the net-proton distribution at $\sqrt{s_{NN}}{\,=\,}27$
and 19.6 GeV (see bottom panel in Figure~\ref{F-PD2}). This drop in and 
of itself can be at least partially reproduced via prosaic effects captured 
in model calculations that do not include any critical point. Theoretical 
calculations of the contributions from critical fluctuations predict  
\cite{Stephanov:2011pb} that if the freezeout $\mu_B$ scans past a 
critical point as the beam energy is lowered, this kurtosis should first 
drop below its Poisson baseline and then rise above it. Both the drop 
and the rise should be largest in central collisions in which the 
quark-gluon plasma droplet is largest and therefore cools most slowly, 
allowing more time for critical fluctuations to develop~\cite{Berdnikov:1999ph}. 
A recent and still preliminary analysis~\cite{CPODKurtosis} 
that includes protons over a larger range in $p_T$ than measured before~\cite{Adamczyk:2013dal},
also shown in the bottom panel of 
Figure~\ref{F-PD2}, 
features a more substantial drop in the net proton kurtosis at $\sqrt{s_{NN}}$= 27 and 19.6 GeV
as well as intriguing hints of a rise above one at 
$\sqrt{s_{NN}}{\,=\,}11.5$ and  7.7\,GeV in central collisions, 
but the uncertainties are at present too 
large to draw conclusions. If this kurtosis does rise at $\sqrt{s_{NN}}$ 
values below 19.6 GeV, it would be difficult to understand in 
conventional terms and thus would be suggestive of a contribution 
from the fluctuations near a critical point. Determining whether this 
is so requires the higher statistics that BES-II will provide, as illustrated
by the grey band in the bottom panel of Figure~\ref{F-PD2}. 

The present data on moments of both the net proton number and the 
net charge at the higher BES-I energies are already very useful, as
they can be compared to lattice calculations of the Taylor expansions 
(in $\mu_B/T$) of the baryon number and charge susceptibilities~\cite{Karsch:2012wm}. 
First versions of this comparison have been
reported recently and are being used to provide an independent
determination of how the freeze-out values of $\mu_B$ and $T$ change
with collision energy~\cite{Bazavov:2012vg,Mukherjee:2013lsa,Borsanyi:2013hza,Borsanyi:2014ewa}. 
However, looking ahead, theoretical 
calculations will need to faithfully
account for the dynamical evolution of the medium formed in the
collision for a full quantitative exploitation of the experimental
data. For the higher statistics BES-II data on the net proton
kurtosis, skewness, and other fluctuation observables at low collision
energies to determine the location of the critical point on the phase
diagram of QCD, if one is discovered, or alternatively to reliably exclude its
existence within the experimentally accessible region of the phase
diagram, a substantial theoretical effort will be needed that couples
the sophisticated hydrodynamic calculations referred to above with a
fluctuating and dynamically evolving chiral  order parameter.

As the following fifth example illustrates, BES-II will also open the
door to measurements that were not yet accessible in the first phase
of the BES program:


{\bf 5.} Dileptons are unique penetrating probes with which to study
the chiral properties of hot and dense matter (see Section~\ref{Sec:EM}). The dielectron
invariant mass distributions measured in the BES-I (in data taken at
$\sqrt{s_{NN}}=$ 200, 62.4, 39 and 19.6 GeV; see Figure~\ref{fig:star-ee}) have shown that there is
a significant enhancement of low mass dileptons below 1 GeV relative
to a hadronic cocktail~\cite{Huck:2014mfa}. The data so far are qualitatively
consistent with a model in which hadron properties are modified in the
medium and there is a partonic contribution as
well~\cite{Rapp:2009yu}. However, data at lower energies with higher
statistics are crucial in order to test the predicted strong
dependence of dilepton yields on baryon density and draw firm
conclusions. The dilepton measurements at and below
$\sqrt{s_{NN}}{\,=\,}19.6$\,GeV that BES-II will provide will yield a
qualitatively new understanding of the chiral properties of QCD matter
with significant baryon density. There are two interesting dilepton
mass windows to be studied at BES-II: the low mass window (300 MeV --
700 MeV) and the high mass window (800 MeV -- 1.5 GeV). The former
will provide indirect information on chiral symmetry restoration via
the interaction of vector mesons with (excited) baryons, while the
latter will probe chiral restoration directly via the mixing between
vector and axial-vector mesons in the hot and dense environment.


Each of these five examples makes it clear that in order to maximize
the physics outcome from BES-I and BES-II, a coherent effort between
experimentalists and theorists working on QCD at nonzero $T$ and
$\mu_B$ is essential.
Indeed,
there has been considerable progress in lattice QCD recently on the
calculation of various QCD susceptibilities
\cite{Borsanyi:2011sw,Bazavov:2012jq} and the QCD equation of state in
the regime where $\mu_B$ is nonzero but sufficiently small compared to
$3\,T_c$ \cite{Borsanyi:2012cr,Hegde:2014wga}. These lattice
calculations provide the necessary inputs for extending 
the kind of sophisticated hydrodynamic calculations (including
initial fluctuations and a late stage hadron cascade) that have been
developed over the past few years to nonzero $\mu_B$. For some purposes, these
calculations additionally require coupling to a fluctuating and evolving
chiral order parameter.
In concert, such developments will provide the critical tools for
obtaining from BES-I and BES-II data answers to fundamental questions
about the phases, the crossover, and perhaps the critical point and
first-order transition, in the QCD phase diagram.

The examples that we have sketched show that BES data,
at present and in the future from BES-II, 
together with the concerted theoretical response that present
data motivates will yield quantitative
understanding of the properties of QCD matter in the crossover region
where QGP turns into hadrons. 
If there is a critical
point with $\mu_B<400$~MeV, BES-II data on fluctuation and flow 
observables together with the theoretical
tools now being developed should yield quantitative evidence for its
presence.  
The span in $T$ and $\mu_B$ that the flexibility of RHIC
makes accessible, along with the technical advantages of
measuring fluctuation observables at a collider mentioned in Section~\ref{Sec:CP},
together with the 
recent and planned
detector and facility upgrades (low energy electron cooling in particular), 
mean that RHIC is uniquely positioned in the world 
to discover a critical point in the
QCD phase diagram if Nature has put this landmark in the
experimentally accessible region. Late in the decade, the FAIR
facility at GSI\cite{FAIR} will extend this search to even higher $\mu_B$ if its
collision energies continue to produce matter at the requisite
temperatures.

%% file: tex/HardProbes.tex
\subsection{Hard Probes of the Quark-Gluon Plasma}
\label{Sec:HardProbes}

High transverse momentum single hadrons, fully reconstructed jets,
open heavy flavor and heavy quarkonia provide information about the
strongly coupled QGP that complements the information provided by the
bulk and collective observables of the soft sector.  The higher
momentum and mass scales of these hard probes drive their production
to the earliest times in the collision; these same properties imply
that these probes don't completely thermalize during the evolution of
the produced medium.  The information imprinted on hard probes by QGP
production mechanisms and its later dynamics can therefore survive
into final state observables, making them uniquely valuable as
investigative tools of the complex emergent physics of the QGP.
However, accessing these very positive attributes of hard probes is
complicated in many cases by small production cross sections or the
need for specialized techniques and detectors.

In this section we first outline the future of jet studies over the
next decade, enabled by major developments in the RHIC and LHC
accelerator facilities and experiments described in
\ref{Sec:FacilitiesFuture}.  After that, we focus on the coming
prospects for open heavy flavor and heavy quarkonia measurements.

\subsubsection{Overview}

Previous studies at RHIC and the LHC have clearly demonstrated the ability 
to reconstruct jet observables in the high multiplicity heavy ion environment. 
Comparisons of hadronic and jet-based measurements to model calculations 
in a perturbative QCD framework have allowed the extraction of the QGP transport coefficient  \qhat\
with an uncertainty of about 50\%. Full jet 
measurements have demonstrated the modification of the dijet and photon-jet
momentum balance due to energy loss in the QGP, and have provided 
the first look at the modification of the jet structure itself due to 
interactions of the hard probe with the medium. These studies demonstrate the 
transport of energy from the jet core to low transverse momenta, close 
to thermal momentum scales, and away from the jet axis towards 
large angles far outside of the typical jet cone definition.

\begin{figure}[t]
\includegraphics[width=1.05\textwidth]{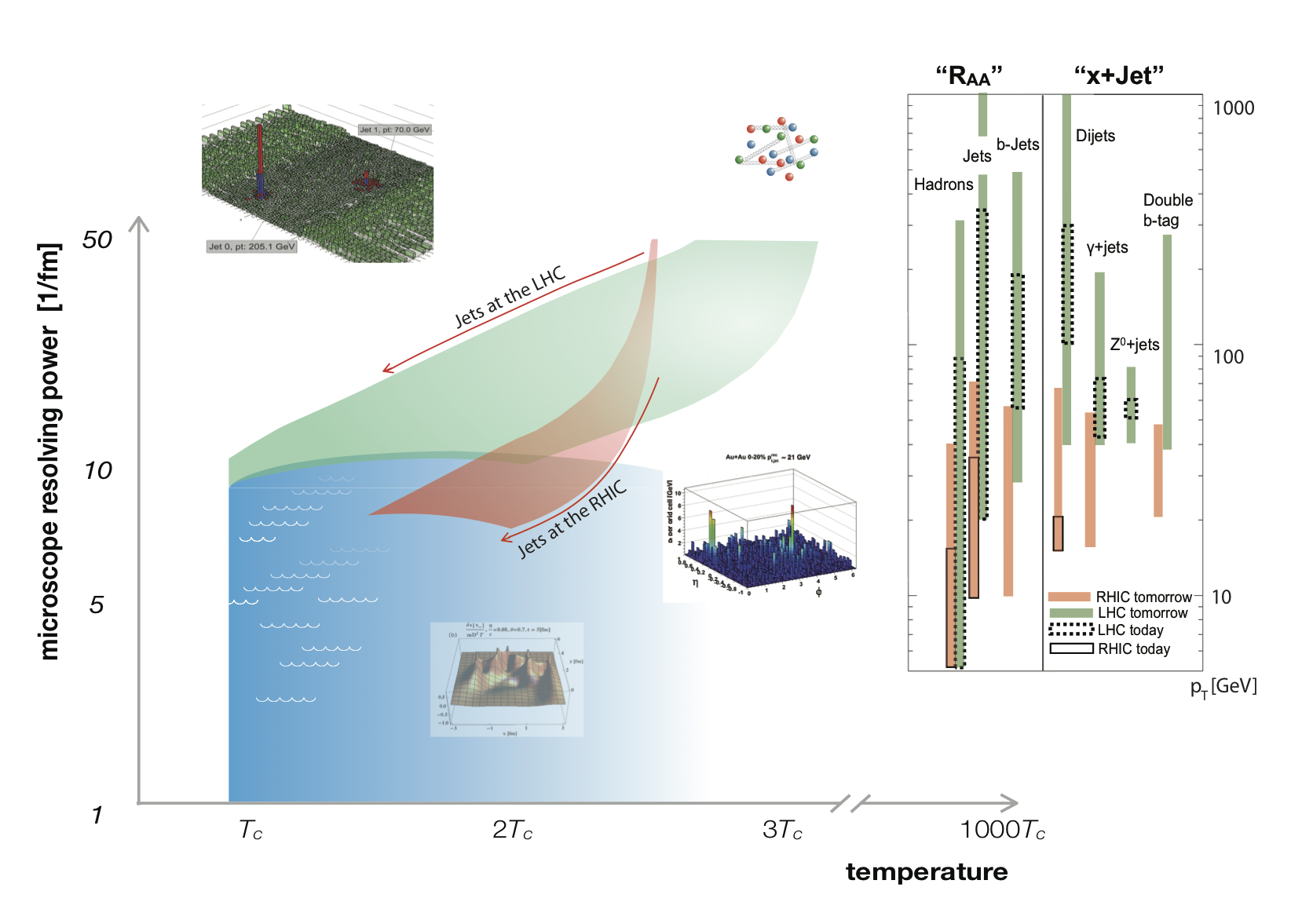}
\caption[Graphical representation of a future physics goal]{Graphical representation of the future physics goal: Hard scattered partons at the LHC and at RHIC evolve through splittings and interaction with the medium, providing sensitivity to QGP dynamics over a wide range of length scales. The kinematic reach of future RHIC and LHC jet observables and their important kinematic overlap, enabled by new instrumentation at RHIC, is shown in an artist rendering.}
\label{Fig:HardProbesFuture}
\end{figure}

Future jet-based studies, built on the achievements at RHIC and the LHC, will 
address fundamental questions about the nature of QGP. These include 
precise measurements of QGP transport coefficients as a function of 
temperature, a detailed characterization of the QGP response to the 
parton energy loss and studies of the modification the jet angular 
and momentum structure as a function of angular and momentum scale. 
In combination, the goal of these studies is to determine the 
microscopic (or quasi-particle) nature of QGP 
and to understand how the macroscopic QGP liquid emerges from the 
underlying QCD degrees of freedom by probing the QGP dynamics over a wide range of length scales, see Figure\ \ref{Fig:HardProbesFuture} for a graphical representation.

This program will be enabled by the evolution of the RHIC and LHC 
accelerator facilities, upgrades to the existing experiments and the 
construction of sPHENIX, a state-of-the-art jet detector at RHIC. In 
parallel, experiment/theory collaborations will be strengthened 
and expanded, to fully utilize the increased precision and range of 
experimental observables. 

\subsubsection{Future jet physics capabilities at RHIC and LHC}
\label{Sec:FutureJetCapabilities}

Section~\ref{Sec:FacilitiesFuture} has provided an outline of the RHIC and LHC
accelerator and experiment upgrade plans. These upgrades benefit jet physics studies
at the two facilities in three major ways:
\begin{enumerate}
\item The statistical precision and kinematic reach for commonly used jet physics observables
is vastly increased, as shown in \ref{Fig:HardProbesFuture}). For single charged hadrons and reconstructed jets, the \pT\ reach will be 
extended by a factor of 2--3, up to 40~GeV for hadrons and 70~GeV for jets at RHIC (see Figure\ \ref{fig:AAphysics_projections}) and 
300~GeV and 1~TeV, respectively, at LHC. The increased lever arm will be crucial in further improving
the extraction of e.g.\ the \qhat\ coefficient from model comparisons. Equally importantly, 
the larger data sets will allow a more detailed determination of \qhat\ as a function
of path length and medium conditions, e.g.\ in very peripheral collisions in the two 
energy regimes.

\item Beyond increasing the kinematic range and statistical precision for 
current ``workhorse'' measurements, the combination of the increased luminosity at RHIC and LHC, 
increased LHC collision energy and the experiment upgrades will move 
the focus of experimental and theoretical studies to rare, highly specific
observables. Key examples are measurements of isolated photon + jet 
correlations at RHIC and LHC, as well as $Z^0$+jet correlations at LHC. 
As a benchmark, the number of recorded photon+jet events at LHC 
with $\ptg > 60$~GeV is expected to reach more than $3 \times 10^5$, compared
to about 3000 in the Run I data sets. Using the photon tag, the initial
energy of the scattered parton is determined on an event-by-event basis to about 15\%.
The photon tag also identifies the partons as quarks and provides their 
initial direction. Furthermore, the comparison of photon+jet events 
at RHIC and the LHC allows the selection of nearly identical initial 
hard scattering configurations embedded in different initial medium conditions
in terms of temperature and energy density (see Figure\ \ref{Fig:HardProbesFuture}). As jets and the medium co-evolve 
from their initial virtuality and conditions to final state hadrons, 
the comparison of various observables between the RHIC and LHC events
will provide key insights into the temperature dependence of the jet-medium
interactions. Future measurements enabled by this program include the 
absolute quark energy loss
as a function of quark energy and path length, modifications of the 
jet momentum and angular structure and the large angle momentum flow
and medium response as a function of event-by-event jet energy loss.
\item Finally, the very high statistics jet samples to be collected at the LHC and 
with a future state-of-the-art jet detector at RHIC will allow analyses based on 
a new generation of jet shape observables. There is intense activity
in the development of generalized jet structure variables at the LHC to maximize
the efficiency of discovery measurements by improving quark/gluon
discrimination and the tagging of boosted objects. Heavy ion studies 
of the modification of the jet momentum and angular structure through 
medium interactions will benefit greatly from these developments.
Present measurements at the LHC have shown the jet structure to be modified 
both within a typical jet cone size (e.g. $R = 0.5$) and beyond, with
the fragmentation products inside the cone shifted towards 
lower $\pt$ and larger angles and an associated transport of most of the ``lost''
jet energy outside of the typical jet cone size. First results 
at RHIC, probing different medium conditions, indicate similarities
in the modifications of the jet momentum structure, but possible
differences in the angular modifications. Using common, well-calibrated
jet shape observables at RHIC and the LHC in different regimes of 
medium conditions will be critical in relating the observed 
modifications to the fundamental properties of the QGP,
in extracting the temperature dependence of QGP transport coefficients
and ultimately in understanding the nature of the medium in the vicinity
of the phase transition and at temperatures much larger than $T_C$.

\end{enumerate}

\subsubsection{Future jet probes of the QGP}
\label{Sec:FutureJetProbes}

The combination of large increases in delivered luminosity over the next
decade, upgrades to the existing LHC detectors and the construction
of a state-of-the-art jet detector at RHIC will enable a coherent 
physics program employing well-calibrated common observables to
study jet modifications and jet-medium interactions over a wide 
range of medium conditions created at RHIC and the LHC, as shown
schematically in Figure\ \ref{Fig:HardProbesFuture}.
Together, RHIC and the LHC will provide a physics program that includes
a precision extraction
of QGP transport coefficients related to jet-medium interactions.
Even more importantly, this program will employ jets as a tool to understand how the 
observed strongly coupled (liquid) nature of the QGP arises from the 
underlying QCD micro-physics by probing the QGP dynamics over a wide range of length scales (see Figure\ \ref{Fig:HardProbesFuture}). 
By conducting these investigations we will move from observing what the properties
of QGP are to understanding how these properties 
arise from the underlying gauge theory.
\pagebreak

\begin{figure}
\includegraphics[width=0.9\textwidth]{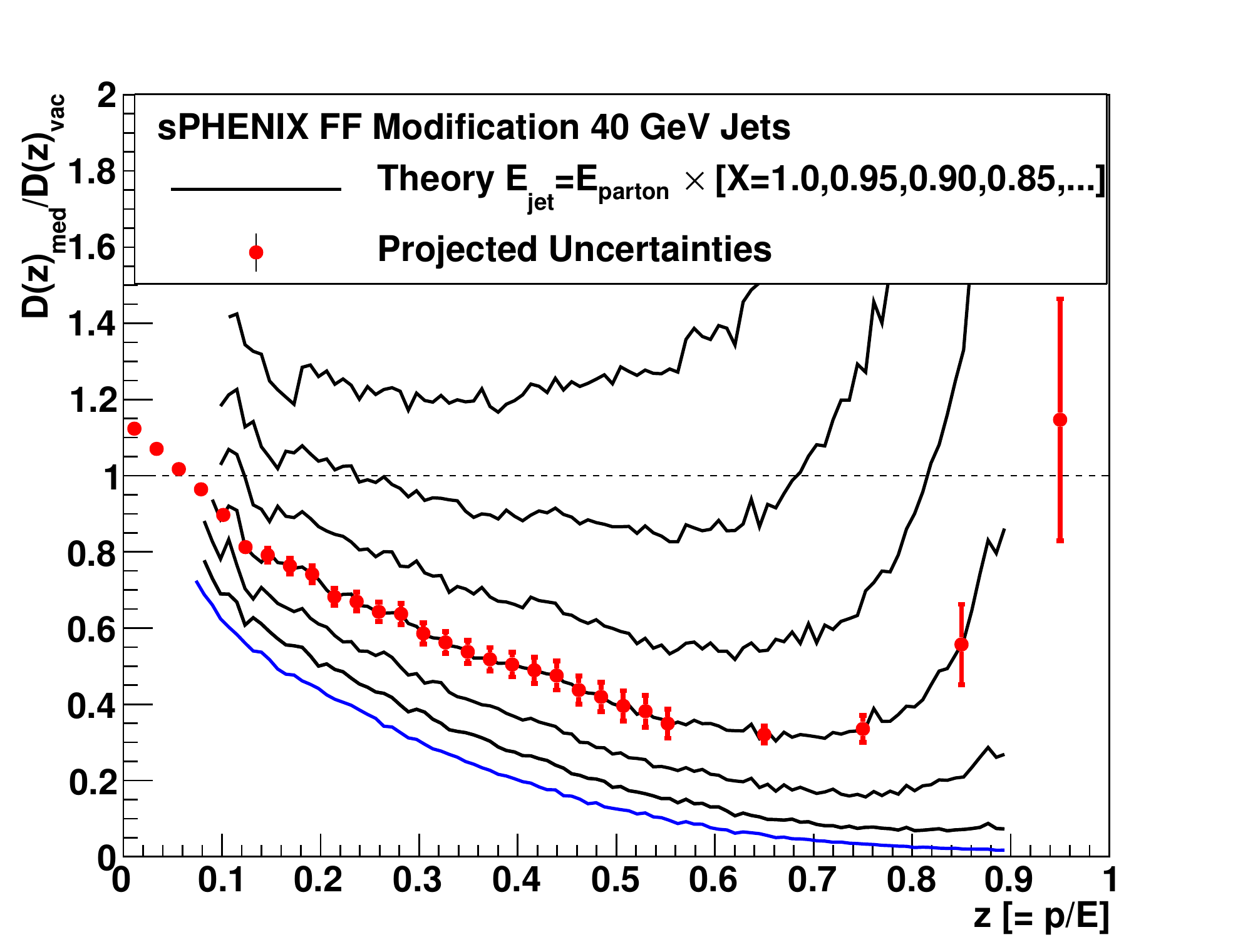}
\caption[Projected sPHENIX statistical uncertainties on modified fragmentation functions]{Modified fragmentation function $D(z)$ in the medium\cite{Armesto:2007dt} expressed as the ratio of the modified $D(z)$ to that assuming vacuum fragmentation. The different black curves show the results of different assumptions for the fraction $X$ of the parton energy retained in the jet cone, with the original prediction corresponding to $X = 1.0$ shown as the lower blue curve. The projected statistical uncertainties are those achievable with sPHENIX for 22 weeks and 10 weeks of Au+Au and p+p data-taking, shown superimposed on the curve for $X = 0.85$. }
\label{Fig:sPHENIXfragfn}
\end{figure}
In detail, this future program includes:
\begin{enumerate}
\item Increased precision in extracting the average \qhat\ and \ehat\ 
transport coefficients, leading to a determination of their scale, energy and temperature dependence.
\item Combined global analysis of multiple observables at RHIC and the LHC 
to extract the temperature dependence of transport coefficients. Both
the RHIC and LHC final states represent an integral of jet-medium
interactions over the evolution of both the jet and the medium from initial
to final state. To disentangle the temperature dependence from
this evolution it will be essential to deploy directly comparable 
observables (theoretically and experimentally) in different 
QGP temperature regimes, in particular with respect to the fraction
of their evolution spend in the vicinity to the phase transition
region. Only a combined effort at RHIC and LHC can address this 
question.
\item Using increased systematic and statistical precision afforded
by new probes (e.g. photon-jet) to identity the medium response to
the modified jet radiation and further elucidate the liquid nature 
of the medium in its response to local perturbations
\item Using precision measurements of modifications of the jet structure 
in angular and momentum space to characterize the microscopic structure 
of the QGP. Jet probes here serve to perform experiments analogous to
Rutherford or deep-inelastic scattering off effective QGP constituents or 
quasi-particles. 
Perhaps the most straightforward signal is the modification of the 
jet fragmentation function $D(z)$, where $z = p/E$ is the momentum fraction 
of a single-particle of momentum $p$ in a jet of energy $E$. 
Other interesting observables include both potential modifications 
to the back-to-back jet scattering distributions, as well as modifications
of the intra-jet angular structure. For the latter, the correlated 
angular and momentum evolution of the jet from the initial scattering 
to the final hadronic structure probes probes a wide
range of scales, opening a window to interactions of jet 
and QGP constituents in between vacuum-like and in-medium cascade 
regimes. To pin down the physics of this intermediate window, 
systematic variations of both the jet conditions 
and medium conditions and dynamics are necessary. Combining RHIC and LHC 
measurements will allow control over initial density and temperature 
(in particular in respect to their vicinity to the critical temperature) 
and expansion dynamics of the system. The different energy regimes and 
tagging of particular initial states (photon+jet, $b$-tagged jets, 
multi-jet events) will allow selection of different or common 
jet populations in relation to different medium conditions. Success in 
this long-term endeavor will require a global analysis of 
a diverse set of RHIC and LHC data in an improved, well controlled
theoretical framework that makes explicit contact with the experimental observables.
\end{enumerate}

\subsubsection{Future challenges in the theory of jet modification}
\label{Sec:FutureJetTheoryChallenges}

In the period subsequent to the 2007 long range plan, there has been a major advance in the theoretical description 
of jet modifications in a dense medium. The ability of experiments both at RHIC and LHC to study full jet modification and 
energy flow with respect to the jet axis has led to the evolution from formalisms focused only on the leading particle towards full jet analyses tools.
Primarily, this has led to the ongoing development of several Monte-Carlo codes: YaJEM~\cite{Renk:2008pp,Renk:2010zx}, JEWEL~\cite{Zapp:2008gi}, MARTINI~\cite{Schenke:2009gb}, Q-PYTHIA~\cite{Armesto:2009fj}, PYQUEN~\cite{Lokhtin:2011qq}, and MATTER~\cite{Majumder:2013re}. While most of these routines are based on weak coupling, there is also a recently developed generator that is based on a hybrid strong and weak coupling approach~\cite{Casalderrey-Solana:2014bpa}. 

All of these approaches are based on one of the established analytical formalisms, and as such, apply to slightly different epochs in the lifespan of a hard jet in a dense medium. As a jet emanates from a hard interaction, it is far off its mass shell, at such hard scales that the jet probes the medium at extremely short distances characterized by its high-$Q^{2}$ 
structure of a dilute gas of partons. In this regime, it is expected that the parton's interactions with the medium are
dominated by radiation of gluons over elastic scattering processes.  
As the virtuality of the partons within a jet begins to drop, different parts of the jet enter different regimes.
The very energetic jet fragments undergo several scatterings per emission in which the multiple scattering prevents their virtuality from falling
below $\hat{q} \tau$, where $\tau$ is the lifetime of 
the parton.
As a result, those hard partons remain weakly coupled with the medium. 
The less energetic partons in the shower decrease in virtuality to the scale of the medium 
(of the order of the local temperature) and become strongly coupled. 
Currently most of the Monte Carlo descriptions apply to only one of these regimes, under the assumption that such a regime dominates the measured observables (with varying approximations regarding the medium). 
Nonetheless, several of these event generators have been tested against a variety of observables, 
such as jet $R_{AA}$,
dijet energy and angular imbalance, 
intra-jet hadron distribution and jet shapes~\cite{Renk:2013rla,Renk:2012cb,Ramos:2014mba,Zapp:2012ak,Young:2012dv,Majumder:2013re}. 
One of the outstanding challenges in this field will be the development of a generator that smoothly interpolates between the various regimes of jet quenching while incorporating a fluctuating medium simulated by an event-by-event viscous fluid dynamical simulation. This will be followed by rigorous testing and validation against all available 
jet data.

Full jet reconstruction, however, involves more than simply a perturbative redistribution of the energy within a jet cone.
Rather, as the energy is deposited within an evolving medium it thermalizes and forms a source of energy and momentum current for the fluid dynamical evolution of the medium. 
A significant fraction of the this energy is carried away from the jet at large angles, with the remainder found 
within the reconstructed jet cone. To date only initial efforts have been made to understand the dynamics of energy deposition and redistribution 
within a fluid medium~\cite{Neufeld:2009ep,Qin:2009uh}, and this remains an open issue. 
In addition to this question of energy redistribution at the medium scale, another outstanding question is the distribution of radiation at the perturbative scale. The ordering of radiation from a hard parton in vacuum is well established, however its modification in the medium is still an open question. There now exist two separate calculations, one in the radiation dominated regime~\cite{Fickinger:2013xwa} which shows no ordering, and one in the 
scattering dominated regime~\cite{Armesto:2011ir}, which shows anti-angular ordering. A resolution of these issues at NLO remains one of the major opportunities in the theory of 
pQCD energy loss. It may indeed turn out that a full understanding of this problem leads to the development of an effective theory of jet modification in dense matter. To date, 
great strides have been made in modifying Soft Collinear Effective Theory (SCET)~\cite{Bauer:2000yr,Bauer:2001yt,Bauer:2002nz} by the addition of off-shell gluon modes with momenta transverse to the collinear modes within a hard jet. 
This modified version of SCET has been uses to calculate the 
propagation, scattering and emission from hard partons in a dense medium~\cite{Idilbi:2008vm, DEramo:2010ak,Ovanesyan:2011xy}. Extending this beyond the radiation dominated regime to the multiple scattering regime remains a future goal. Several model calculations~\cite{CasalderreySolana:2011rq,Qin:2010mn} have already laid the groundwork for what a fully developed jet modification theory should look like.

Developing in parallel with the theoretical description of parton showers in a dense medium is the improved description of transport coefficients, which are the actual measurable quantities that describe the medium probed by hard jets. A great deal of work has gone into determining the temperature dependence of the transverse diffusion coefficient $\hat{q}$. 
More recently, efforts have been made to determine the longitudinal energy loss coefficient $\hat{e}$. 
While light flavor energy loss is known to be weakly dependent on $\hat{e}$, this is not the case for 
heavy flavors. 
Significant theoretical activity now focuses on determining the dynamics of heavy flavor energy loss 
and the sensitivity of these and other heavy flavor observables
to $\hat{e}$~\cite{Qin:2009gw,Djordjevic:2013pba,Abir:2014sxa}. 
Efforts to extend this to $b-$tagged jets are also being carried out~\cite{Huang:2013vaa}. 
Future developments in the theory of heavy flavor dynamics in the medium serve not only as a consistency check for the pQCD based formalism of jet modification, but also provide the 
primary means to determine the drag coefficient $\hat{e}$.

While the dependencies of the transport coefficients on temperature reveal the dynamics of the medium as seen by the jet, they also depend on the scale and energy of the jet. 
Recently there has been a series of developments to quantify this scale and energy dependence of transport coefficients by carrying out NLO calculations of these coefficients, 
most notably of $\hat{q}$. Several calculations, once again performed in different regimes of jet quenching, have obtained rather different dependencies on energy and 
scale~\cite{Liou:2013qya,Kang:2013raa,Blaizot:2014bha,Iancu:2014kga}. Much theoretical effort is currently being devoted to a resolution of these differences. 
In all of the current jet quenching calculations, either of leading particles or of full jets, the normalization of the transport coefficients is determined
empirically by fitting to one data set. 
While this emphasizes the absolute requirement of parallel theoretical and experimental investigations,
it also demonstrates that the current calculations of jet modifications do not result directly from the QCD Lagrangian. In an effort to 
resolve this, several groups have considered the possibility of evaluating such coefficients 
on the lattice~\cite{Majumder:2012sh,Panero:2013pla,Ji:2013dva}. These represent 
extremely difficult calculations which, however, hold the promise of determining the transport coefficients with no input other than the local temperature. 
Future calculations of transport coefficients on the lattice, combined with calculations of the 
scale and energy dependence described above will allow for a rigorous and first principles test of the entire formalism of jet modification.

\subsubsection{Future Quarkonia measurements at RHIC and the LHC}
\label{Sec:FutureQuarkonia}

%
%

\begin{figure}[!pht]
  \centering
      \includegraphics[width=\textwidth]{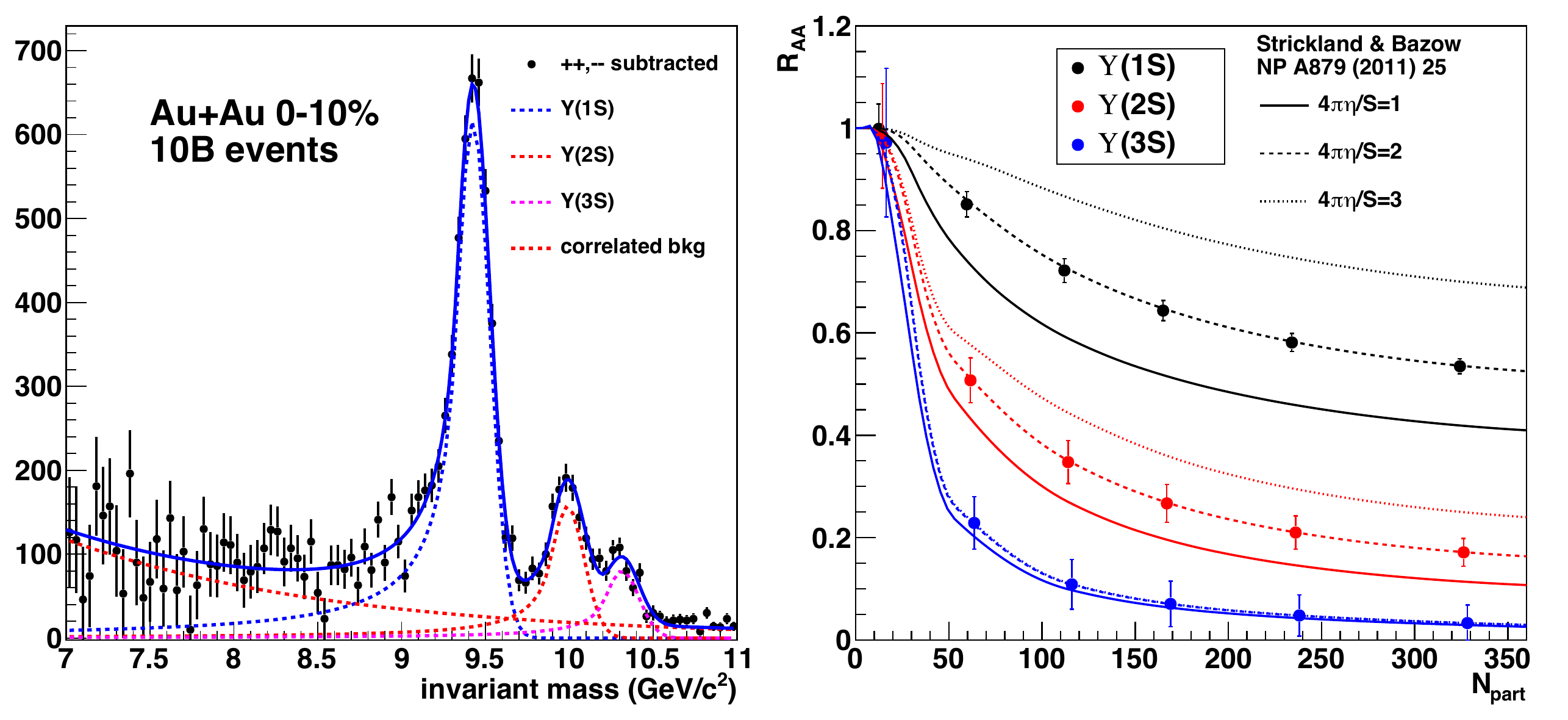}
      \caption[Projected quarkonia results from a 20 week \AuAu\ run
        with sPHENIX]{Projected quarkonia results from a 20 week \AuAu\ run
        with sPHENIX.  (left) The di-electron invariant mass
        distribution for 0--10\% central \AuAu\ events after the
        combinatorial background has been removed by subtracting all
        like-sign pairs. (right) Estimate of the statistical
        precision of a measurement of the $\Upsilon$ states assuming
        that the measured $R_{AA}$ is equal to the results of a
        recent theory calculation~\cite{Strickland:2011aa}.}
\label{Fig:upsilon_auau_0-10pc}
\label{Fig:upsilon_raa}
\end{figure}

In addition to the jet capabilities added at RHIC by the sPHENIX detector upgrade, 
outlined in Section~\ref{Sec:FacilitiesFuture}, high precision
measurements of the modification of the $\Upsilon(1S)$, $\Upsilon(2S)$ and $\Upsilon(3S)$
states in A+A and $p$+A collisions will also be provided by sPHENIX. The key features 
of the upgrade that enable this are:

\begin{itemize}

\item The 1.5 T magnetic field of the Babar magnet, combined with precise tracking, will 
provide 100 MeV mass resolution for measurements of $\Upsilon \rightarrow e^+e^-$ decays
at mid rapidity.

\item Good signal to background performance is obtained by using low mass tracking to limit 
radiative losses for electron tracks and the Electromagnetic and Hadronic calorimeters to 
reject combinatoric background from charged hadrons.

\item Increased RHIC luminosity combined with the large acceptance of sPHENIX
and long RHIC running times provide the statistical precision for the $\Upsilon$ measurements at RHIC that will
tightly constrain theoretical models of the modification. These measurements will 
complement similar high precision measurements at higher temperature from LHC 
experiments that will be available on a similar time scale (i.e. by the end of LHC Run 3) 
due to LHC machine and detector upgrades, and accumulated luminosity.

\end{itemize}

Critical variables to manipulate when probing the QGP are the temperature of
the QGP and the length scale probed in the medium. For this reason, measurements 
at two widely different temperatures of the three Upsilon states, which span a large range of 
binding energies and sizes, are ideal. However for comparisons between data at RHIC and the LHC
 to be effective, high precision is required at both facilities.

The planned program of measurements at RHIC starting in 2021 includes \pp\,
\pAu\ and \AuAu\ measurements. A very important feature for $\Upsilon$ 
measurements with sPHENIX is that the increases in \pp\ luminosity at RHIC will permit a
measurement of the \pp\ reference cross section having similar precision to that which can be
attained in \AuAu\ and \pAu\ collisions. The projected invariant mass spectrum for central collisions
from a 10 week \AuAu\ run is shown in Figure~\ref{Fig:upsilon_auau_0-10pc}. This spectrum is shown
after the combinatorial background has been subtracted, and assumes no modification of the yields relative
to \pp\ collisions.

The projected statistical precision for measuring the nuclear suppression factor $R_{AA}$ in a 20 week \AuAu\ run, using \pp\
reference data also from a 10 week run, is illustrated in Figure~\ref{Fig:upsilon_raa}. 
For the sake of this illustration it is assumed that the suppression for each state
is equal to that from a theory calculation~\cite{Strickland:2011aa} in which the shear viscosity to
entropy density ratio is a parameter. The data are expected to provide good constraints on models.

At the LHC, CMS has measured the $\Upsilon$ modification in $\sqrt{s_{NN}}$=2.76 GeV Pb+Pb
collisions with mass resolution that cleanly resolves the three states. It is expected that CMS will 
accumulate (at $\sqrt{s_{NN}}$=5.5 TeV) by the end of Run 3 an $\Upsilon$ data set that is roughly 
100 times larger than their existing one, yielding statistical precision that is even better than that expected 
from sPHENIX. 

The different color screening environments caused by the different temperatures attained in collisions at RHIC 
and LHC, combined with large differences in the time evolution of the QGP and in the underlying bottom
production rates, make them distinctly different laboratories for studying the effect of the plasma on the 
$\Upsilon$ states. The combination of high precision $\Upsilon$ data from the LHC and RHIC will 
constrain theoretical models in ways that data measured at only one energy could not.

%% file: tex/Theory.tex
\subsection{Towards Quantitative Understanding: Opportunities and Challenges in Theory}
\label{Sec:Theory}

Part of Recommendation IV of the 2007 Long Range Plan was the appreciation that 
{\em ``achieving a quantitative understanding of the properties of the quark-gluon plasma 
also requires new investments in modeling of heavy-ion collisions, in analytical approaches, and in large-scale computing.''} Since then there has been tremendous progress along these lines.
A large and diverse worldwide theory community is working on the challenges
posed by the discoveries of the experimental heavy ion programs at RHIC and at the LHC.
This community develops theoretical tools suited for the upcoming era of detailed experimental
investigation. It includes, amongst others, lattice QCD groups producing
{\em ab initio} calculations of QCD thermodynamics
at  finite temperature and density, nuclear and high energy physicists 
working on the embedding of hard partonic processes in a dense nuclear environment, groups advancing
the development of relativistic fluid dynamic simulations of heavy ion collisions and the interfacing of these
simulations with hadronic cascades, field theorists aiming at developing a description from first principles of
the initial conditions of high parton density and their (non-equilibrium) evolution, 
as well as people developing many-body approaches to evaluate spectral and transport properties of QCD matter,
and string theorists contributing
to the exploration of novel strong coupling techniques suited for the description of strongly coupled, 
nearly perfectly liquid, non-abelian plasmas. 
This diverse theory community is active and forward looking;
it supports, advances and motivates a multifaceted
experimental program, and does so 
with a long perspective. It develops improved 
phenomenological tools that address with increased precision and broadened versatility the diverse needs of 
the experimental program and mediates its impacts that branch out into neighboring fields of theoretical physics,
including high energy physics, string theory, condensed matter physics 
and astrophysics/cosmology. Here, we highlight only a few 
important recent developments that support these general statements:

\begin{itemize}

\item
Convergence has been reached in lattice QCD calculations of the temperature for the crossover 
transition in strongly interacting matter which has now been established 
at $145\,\mathrm{MeV}{\,<\,}T_c{\,<\,}163\,\mathrm{MeV}]$~\cite{Aoki:2006br,Aoki:2009sc,Bazavov:2011nk,Bazavov:2014pvz}. Continuum extrapolated results for the equation of state, the speed of sound and many other properties 
of strong interaction matter have also been provided \cite{Borsanyi:2013bia,Bazavov:2014pvz}.

\item 
The modeling of the space-time evolution of heavy-ion collisions has become increasingly reliable. (2+1)-dimensional, and subsequently, (3+1)-dimensional relativistic viscous fluid dynamics computations have been performed. 
All such computations use an equation of state extracted from lattice QCD.
Viscous relativistic fluid dynamic (3+1)-dimensional
simulation tools have been developed and subjected to a broad set of 
theoretical precision tests. These tools are instrumental in the ongoing program of extracting 
material properties of the produced  
quark-gluon plasma
from the experimentally observed flow harmonics and reaction plane correlations. 
Within the last five years, in a community-wide effort coordinated by, amongst others,
the TECHQM\cite{TECHQM} initiative, these simulation codes were validated against each other. The range of
applicability of these simulations continues to be pushed to further classes of experimental observables.
Still, already with the limited data/theory comparison tools that have so far been brought to bear on the large sets of experimental data collected at RHIC and LHC, the specific shear viscosity $\eta/s$ of QCD matter created at RHIC could be constrained to be approximately 50\% larger than the limiting value $1/4\pi=0.08$ obtained in strongly coupled plasmas
with a dual gravitational description~\cite{Kovtun:2004de}, 
and to be about 2.5 times larger than this value at the LHC, see e.g.~Ref.~\cite{Gale:2012rq}. 

\item 
The JET collaboration\cite{JET} has coordinated a similarly broad cross-evaluation of the tools available
for the description of jet quenching in hadron spectra,
and have undertaken a
consolidation of the results of different approaches to 
determining the transport properties of jets as they traverse the strongly correlated quark-gluon plasma.
The range of values for the jet quenching parameter $\hat{q}/T^3$ obtained from  
theory-data comparisons has been narrowed to
$2{\,<\,}\hat{q}/T^3{\,<\,}6$ within the temperature range probed by RHIC and the LHC~\cite{Burke:2013yra}.
At the same time, a significant number of 
tools were developed for the simulation of full medium-modified parton showers suited for the modeling
of reconstructed jets. In the coming years, these tools will be the basis for a detailed analysis of jet-medium
interactions.  

\item 
There is a community-wide effort devoted to extending CGC calculations from LO
 (current phenomenology) to 
NLO~\cite{Balitsky:2008zza,Chirilli:2011km,Stasto:2013cha,Beuf:2014uia,Kang:2014lha}.
Doing NLO calculations in the presence of a non-perturbatively large parton density requires overcoming
qualitatively novel, conceptually challenging issues that are not present in standard {\em in vacuo} NLO calculations in QCD. Within the last
year, the key issues in this program have been addressed by different groups in independent but consistent
approaches, and the field is now rapidly  advancing these calculations of higher precision to a practically
usable level. 

\item 
In recent years, there have been significant advances in understanding how thermalization occurs in the
initially overoccupied and strongly expanding systems created in heavy ion 
collisions~\cite{Berges:2013eia,Gelis:2013rba,Kurkela:2014tea}. While some of these
developments are still on a conceptual field theoretical level, there is by now the exciting realization that the thermalization
processes identified in these studies share many commonalities with the problem of dynamically describing the
quenching of jets in dense plasmas~\cite{Blaizot:2013hx,Kurkela:2014tla}. 
This is likely to open new possibilities for understanding via the detailed measurements
of jet quenching how non-abelian equilibration processes occur in primordial plasmas. 

\item 
Systematic efforts are being pursued to unravel key properties of QCD matter with heavy-flavor particles. The 
construction of heavy-quark effective theories benefits from increasingly precise information from thermal lattice QCD, to 
evaluate dynamical quantities suitable for phenomenology in heavy-ion collisions (heavy-flavor diffusion coefficient, 
quarkonium spectral properties). This will enable precision tests of low momentum heavy-flavor observables, providing a 
unique window on how in-medium QCD forces vary with temperature.

\item
Much progress has been made towards a systematic understanding from first principles of the 
properties of strongly interacting matter at non-zero baryon number density. Such studies 
rely heavily on the development of theoretical concepts on critical behavior 
signaled by conserved charge fluctuation~\cite{Stephanov:1998dy,Ejiri:2005wq,Stephanov:2011pb}. 
They are accessible to lattice QCD calculations which opens up the possibility, via dynamical modeling, 
for a systematic comparison of experimental fluctuation observables with calculations performed in 
QCD~\cite{Karsch:2012wm,Bazavov:2012vg,Mukherjee:2013lsa,Borsanyi:2013hza,Borsanyi:2014ewa}. This 
will greatly profit from the steady development of computational facilities which are soon expected to deliver 
sustained petaflop/s performance for lattice QCD calculations.

\end{itemize}


The significant advances listed above document how theory addresses the challenge of keeping pace with the
experimental development towards more complete and more precise exploration of the hot and dense rapidly 
evolving systems produced in heavy ion collisions. We emphasize that all these research directions show strong 
potential for further theoretical development and improved interfacing with future experimental analyses.
Some of the challenging issues over which we need to get better control include: 
i) the pre-equilibrium ``glasma" 
dynamics of coherent gluon fields, and the approach to thermalization; 
ii) the extraction of the values and temperature dependences of transport parameters that reflect the many-body QCD dynamics in deconfined matter; 
iii) the initial conditions at lower collision energies where the Glasma framework breaks down; 
iv) the proper inclusion of the physics of hydrodynamic fluctuations; 
v) an improved treatment of hadron freeze-out and the transition from hydrodynamics  to transport theory, in particular the treatment of viscous corrections that can influence the extraction from data of the physics during the earlier collision stages. 
Quantitative improvements in these aspects of the dynamical modeling of a heavy-ion collision will lead to increased precision in the extraction of the underlying many-body QCD physics that governs the various collision stages. Additional conceptual advances in our understanding of QCD in matter at extreme temperatures and densities are required to answer a number of further outstanding questions. We  list a few of them:

 
\begin{itemize}

\item
A complete quantitative understanding of the properties of the nuclear wave functions that are resolved in nucleus-nucleus and proton-nucleus collisions remains elusive to date. Progress requires the extension of computations of the energy evolution of these wave functions in the Color Glass Condensate (CGC) framework to next-to-leading logarithmic accuracy as described above, matching these to next-to-leading order perturbative QCD computations at large momenta, and pushing the development of these
calculations into predictive tools. 
Simultaneously, conceptual questions regarding the factorization and universality of distributions need to be addressed for quantitative progress. These ideas will be tested in upcoming proton-nucleus collisions at RHIC and the LHC, and with high precision at a future EIC. 

\item
How the glasma thermalizes to the quark-gluon plasma is not well understood. There has been significant progress in employing classical statistical methods and kinetic theory to the early stage dynamics --- however, these rely on extrapolations of weak coupling dynamics to realistically strong couplings. Significant insight is also provided from extrapolations in the other direction --- from large couplings -- using the holographic AdS/CFT correspondence between strongly coupled ${\cal N}{\,=\,}4$ supersymmetric Yang-Mills theory in four dimensions and weakly coupled gravity in an AdS$_5{\times}$S$_5$ space. Significant numerical and analytical progress can be anticipated in this fast evolving field of non-equilibrium non-Abelian plasmas, with progress on
the question of how
characteristic features of thermalization processes can be constrained in an interplay between experimental and
 theoretical developments. 

\item
A novel development in recent years has been the theoretical study of the possible role of quantum anomalies in heavy-ion collisions. A particular example is the Chiral Magnetic Effect (CME), which explores the phenomenological consequences of topological transitions in the large magnetic fields created at early times in heavy-ion collisions. How the sphaleron transitions that generate topological charge occur out of equilibrium is an outstanding question that can be addressed by both weak coupling and holographic methods. Further, the effects of these charges can be propagated to late times via anomalous hydrodynamics. While there have been hints of the CME in experiments, conventional explanations of these data exist as well. For the future beam energy scan at RHIC, quantifying the predictions regarding signatures of quantum anomalies is crucial. This requires inclusion of the anomalies into the standard 
hydrodynamical framework. We note that the study of the CME has strong cross-disciplinary appeal, with applications in a number of strongly correlated condensed matter systems. 


\item
As observed in Section~\ref{Sec:HardProbes}, progress has been made in quantifying the jet quenching parameter $\hat{q}$, which characterizes an important feature of the transverse response of the quark-gluon medium. However, significant challenges persist.  Another important transport parameter $\hat{e}$, characterizing the longitudinal drag of the medium on the hard probe, also needs to be quantified. Much recent theoretical effort has gone into extending the splitting kernel for gluon radiation by a hard parton traversing a dense medium to next-to-leading-order accuracy. In this context Soft Collinear Effective Theory (SCET), imported from high energy theory, has proven a promising theoretical tool whose potential needs to be further explored. There have been recent theoretical developments in understanding how parton showers develop in the quark-gluon medium; confronting these with the available jet fragmentation data requires their implementation in Monte-Carlo codes coupled to a dynamically evolving medium, 
and
the community-wide validation of  these jet quenching 
event generators.
There have been recent attempts to compute the jet quenching parameter using lattice techniques; while very challenging, such studies provide a novel direction to extract information on the non-perturbative dynamics of the strongly correlated quark-gluon plasma.

\item
Quarkonia and heavy flavor, like jets, are hard probes that provide essential information on the quark-gluon plasma on varied length scales. Further, the two probes find common ground in studies of b-tagged and c-tagged jets. In proton-proton and proton-nucleus collisions, non-relativistic-QCD (NRQCD) computations are now standard, and these have been extended to nucleus-nucleus collisions, even to next-to-leading order accuracy. Lattice studies extracting quarkonium and heavy-light meson spectral functions have increased in sophistication, and clear predictions for the sequential melting of quarkonium states exist and need to be confronted with experiment. The direct connection to experiment requires, however, considerable dynamical modeling effort. 
For instance, the question of how the fluid dynamic evolution can be
interfaced with microscopic probes that are not part of the fluid, such as charm and beauty quarks or jets, and
how the yield of electromagnetic processes can be determined with satisfactory precision within this framework, 
are questions of high phenomenological relevance for the experimental program in the coming years, and
the community is turning now to them. 

\item
An outstanding intellectual challenge in the field is to map out the QCD phase diagram. 
We have described the path toward this goal that starts from experimental measurements made
in a beam energy scan in Section~\ref{Sec:CP}.
While the lattice offers an {\it ab initio} approach, its successful implementation is beset by the well known sign problem, which is also experienced in other branches of physics. Nonetheless, approaches employing reweighting and Taylor expansion techniques have become more advanced and are now able to explore the equation of state and freeze-out 
conditions at baryon chemical potentials $\mu_B/T\le 2$. This covers a large part of the energy range currently explored in the 
RHIC Beam Energy Scan and suggests that a possible critical endpoint may only be found at beam energies less than 20~GeV. 
Other promising approaches include the complex Langevin approach \cite{Aarts:2009uq,Aarts:2014kja} and the integration over a Lefschetz thimble \cite{Cristoforetti:2012su,Aarts:2014nxa}. There has been considerable work outlining the phenomenological consequences of a critical point in the phase diagram. However, quantitative modeling of how critical fluctuations affect the measured values of the relevant observables will require the concerted theoretical effort sketched in Section~\ref{Sec:CP}.

\end{itemize}




Continued support of these theory initiatives is needed to optimally exploit the
opportunities arising from the continued experimental analysis of heavy ion collisions 
and to interface the insights so obtained with the widest possible cross-section of
the worldwide physics community.
Achieving the impressive intellectual achievements we have 
outlined, and meeting the challenges ahead, depend strongly on the development of the theory of 
strongly interacting matter which involves advances in heavy ion phenomenology, 
perturbative QCD, lattice QCD, holographic calculations of equilibration in 
strongly coupled systems, and effective field theories for QCD as well as the strong synergy with 
overlapping and related areas in high energy physics, condensed matter physics, cold atom physics, string theory and studies of complex dynamical systems. 

Continued advances will also require significantly increased computational resources. 
While lattice QCD continues to play a crucial role by delivering first-principles answers to important questions that require a non-perturbative approach, it does not yet have the ability to address dynamic systems.
Questions such as how the matter formed in heavy ion collisions reaches local thermal equilibrium, 
and how it subsequently evolves to the final state observed in experiments
require sophisticated frameworks that incorporate realistic initial conditions, the interactions of hard processes with the medium, 
and 3+1 viscous hydrodynamics coupled to hadronic transport codes.
Over the last few years, 
the community has developed an arsenal of highly sophisticated dynamical evolution codes 
that simulate the underlying physical mechanisms with unprecedented accuracy 
to provide quantitative predictions for all experimentally accessible observables, 
but at the expense of a huge numerical effort.
In most cases the limiting factor in comparing the output of such calculations to the data
is the accuracy obtainable with the currrently available computational resources rather than the statistical precision of the experimental data. 
Addressing this requires new investments, especially in capacity computing that supplement 
the necessary expansion and upgrades of leadership-class computing facilities.
Specific information on the needed resources for both lattice QCD and dynamical modeling may be found in the report of the Computational Nuclear Physics Meeting Writing Committee\cite{CompReport}, whose 
recommendations we fully endorse.

%% file: tex/Summary.tex
\section{Summary}
\label{Sec:Summary}
The previous sections of this white paper have outlined the enormous scientific opportunities in the study of thermal QCD that await exploration in the next decade. Realizing these opportunities on a timely basis will require a dedicated program, key elements of which would include

\begin{enumerate}

\item Detector upgrades to enable 
\begin{itemize}
   \item The second phase of the Beam Energy Scan program at RHIC, utilizing the planned increase in low-energy luminosity from electron cooling at RHIC (see Section~\ref{Sec:RHICUpgrades}),
    to explore the phase diagram of nuclear matter,
    measure the temperature and chemical potential dependence of transport properties,
     and continue the search for the QCD critical point. 
   \item State of the art jet measurements at RHIC to understand the underlying degrees of freedom that create the near-perfect liquidity of the QGP.
   \item Extension of the heavy ion capabilities of the LHC detectors in order to study the QGP at the highest temperatures and densities. 
\end{itemize}

\item Commitments to the RHIC campaigns and the LHC heavy ion runs outlined in Section~\ref{Sec:FacilitiesFuture} to ensure timely availability of new data as the enhanced and upgraded detectors become available. 

\item A strong theory effort containing the following items:

\begin{itemize}
\item Strong continued support of the core nuclear theory program supporting university PI's, national lab groups and the national Institute for Nuclear Theory (INT), which in concert generate key ideas that drive the field and train the next generation of students and post-doctoral fellows.

\item Strong continued support of the DOE Early Career Award (ECA) program in Nuclear Theory, as well as the NSF Early Career Development (CAREER) and Presidential Early Career (PECASE) award programs to recognize and promote the careers of the most outstanding young nuclear theorists. 

\item Strong support of expanded computational efforts in nuclear physics, as outlined in the Computational Nuclear Physics white paper. 

\item Continuation and expansion of the Topical Research Collaboration program, since 
thermal QCD features several outstanding challenges that require the synthesis of a broad range of expertise, and which could strongly benefit from an expansion of the Topical Collaboration program.
\end{itemize}
\end{enumerate}

Pursuing this program over the next decade will consolidate our knowledge of the thermal properties of QCD, the only gauge theory of nature amenable to experimental study in both the strongly and weakly coupled regimes. In addition, it is likely that the new insights that will emerge from this 
process will surprise us, just as the 
initial studies of truly relativistic heavy ion collisions 
led to paradigm shifts in our understanding 
of fundamental aspects of QCD.

%% file: tex/Acknowledgements.tex
\section{Acknowledgements}
\label{Sec:Acknowledgement}
The members of the Hot QCD Writing Group wish to acknowledge the thoughtful comments, remarks, input and materials provided by many members of our community. Special thanks go to 
Abhay Deshpande,
Carl Gagliardi,
Frank Geurts, 
Frithof Karsch,
Anthony Kesich,  Krishna Kumar, Marco van Leeuwen, 
Zein-Eddine Meziani,
Richard Milner,
Berndt M\"{u}ller,
Michael Murray, 
Jamie Nagle, 
Jianwei Qiu,
Lijuan Ruan, 
Daniel Tapia Takaki, 
Julia Velkovska, 
Raju Venugopalan
and Zhangbu Xu.
We would also like to take this opportunity to thank 
Jim Napolitano (Co-Chair),
Bernd Surrow (Co-Chair),
Jeff Martoff,
Andreas Metz,
Zein-Eddine Meziani
and 
Nikos Spaveris 
for their very effective organization and hosting
of the QCD Town Meeting at Temple University in September 2014.